\documentclass[secnumeq]{FBSart}
\usepackage{amsfonts}
\usepackage{amssymb}
\usepackage{amsmath}
\usepackage{epsfig,bm}

\newcommand{\vi}[1]{\mbox{\boldmath $#1$}}
\newcommand{\vis}[1]{\mbox{\boldmath ${\scriptstyle #1}$}}

\title{Global-Vector Representation of the Angular Motion of
Few-Particle Systems II}
\author{Y. Suzuki\instnr{1}, 
W. Horiuchi\instnr{2}\thanks{Supported by the JSPS Research 
Fellowships for Young Scientists}, M. Orabi\instnr{2,3}, and K. Arai\instnr{4}}
\instlist{Department of Physics, and Graduate School of 
Science and Technology, Niigata University, Niigata
950-2181, Japan 
\and Graduate School of 
Science and Technology, Niigata University, Niigata
950-2181, Japan
\and Department of Physics, Cairo University, Giza 12613, Egypt
\and Division of General Education, Nagaoka National College of
Technology, Nagaoka 940-8532, Japan}

\runningauthor{Y. Suzuki, W. Horiuchi, M. Orabi, and K. Arai}
\runningtitle{Global-Vector Representation of the Angular Motion of
Few-Particle Systems II}
\begin{document}

\maketitle

\begin{abstract}
The angular motion of a few-body system is described with global 
vectors which depend on the positions of the particles. The previous 
study using a single global vector 
is extended to make it possible to describe both natural and unnatural 
parity states. Numerical examples include three- and four-nucleon 
systems interacting via nucleon-nucleon 
potentials of AV8 type and a 3$\alpha$ 
system with a nonlocal $\alpha\alpha$ potential. The results 
using the explicitly correlated Gaussian basis with 
the global vectors are shown to be in good agreement with those of 
other methods. A unique role of the unnatural parity component, 
caused by the tensor force, is clarified 
in the $0^-_1$ state of $^4$He. Two-particle 
correlation function is calculated in the coordinate and momentum spaces 
to show different characteristics of the interactions employed. 
\end{abstract}

\section{Introduction}

Realistic nucleon-nucleon potentials contain strong tensor components. 
The tensor force induces the coupling of different orbital and spin 
angular momenta in the wave function of a few-nucleon system. Because of 
this property a variational solution of the system faces 
complexities in describing the orbital motion. There are two widely 
used methods to expand the orbital part of the wave function. One is a 
hyperspherical harmonics (HH) expansion~\cite{viviani1,viviani2}, 
and the other is a 
successive coupling of partial waves for the relative motion of the 
particles~\cite{gloeckle,kamimura,oba,few.electron,bromley}. 
A converged solution is attained by increasing the orbital angular momenta. 

Another representation for the orbital motion, proposed in 
refs.~\cite{vs95,book}, is to use a global vector (GV) 
which is defined by a 
linear combination of the relative coordinates. The coefficients of the 
combination determine the vector responsible for the rotation of 
the system. The spherical part 
of the orbital motion is expressed in terms of an 
explicitly correlated Gaussian (CG). 
The efficiency of the GV representation (GVR) has been tested in many 
examples~\cite{suv98,vsu98,vus98,nas02} including not only nuclear 
but also atomic and 
hypernuclear physics problems. Its accuracy 
was confirmed in ref.~\cite{vsu98} 
by comparing with the partial-wave expansion (PWE) 
for $^3$H and $\alpha$-particle interacting 
with central and tensor forces. In these studies only 
a single GV is used, and its application is thus restricted to 
natural parity states. 

The purpose of this paper is to extend the GVR to 
describe the orbital motion with unnatural parity and to test the 
power of its representation. The unnatural parity state 
is usually unfavorable to gain the energies of low-lying levels, but it 
is needed to reach a precise energy particularly when the noncentral 
forces are taken into account. For example, the ground state 
of $^4$He can be specified by three channels with $(L, S)$=(0,0), (1,1) 
and (2,2), where $L$ and $S$ are respectively the total orbital and spin 
angular momenta. Here the channel with $L$=1 and positive parity 
is unnatural, and its mixing probability is of order only 
0.37\,\%~\cite{benchmark}. However, the $L$=1 unnatural component 
gives rise to a contribution of more than 1\,MeV to the ground state 
energy. 

It is not a surprise that the interaction between composite 
particles like the nucleon becomes nonlocal~\cite{ppnp}, 
and indeed there is now 
available a new class of nonlocal nucleon-nucleon interactions, e.g.,  
N$^3$LO~\cite{entem} and low-$k$~\cite{bogner} potentials. 
The nonlocal operator also appears when one needs to eliminate some 
redundant or Pauli-forbidden states from the relative motion 
between composite particles~\cite{kukulin}. Thus we take a 
certain class of 
nonlocal operators into consideration in the present paper. 

A simple method of incorporating the unnatural parity state is 
to introduce two GVs. This was performed in the case of 
central forces and successfully applied to a search for excited states of 
a positronium molecule~\cite{su00}. Here we develop 
a method of calculating the matrix elements of not only noncentral 
forces but also other local operators of physical interest such 
as a multipole density and some nonlocal operators.  

The paper is organized as follows.  The CG 
basis with two GVs is defined in Sect.~\ref{sect.2} and 
its generating function is introduced. A nice property 
of the CG basis is that its Fourier transform is again the CG in momentum 
variables. A basic method  
of calculating matrix elements is explained in Sect.~\ref{sect.3}. 
The matrix element of a nonlocal operator between the generating
functions is given in Sect.~\ref{sect.4}. In Sect.~\ref{sect.5} we 
present results calculated for three- and four-nucleon systems 
using the two GVs and compare them to other calculations. 
As an example of the nonlocal potential, we consider a 3$\alpha$ system 
interacting via the nonlocal $\alpha$-$\alpha$ potential derived 
from the resonating group method (RGM)~\cite{tang}. A summary 
is drawn in Sect.~\ref{sect.6}. 
Formulas used in the present calculation are collected in
Appendices. We give the CG basis in the momentum space in Appendix A, 
present the matrix elements of physically important 
operators in Appendix B, discuss the integral transform of the CG basis 
in Appendix C, and derive the matrix elements for 
some nonlocal operators in Appendix D.

\section{Correlated Gaussian basis with double global vectors}
\label{sect.2}

The basis function in $LS$ coupling scheme takes a form  
\begin{equation}
\Phi_{(LS)JM_JTM_T}=[\psi_L^{(\rm space)}\psi_S^{(\rm spin)}]_{JM_J}
\psi_{TM_T}^{(\rm isospin)},
\label{basis.LS}
\end{equation}
where the square bracket $[\ldots]$ stands for the angular momentum
coupling. The spin and isospin parts are expanded using the basis 
of successive coupling, e.g.,  
\begin{equation}
\psi_{SM_S}^{(\rm spin)}=\big|[\cdots [[[\textstyle{\frac{1}{2}}
\textstyle{\frac{1}{2}}]_{S_{12}}
\textstyle{\frac{1}{2}}]_{S_{123}}]\cdots]_{SM_S}\rangle.
\end{equation}
Here the set of intermediate spins $(S_{12},S_{123},\ldots)$ takes all 
possible values for a given $S$. 
The orbital part $\psi_{LM}^{(\rm space)}$ is expanded in terms of 
the CG basis which is explained in detail below. The symmetry of 
identical particles has to be imposed on the basis function.

\subsection{Generating function for correlated Gaussian basis}
Assuming that the system consists of $N$ particles, we denote 
the single-particle coordinate of particle $i$ by ${\vi r}_i$. 
It is convenient to introduce 
relative and center of mass coordinates, denoted ${\vi x}_i$, 
to describe intrinsic excitations of the system. Both coordinates 
are related to each other by a linear 
transformation
\begin{equation}
{\vi x}_i=\sum_{j=1}^{N}U_{ij}{\vi r}_j,\ \ \ \ \ 
{\vi r}_i=\sum_{j=1}^{N}(U^{-1})_{ij}{\vi x}_j,
\label{def.matU}
\end{equation}
where ${\vi x}_N$ is chosen to be the center of mass coordinate. 
Let ${\vi x}$ denote an $(N\!-\!1)$-dimensional column vector,  
excluding ${\vi x}_N$, whose $i$th element is a usual 3-dimensional 
vector ${\vi x}_i$. 
A choice of ${\vi x}$ often employed is the Jacobi set of coordinates, 
but may be any other set of relative coordinates. 

A natural parity basis with the orbital angular momentum $LM$ 
and parity $(-1)^L$ is described as 
\begin{equation}
{\rm exp}\left(-{\frac{1}{2}}{\widetilde{{\vi x}}} A {\vi x}\right)
{\cal Y}_{LM}({\widetilde{u_1}}{\vi x}),
\label{cg1}
\end{equation}
where 
${\cal Y}_{LM}({\widetilde{u_1}}{\vi x})$ is a solid spherical harmonics 
\begin{equation}
{\cal Y}_{LM}({\widetilde{u_1}}{\vi x})=|{\widetilde{u_1}}{\vi x}|^{L}
Y_{LM}(\widehat{{\widetilde{u_1}}{\vi x}})
\label{ylm}
\end{equation}
which is specified by a single GV, 
$\widetilde{u_1}{\vi x}\!=\!\sum_{i=1}^{N-1}u_{1_i}{\vi x}_i$. 
The symbol $\widetilde{\ \ }$ denotes the transpose of a matrix. 
In Eq.~(\ref{cg1}) $A$ is an $(N\!-\!1)\!\times \!(N\!-\!1)$ 
positive-definite, 
symmetric matrix, and ${\widetilde{{\vi x}}} A {\vi x}$ 
is a short-hand notation for 
$\sum_{i,j=1}^{N-1}A_{ij}{\vi x}_i\cdot {\vi x}_j$.
The matrix $A$ and the $(N\!-\!1)$-dimensional column vector 
$u_1$ are parameters to characterize the ``shape'' of the basis 
function. The function (\ref{cg1}) is a natural extension of 
$\exp(-\frac{1}{2}ar^2){\cal Y}_{\ell m}(\vi r)$ for a single variable 
case to many-particle system. 

Another extension commonly used 
is a successive coupling of the partial waves corresponding to the 
respective coordinates, namely, $\exp(-\frac{1}{2}a_1{\vi x}_1^2
\!-\!\frac{1}{2}a_2{\vi x}_2^2\!-\!\cdots)$ 
$[\cdots[[{\cal Y}_{\ell_1}({\vi x}_1)
{\cal Y}_{\ell_2}({\vi x}_2)]_{L_{12}}{\cal Y}_{\ell_3}({\vi
x}_3)]_{L_{123}}
\cdots]_{LM}$. The basis function (\ref{cg1}) 
was compared to that of PWE~\cite{vsu98} 
and the GVR is found to be an excellent alternative to describe the 
rotational motion. The angular part 
was extended in refs.~\cite{suv98,vsu98} to 
$|{\widetilde{u_1}}{\vi x}|^{2K}{\cal Y}_{LM}({\widetilde{u_1}}{\vi
x})$, which was found to be advantageous to cope with 
short-ranged repulsive forces.  We use the form 
(\ref{ylm}) in this paper, however, since it 
greatly simplifies the calculation of matrix elements. 

An unnatural parity basis with the angular momentum 
$LM$ and parity $(-1)^{L+1}$ is described using 
two GVs as 
\begin{equation}
{\rm exp}\left(-{\frac{1}{2}}{\widetilde{{\vi x}}} A {\vi x}\right)
[{\cal Y}_{L}({\widetilde{u_1}}{\vi x})
{\cal Y}_{1}({\widetilde{u_2}}{\vi x})]_{LM},
\label{cg2}
\end{equation}
where $u_2$ is an $(N\!-\!1)$-dimensional column vector which defines the 
second GV. Both forms of  
Eqs.~(\ref{cg1}) and (\ref{cg2}) are unified as 
\begin{equation}
F_{(L_1L_2)LM}(u_1,u_2,A,{\vi x})
= {\rm exp}\left(-{\frac{1}{2}}{\widetilde{{\vi x}}} A {\vi x}\right)
[{\cal Y}_{L_1}({\widetilde{u_1}}{\vi x}) 
{\cal Y}_{L_2}({\widetilde{u_2}}{\vi x})]_{LM}.
\label{cg}
\end{equation}
That is, $L_1\!=\!L,\, L_2\!=\!0$ for the natural parity case  
and $L_1\!=\!L,\, L_2\!=\!1$ for the unnatural parity case.    
We stress that some important symmetry 
properties such as translation invariance and rotation invariance are 
already built in the basis (\ref{cg}). 
We show in Appendix A that 
the Fourier transform of Eq.~(\ref{cg}) is again a combination of 
CG in momentum space. 

The basis function introduced above provides us with a trial function for 
all states but $L\!=\!0$ and a negative parity. 
The angular part of the basis for such a particular 
case must contain at least 
three vectors like ${\vi x}_i\cdot({\vi x}_j\times {\vi x}_k)$~\cite{su00}. 

We present calculation formulas of 
matrix elements for the basis (\ref{cg}) with arbitrary $L_1$ and 
$L_2$ values for a given $L$. 
Formulas for some simple operators are already given 
in refs.~\cite{vs95,book,suv98,vus98} for natural parity and in
ref.~\cite{su00} for unnatural parity. In the appendices 
we give formulas for various operators including noncentral 
potentials as well as those nonlocal kernels which appear in the 
RGM formulation for nuclear cluster models~\cite{tang}. 

Our method 
is based on the use of the generating function $g$ for the CG: 
\begin{equation}
g({\vi s}; A, {\vi x})=\exp\Big(-{\frac{1}{2}}{\widetilde{\vi x}}A{\vi x}+
\widetilde{\vi s}{\vi x}\Big),
\label{DEFGFN}
\end{equation}
where ${\vi s}$ is an $(N\!-\!1)$-dimensional column 
vector whose $i$th element is a 3-dimensional vector ${\vi s}_i$. 
By expressing ${\vi s}_i$ 
with 3-dimensional unit vectors ${\vi e}_1$ and ${\vi e}_2$ as 
${\vi s}_i\!=\!{\lambda}_1{\vi e}_1 u_{1_i}\!+\!{\lambda}_2{\vi e}_2 u_{2_i}$, 
the basis function (\ref{cg}) is generated as follows:
\begin{eqnarray}
&& F_{(L_1L_2)LM}(u_1,u_2,A,{\vi x}) 
= {\frac{B_{L_1}B_{L_2}}{ L_1!L_2!}}\int\!\int 
d{\vi e}_1\, d{\vi e}_2 \, [Y_{L_1}({\vi e}_1) 
Y_{L_2}({\vi e}_2)]_{LM} \nonumber \\
& &\qquad \qquad \qquad \qquad \times\,  {\frac{\partial^{L_1+L_2}}
 {\partial{\lambda}_1^{L_1}\partial{\lambda}_2^{L_2}}}\, 
g({\lambda}_1{\vi e}_1 u_1\!+\!{\lambda}_2{\vi e}_2 u_2; A,{\vi x})
\Big\vert_{\lambda_1=\lambda_2=0}, 
\label{gfn}
\end{eqnarray}
where  
\begin{equation}
B_L={\frac{(2L+1)!!}{ 4 \pi}}.
\label{def.bl}
\end{equation} 
When $g({\lambda}_1{\vi e}_1 u_1\!+\!{\lambda}_2{\vi e}_2 u_2; A,{\vi x})$ 
is expanded in powers of $\lambda_1$, only the term of degree 
$\lambda_{1}^{L_1}$ contributes in Eq.~(\ref{gfn}), and 
this term contains the $L_1$-th degree of ${\vi e}_1$ 
because $\lambda_1$ and 
${\vi e}_1$ always appear simultaneously. In order for the term 
to contribute to the integration over ${\vi e}_1$, these $L_1$ vectors of 
${\vi e}_1$ must couple to the angular momentum $L_1$, that is they 
are uniquely coupled to the maximum possible angular momentum. 
The same applies to $\lambda_2$ as well. 

\subsection{Coordinate transformation}

Suppose that the coordinate set ${\vi x}$ is transformed to a new set of 
relative coordinates ${\vi y}$ through ${\vi y}\!=\!{\cal P}{\vi x}$ with 
an $(N\!-\!1)\!\times \!(N\!-\!1)$ matrix ${\cal P}$. The 
transformation of this kind is needed, e.g., when the permutation 
symmetry of 
the constituent particles is imposed on the basis function or different 
coordinate sets are used to describe particular correlated 
motion~\cite{kamimura}. The CG basis 
function in the coordinate set ${\vi y}$ is rewritten as 
\begin{eqnarray}
F_{(L_1L_2)LM}(u_1,u_2,A,{\vi y})
= F_{(L_1L_2)LM}({u_{\cal P}}_1,{u_{\cal P}}_2,A_{\cal P},{\vi x})
\label{trans.cg}
\end{eqnarray}
with
\begin{equation}
A_{\cal P}=\widetilde{\cal P}A{\cal P},\ \ \ \ \  
{u_{\cal P}}_1=\widetilde{\cal P}u_1,\ \ \ \ \  
{u_{\cal P}}_2=\widetilde{\cal P}u_2.
\label{aup.relation}
\end{equation}
The coordinate transformation ends up redefining 
$A$, $u_1$ and $u_2$ as above, and we do not need to introduce 
different coordinate sets. This property that  
the functional form of the CG remains unchanged under the coordinate 
transformation enables one 
to unify the method of calculating the matrix element. 

\section{Calculation of matrix elements}
\label{sect.3}

We introduce the following abbreviated notation for a matrix 
element 
\begin{equation}
\langle F'\vert {\cal O}\vert F\rangle \equiv
\langle F_{(L_3L_4)L'M'}({u_3},{u_4},A',{\vi x})\vert
 {\cal O}\vert F_{(L_1L_2)LM}(u_1,u_2,A,{\vi x})\rangle,
\label{meofop}
\end{equation}
where ${\cal O}$ stands for an operator of interest.  
Using Eq.~(\ref{gfn}) in Eq.~(\ref{meofop}) enables one to relate 
the matrix element to that between the generating functions: 
\begin{eqnarray}
\langle F'\vert {\cal O}\vert F\rangle 
&\!=\!& \! \left(\prod_{i=1}^{4}{\frac{B_{L_i}}{L_i!}}
\int d{\vi e}_i\right)
([Y_{L_3}({\vi e}_3) Y_{L_4}({\vi e}_4)]_{L'M'})^{*}
[Y_{L_1}({\vi e}_1)  Y_{L_2}({\vi e}_2)]_{LM} 
\nonumber \\
&\!\times\!& \!\left(\prod_{i=1}^4 {\frac{\partial^{L_i}}{\partial{{\lambda}_i}^{L_i}}}\right)
\langle g({\vi s}'; A',{\vi x})\vert {\cal O} \vert g({\vi s}; A,{\vi x})
\rangle\Big|_{\lambda_1=\lambda_2=\lambda_3=\lambda_4=0}
\label{grandformula}
\end{eqnarray}
with
\begin{equation}
{\vi s}={\lambda}_1{\vi e}_1 u_1\!+\!{\lambda}_2{\vi e}_2 u_2,\ \ \ \ \ 
{\vi s}'={\lambda}_3{\vi e}_3 u_3\!+\!{\lambda}_4{\vi e}_4 u_4.
\end{equation}

The calculation of the matrix element consists of three stages: 
\begin{enumerate}
\vspace{-2mm}
\item Calculate the matrix element, 
 
$\quad {\cal M}=\langle g({\lambda}_3{\vi e}_3 u_3\!+\!{\lambda}_4{\vi e}_4
      u_4; A',{\vi x})\vert {\cal O}
\vert g({\lambda}_1{\vi e}_1 u_1\!+\!{\lambda}_2{\vi e}_2 u_2; A,{\vi
      x})\rangle$, 

between the generating functions.
\vspace{-2mm}
\item Expand ${\cal M}$ in powers of $\lambda_i$ and keep only 
those terms of degree $L_i$ for each $i$. The terms should contain 
the unit vector ${\vi e}_i$ of degree $L_i$ as well. One may 
drop any term with $\lambda_i^2 {\vi e}_i\!\cdot\!{\vi e}_i\!
=\!\lambda_i^2$ etc.
\vspace{-2mm}
\item Perform the differentiation and integration prescribed in 
Eq.~(\ref{grandformula}).
\end{enumerate}
Formulas for most important matrix elements ${\cal M}$ are tabulated in
ref.~\cite{book}.  
The angle integration in stage 3 can be performed using a formula   
\begin{eqnarray}
& &\int\! \cdots\! \int \prod_{i=1}^4\,d{\vi e}_i
([Y_{L_3}({\vi e}_3) Y_{L_4}({\vi e}_4)]_{L'M'})^{*}
[Y_{L_1}({\vi e}_1)  Y_{L_2}({\vi e}_2)]_{LM} 
\nonumber \\
& & \qquad \qquad \qquad \times [[Y_{L_1}({\vi e}_1)  Y_{L_2}({\vi e}_2)]_{L}
[Y_{L_3}({\vi e}_3) Y_{L_4}({\vi e}_4)]_{L'}]_{\kappa \mu}
\nonumber \\
& &\, = (-1)^{L-L'+\kappa+L_1+L_2}
\sqrt{\frac{2\kappa+1}{2L'+1}}\langle LM\kappa \mu|L'M'\rangle.
\end{eqnarray}
Appendix B collects formulas for the matrix elements of some local operators.

\section{Basic matrix elements for nonlocal operators} 
\label{sect.4}

Let us consider the matrix element of a 
nonlocal operator acting between particles 
$k$ and $l$. The relative distance vector, ${\vi r}_k\!-\!{\vi r}_l$, 
can be expressed 
as a linear combination of ${\vi x}_i$ (excluding the 
center of mass coordinate ${\vi x}_N$) as 
\begin{equation}
{\vi r}_k-{\vi r}_l=\sum_{i=1}^{N-1}w_i{\vi x}_i=\widetilde{w}{\vi x},
\label{def.rel.dist.vec}
\end{equation}
where the $(N\!-\!1)$-dimensional column vector $w$ is given by 
\begin{equation}
w_i=(U^{-1})_{ki}-(U^{-1})_{li}.
\label{w.expression}
\end{equation}
See Eq.~(\ref{def.matU}). The nonlocal operator may in general be expressed as 
\begin{equation}
 V_{{\rm NL}}=
\int\!\int d{\vi r}\,d{\vi r}' \, V({\vi r}',{\vi r})
\, \vert \delta(\widetilde{w}{\vi x}-{\vi r}')\rangle
\langle \delta(\widetilde{w}{\vi x}-{\vi r}) \vert,
\label{def.Vnl}
\end{equation}
where $V({\vi r}',{\vi r})$ determines the form factor of the nonlocal 
operator.

The operator of type (\ref{def.Vnl}) was applied to 
the study of $\alpha$-cluster 
condensation in $^{12}$C and $^{16}$O~\cite{takahashi}. 
Evaluating the matrix element of the nonlocal operator requires 
deliberation because the coordinate $\widetilde{w}{\vi x}$ has to be 
singled out from the set of $(N\!-\!1)$ coordinates ${\vi x}_i$, and
because the rest of other $(N\!-\!2)$ coordinates must be properly 
defined on the condition that they are all independent 
of $\widetilde{w}{\vi x}$. Though the choice of these $(N\!-\!2)$ 
coordinates is not unique, 
the matrix element should not depend on its choice. 
This problem is discussed in Appendix C. 

As derived in Appendix~\ref{app.D1}, the basic matrix element reads 
\begin{eqnarray}
& &\!\! \langle g({\vi s}'; A',{\vi x}) \mid V_{{\rm NL}}\mid 
g({\vi s}; A,{\vi x}) \rangle 
=\left(\frac{(2\pi)^{N-2}c}{{\rm det}B}
\right)^{\frac{3}{2}}{\rm exp}\left(\frac{1}{2}(\widetilde{\vi
s}+\widetilde{{\vi s}'})J^{-1}({\vi s}+{\vi s}')\right)
\nonumber \\
& &\qquad \qquad \times \int\!\int  d{\vi r}'d{\vi r}V({\vi r}',{\vi r})\, 
{\rm exp}\left(-{\frac{1}{2}}\alpha'{\vi r}'^2
-{\frac{1}{2}}\alpha{\vi r}^2-\beta{\vi r}'\cdot{\vi r}\right)
\nonumber \\
& &\qquad \qquad \times \exp\left(\big[
(\widetilde{\zeta}-\widetilde{\eta'}){\vi s}'
-\widetilde{\eta'}{\vi s}\big]\cdot{\vi r}'+
\big[(\widetilde{\zeta}-\widetilde{\eta}){\vi s}
-\widetilde{\eta}{\vi s}'\big]\cdot{\vi r}\right),
\label{gndformula}
\end{eqnarray}
where 
\begin{equation}
B=A+A',\ \ \ \ \ c=({\widetilde{w}B^{-1}w})^{-1},\ \ \ \ \ J^{-1}=B^{-1}-c B^{-1}w\widetilde{w}B^{-1},
\label{def.c.and.J}
\end{equation}
and 
\begin{eqnarray}
& &a=\widetilde{\zeta}A\zeta,\ \ \ \ \ a'=\widetilde{\zeta}A'\zeta,
\ \ \ \ \ \eta=J^{-1}A\zeta,\ \ \ \ \ \eta'=J^{-1}A'\zeta, 
\nonumber \\
& &\alpha=a-\widetilde{\zeta}A\eta,\ \ \ \ \ 
\alpha'=a'-\widetilde{\zeta}A'\eta',\ \ \ \ \ 
\beta=-\frac{1}{2}\widetilde{\zeta}(A\eta'+A'\eta).
\label{def.alpha.beta}
\end{eqnarray}
Here $\zeta$ is an $(N\!-\!1)$-dimensional column vector that is 
determined uniquely from $w$ defined in  
Eq.~(\ref{def.rel.dist.vec}). See Appendix C.2.

We perform the 
integration in Eq.~(\ref{gndformula}) for a given $V({\vi r}',{\vi r})$ 
and follow the procedure described in Sect.~\ref{sect.3}.
Some examples are collected in Appendix D.

\section{Numerical results}
\label{sect.5}

\subsection{Spectroscopic properties of few-nucleon systems}

A few-nucleon system interacting via a realistic two-body force offers 
good examples of testing the CG basis function with the GVR. 
The $N$-nucleon system is specified by the following Hamiltonian
\begin{equation}
H=\sum_{i=1}^N T_i-T_{\text{cm}}+\sum_{j>i=1}^N v_{ij}.
\end{equation}
Three-body forces are neglected though their inclusion does not cause 
any problem. 
The operator $T_i$ is a kinetic energy and the center of mass
kinetic energy $T_{\text{cm}}$ is subtracted in the Hamiltonian. 
A realistic nucleon-nucleon force is characterized by 
a short-ranged repulsion and a long-ranged tensor force.
To clarify the role of these properties, we employ three types of 
potential models as $v_{ij}$. 
The first is the Minnesota (MN) potential~\cite{minnesota} which contains
rather mild short-ranged repulsion and renormalizes the effect of 
the tensor force into its central term. 
No spin-orbit component of the MN potential is included.
The second is the AV8$^\prime$ potential~\cite{av8p} which is obtained 
from the Argonne V18 potential with a suitable 
renormalization procedure. The AV8$^\prime$ potential consists of 
central, tensor and spin-orbit terms. 
A few-body calculation with this potential shows slow convergence because
of its strong short-ranged repulsion, 
and in addition the radial form factor of AV8$^\prime$ makes the evaluation of 
matrix elements time-consuming in the present approach. The third is the 
G3RS potential~\cite{tamagaki} 
whose radial form is given as a combination of Gaussians 
with three ranges, which makes the numerical calculation much faster.  
Other features of the G3RS potential compared to AV8$^\prime$ are that 
the central force is deep, the tensor force is weak and 
the repulsion at the origin is mild. 
The original G3RS potential contains $\bm{L}^2$ and quadratic 
$\bm{L}\!\cdot\!\bm{S}$ terms in even partial waves.
The contribution of these terms is small~\cite{horiuchi}, and 
they are ignored in the present calculation. 

The form of $v$ is written as 
\begin{align}
v_{12}=V_{\rm c}(r)+V_{\text{Coul}}(r)P_{1\pi}P_{2\pi}
+V_{\rm t}(r) S_{12}+V_{\rm b}(r) \bm{L}\cdot\bm{S},
\end{align}
where $S_{12}$ and $\bm{L}\!\cdot\!\bm{S}$ stand 
for the tensor and spin-orbit operators and 
$P_{i\pi}\!=\!1$ for proton and 0 for neutron. 
The neutron projection operator $P_{i\nu}$ is defined in a similar way. 
The inputs used in this subsection are 
$\hbar^2/m_N$=41.47106 MeV\,fm$^2$
and $e^2$=1.440 MeV\,fm. Here $m_N$ is the nucleon mass, and the 
mass difference between proton and neutron is ignored. 
No isospin mixture is taken into account. 
The $u$ parameter of the MN potential is set to
$u$=1. 

The wave function of the system is expressed as a combination of different 
channel components specified with $L$ and $S$ (see Eq.~(\ref{basis.LS})):
\begin{align}
\Psi=\sum_{LS}X_{LS}\Psi_{(LS)JM_JTM_T},
\end{align}
where $\Psi_{(LS)JM_JTM_T}$ is assumed to be normalized. The squared 
coefficient 
\begin{align}
P(L, S)=(X_{LS})^2
\end{align}
denotes the probability of finding the system in the channel $(L, S)$.
Possible channels for the ground states of $N\!\leq \!4$ systems
are $(L,S)$=(0,\,1), (2,\,1) for deuteron, 
(0,\,1/2), (2,\,3/2), (1,\,1/2), (1,\,3/2) for $^3$H and $^3$He, and 
(0,\,0), (2,\,2), (1,\,1) for $\alpha$-particle.
Among these channels, (1,\,1/2) and (1,\,3/2) for $^3$H ($^3$He) 
and (1,\,1) for $\alpha$-particle are unnatural parity. 
Table~\ref{deuteron} compares the deuteron properties calculated 
with the three potentials.  Comparing the result 
between AV8$^\prime$ and G3RS, we see that 
the AV8$^\prime$ potential is stronger in the tensor and spin-orbit 
components but weaker in the central component than the 
G3RS potential, as already noted above. 

\begin{table}[b]
\caption{Comparison of deuteron properties calculated with the 
different potentials. The energy and point nucleon root 
mean square (rms) radius are given in MeV and fm, 
respectively. The probability 
$P(L,\,S)$ is given in \%.}
\begin{tabular}{ccccc}
\hline
Potential&&MN&G3RS&AV8$^\prime$\\
\hline
$E$                            &&$-2.202$ &$-2.277$ &$-2.242$ \\
$\left<T\right>$               &&10.487	 &16.478   &19.881	  \\
$\left<V_{\rm c}\right>$       &&$-12.689$&$-7.294$ &$-4.458$ \\
$\left<V_{\rm t}\right>$       &&--	 &$-11.460$&$-16.641$\\
$\left<V_{\rm b}\right>$       &&--	 &--	  &$-1.024$ \\
$\sqrt{\left<r^2\right>}$      &&1.952    &1.979    &1.961    \\
$P(0,1)$                       &&100     &95.22   &94.23\\
$P(2,1)$                       &&--      &4.78    &5.77\\
\hline
\end{tabular}
\label{deuteron}
\end{table}

The orbital part of the full wave function is expanded 
in terms of the CG (\ref{cg}). The CG for the natural parity state 
has $L_1$=$L$ and $L_2$=0, and contains the 
parameters $A$ and $u_1$, while 
the one for the unnatural parity state has $L_1$=$L$ and $L_2$=1, and 
contains $A$, $u_1$ and $u_2$. These parameters 
are selected by the stochastic variational method (SVM)~\cite{vs95,book} 
as follows: 
First, we randomly choose a channel specified by $L_i,\,L,\,S$ 
from among those $(L,S)$ channels 
which are included in the calculation. For a given $S$ value, 
a set of intermediate spins is randomly chosen. For example, 
for three-nucleon system with $S$=1/2,  
$|[[\frac{1}{2}\frac{1}{2}]_0\frac{1}{2}]_{1/2\, M_S}\rangle$ 
or $|[[\frac{1}{2}\frac{1}{2}]_1\frac{1}{2}]_{1/2\, M_S}\rangle$ 
is randomly chosen. A set of intermediate isospins is 
chosen randomly as well. 
After choosing these discrete sets, we determine 
the nonlinear parameters $A,\,u_i$ following the SVM procedure. 
The elements of $A$ are chosen randomly from a physically
important interval. 
The diagonal and off-diagonal elements of $A$ describe the 
correlated motion of the system. 
The parameter $u_i$ is also chosen randomly  
under the condition $\widetilde{u_i}u_i$=1. This parameter 
serves to express the partial waves needed to 
represent the rotational motion but its role is usually not as 
important as that played by $A$. 
Therefore we make more efforts to select a suitable $A$. 

We decrease the energy first by increasing 
the number of basis functions one by one up to a certain 
dimension. Keeping the number of basis functions fixed, we then 
switch to a refinement process~\cite{vs95,book} 
in which each basis function already selected 
is tested against other randomly chosen candidates.
We repeat these two optimization procedures until a fair 
convergence is attained. 
When the number of basis functions becomes large enough to 
get a converged solution,
it is convenient to rearrange the selected basis set according to 
importance. By the importance we mean the following. 
Suppose that the number of the basis functions is $K$. The first 
basis $\Phi_1$ is the one that gives the lowest energy 
among the $K$ functions. The second basis $\Phi_2$ is the one 
that gives the lowest energy together with $\Phi_1$ among the 
$K\!-\!1$ functions excluding $\Phi_1$. 
This ordering is continued until 
the last basis $\Phi_K$ is determined. 
After rearranging the basis functions in this way, 
we often have those $\Phi_i$ 
which play a very minor role in lowering energy. This is the 
case particularly when $i$ is close to $K$.  
These inactive bases may be discarded from the basis set 
to save the basis dimension. After this contraction of the basis set, 
we may again enlarge the basis dimension to search for better basis 
functions.

\begin{table}[t]
\caption{The ground state properties of three- and four-nucleon 
systems. The energy and rms radius are given in MeV and fm. 
The probability $P(L,\,S)$ is given in \%. The basis dimensions for
 $^4$He are 400 (G3RS) and 600 (AV8$^{\prime}$) in GVR, and 
reduced to half in PWE, respectively. }
\label{spectra}
\begin{tabular}{cccccccccc}
\hline
Potential&&MN&&\multicolumn{2}{c}{G3RS}&&\multicolumn{3}{c}{AV8$^\prime$}\\
\cline{1-1}\cline{3-3}\cline{5-6}\cline{8-10}
Method&&GVR&&GVR&PWE&&GVR&PWE&Ref.~\cite{exotic}\\
\hline
$^3$H$(\frac{1}{2}^+)$&&&&&&&&\\
\hline
$E$                            &&$-8.38$ &&$-7.73$&$-7.72$&&$-7.76$&$-7.76$&$-7.767$
\\
$\left<T\right>$               &&27.21	 &&40.24   &40.22&&47.59   &47.57&$47.615$\\
$\left<V_{\rm c}\right>$       &&$-35.59$&&$-26.80$&$-26.79$&&$-22.50$&$-22.49$&$-22.512$\\
$\left<V_{\rm t}\right>$       &&--	 &&$-21.13$&$-21.13$&&$-30.85$&$-30.84$&$-30.867$\\
$\left<V_{\rm b}\right>$       &&--	 &&$-0.03$ &$-0.03$ &&$-2.00$ &$-2.00$&$-2.003$\\
$\sqrt{\left<r^2\right>}$      &&1.71    &&1.79    &1.79    &&1.75    &1.75&\\
$P(0,1/2)$                     &&100     &&92.95   &92.94&&91.38&91.37&91.35\\
$P(2,3/2)$                     &&--      &&7.01    &7.02 &&8.55&8.57&8.58\\
$P(1,1/2)$                     &&--      &&0.03    &0.03 &&0.04&0.04&
\raisebox{-.7em}[0pt][0pt]{\}0.07}\\
$P(1,3/2)$                     &&--      &&0.02    &0.02 &&0.02&0.02&\\
\hline
$^3$He$(\frac{1}{2}^+)$&&&&&&&&&\\
\hline
$E$                            &&$-7.71$   &&$-7.08$&$-7.08$&&$-7.10$&$-7.10$&\\
$\left<T\right>$               &&26.69     &&39.46&39.43&&46.68&46.67 &\\
$\left<V_{\rm c}\right>$       &&$-35.06$  &&$-26.26$&$-26.24$&&$-22.00$&$-21.98$&\\
$\left<V_{\text{Coul}}\right>$ &&0.67	   &&0.64    &0.64&&0.65&0.65 &\\
$\left<V_{\rm t}\right>$       &&--	   &&$-20.89$&$-20.88$&&$-30.47$&$-30.47$&\\
$\left<V_{\rm b}\right>$       &&--	   &&$-0.03$&$-0.03$&&$-1.97$&$-1.97$&\\
$\sqrt{\left<r^2\right>}$      &&1.74      &&1.82   &1.82&&1.79&1.79&\\
$P(0,1/2)$                     &&100       &&92.98  &92.96&&91.42&91.41 &\\
$P(2,3/2)$                     &&--        &&6.98   &6.99&&8.51 &8.53 &\\
$P(1,1/2)$                     &&--        &&0.03   &0.03&&0.04 &0.04 &\\
$P(1,3/2)$                     &&--        &&0.02   &0.02&&0.02 &0.02 &\\
\hline
$^4$He$(0^+)$&&&&&&&&&\\
\hline
$E$                            &&$-29.94$&&$-25.29$&$-25.26$&&$-25.08$&$-25.05$&\\
$\left<T\right>$               &&58.08	 &&86.93   &86.77&&101.59   &101.36&\\
$\left<V_{\rm c}\right>$       &&$-88.86$&&$-66.24$&$-66.11$&&$-54.93$&$-54.73$&\\
$\left<V_{\text{Coul}}\right>$ &&0.83	 &&0.76	   &0.76&&0.77    &0.77&\\
$\left<V_{\rm t}\right>$       &&--	 &&$-46.62$&$-46.55$&&$-67.85$&$-67.79$&\\
$\left<V_{\rm b}\right>$       &&--	 &&$-0.13$ &$-0.12$&&$-4.65$ &$-4.66$&\\
$\sqrt{\left<r^2\right>}$      &&1.41    &&1.51    &1.51&&1.49  &1.49&\\
$P(0,0)$                       &&100     &&88.46&88.50&&85.76&85.79&\\
$P(2,2)$                       &&--      &&11.30&11.26&&13.87&13.85&\\
$P(1,1)$                       &&--      &&0.25&0.24&&0.36&0.36&\\
\hline
\end{tabular}
\end{table}

Table~\ref{spectra} lists the energies of three- and four-nucleon 
systems and the contributions of the respective terms of the 
Hamiltonian together with the nucleon rms radii. The GVR performance 
is tested by comparing to other calculations, particularly 
PWE calculations. The basis function in PWE is also Gaussian but no 
explicit correlated terms between the different coordinates are
included, so that the matrix $A$ is always chosen to be diagonal. 
A correlated motion is, however, accounted for by expanding the 
trial wave function in different coordinate sets. For example, both 
coordinates of K- and H-types are employed to obtain the solution 
for $\alpha$-particle. Thus this PWE calculation is the same 
as the SVM of ref.~\cite{benchmark}, and its accuracy is well tested. 
A noteworthy difference between GVR and 
PWE is that the latter expresses the angular part of the wave 
function by successively 
coupling the partial waves $\ell_i$ to the resultant $L$. In the present 
calculation $\ell_i$ is restricted to $\ell_i\!\le\!2$. 
The total energy of $^3$H calculated with AV8$^\prime$ 
and its decomposition to  
each term agrees with the results of PWE and Faddeev~\cite{exotic}. 
The agreement between the GVR and PWE calculations is also very good 
for the properties of the $\alpha$-particle. 
The four-nucleon system is solved by the different 
methods~\cite{benchmark} using the AV8$^\prime$ 
potential with the Coulomb force being neglected. 
If the Coulomb contribution of the present calculation is omitted,
the GVR energy becomes $-25.85$\,MeV, which agrees with the lowest 
energy of the benchmark calculations within 70\,keV. 
These results confirm that the CG basis with 
GVR is versatile enough to provide us with such 
accurate solutions that are comparable to state-of-the-art calculations 
available in literatures.  

We note from the comparison between AV8$^\prime$ and 
G3RS results that the solution with AV8$^\prime$ has 
large kinetic energy corresponding to the strong short-ranged repulsion. 
The tensor contribution is larger than the central 
force contribution in AV8$^\prime$. On the contrary, the G3RS 
potential gives larger 
attraction in the central contribution than in the tensor 
contribution. Though they give different potential contributions, both 
lead to almost the same total energy and nucleon rms radius. 
These features of the realistic potentials are very consistent with 
those of Table~\ref{deuteron} listed for the deuteron. 

Now we discuss how well
the GVR can incorporate unnatural parity components. 
The unnatural parity state is usually unfavored because it has a larger 
kinetic energy than the natural parity state and its 
diagonal matrix element becomes rather high. If the off-diagonal matrix 
elements between the unnatural parity and natural parity states are 
large enough to compensate the loss of the kinetic energy, 
the mixing of the unnatural parity state becomes important. 
Table~\ref{decom1} lists the decomposition of 
the energies of $^3$H and $\alpha$-particle according to the 
$(L, S)$ channel. The value for the diagonal channel denotes 
$(X_{LS})^2\langle \Psi_{(LS)JM_JTM_T}|H|\Psi_{(LS)JM_JTM_T}\rangle$, 
while the value between the different channels is $2X_{LS}X_{L'S'}
\langle \Psi_{(LS)JM_JTM_T}|H|\Psi_{(L'S')JM_JTM_T}\rangle$. Note 
that, in the case of $^3$H, the contributions from the 
two spin channels of $S$=1/2 
and 3/2 with the unnatural parity of $L$=1 are summed together. 
The probability of 
the unnatural parity state is negligibly small in $^3$H, as shown in 
Table~\ref{spectra}. 
The probability of the unnatural parity state $P(1,1)$ for 
$\alpha$-particle increases to 0.25\,\% for G3RS and 0.36\,\% 
for AV8$^\prime$. 
With this mixing of the unnatural parity component the tensor coupling 
with the (22) channel becomes important and thus the $\alpha$-particle 
gains energy by about 0.8\,MeV for G3RS and 1.6\,MeV for AV8$^\prime$. 

\begin{table}[b]
\caption{Total energies, in MeV, of $^3$H and $\alpha$-particle 
decomposed into different $(L,\,S)$ contributions. }
\label{decom1}
\begin{tabular}{crrrrrrr}
\hline
&\multicolumn{3}{c}{G3RS}&&\multicolumn{3}{c}{AV8$^\prime$}\\
\cline{2-4}\cline{6-8}
$^3$H$(\frac{1}{2}^+)$&&&&\\
\hline
  &\multicolumn{1}{c}{$(0,\frac{1}{2})$}&\multicolumn{1}{c}{$(2,\frac{3}{2})$}&\multicolumn{1}{c}{$(1,\frac{1}{2}$+$\frac{3}{2})$}&
&\multicolumn{1}{c}{$(0,\frac{1}{2})$}&\multicolumn{1}{c}{$(2,\frac{3}{2})$}&\multicolumn{1}{c}{$(1,\frac{1}{2}$+$\frac{3}{2})$}\\
\hline
$(0,\frac{1}{2})$&4.09&$-$22.54&$-$0.00&&9.72&$-$33.60&$-$0.03\\
$(2,\frac{3}{2})$&    &10.85   &$-$0.23&& &16.35&$-$0.42\\
$(1,\frac{1}{2}$+$\frac{3}{2})$&     &        &0.11&&&&0.22\\
\hline
$^4$He$(0^+_1)$&&&&\\
\hline
    &(0,0)&(2,2)&(1,1)&&(0,0)&(2,2)&(1,1)\\
\hline				  
(0,0)&0.95   &$-$46.65&$-$0.01  &&12.94&$-$68.67&$-$0.21\\
(2,2)&        &21.24   &$-$1.56 &&     &32.31&$-$2.90\\
(1,1)&        &        &0.72    &&     &     &1.47\\   
\hline
\end{tabular}
\end{table}

The $0^-_1$ excited state of $^{4}$He, located at 7.19\,MeV below 
the $p$+$p$+$n$+$n$ threshold,  
is a very good example to demonstrate the 
importance of the unnatural parity state. 
The $0_1^-$ state consists of two channels, $(L, S)$=$(1,\,1)$ 
(natural parity) and (2,\,2) (unnatural parity).
The matrix element of each operator in the Hamiltonian 
is listed in Table~\ref{decom2} for the $0_1^-$ state, 
with the column-row index of the matrix being labeled by 
the channel, $(1,1)$ or $(2,2)$. 
The values in parentheses stand for 
the matrix elements, 
$\langle\Psi_{(LS)JM_JTM_T}|H|\Psi_{(L'S')JM_JTM_T}\rangle$ etc. 
The table confirms that the $0^-_1$ state cannot be predicted well below 
the $p$+$p$+$n$+$n$ threshold if the coupling of the two channels 
is neglected. The diagonal matrix element 
of the Hamiltonian is 0.64\,MeV in (1,\,1) channel and 
6.46\,MeV in (2,\,2) channel. The unnatural parity channel 
gives large positive energy because its kinetic energy is very 
large. However, the coupling matrix element 
between (1,\,1) and (2,\,2) channels 
amounts to $-$13.50\,MeV, which locates 
the $0_1^-$ state at $-$6.40\,MeV in good agreement 
with experiment. 
This large coupling matrix element is brought 
about by the tensor force and its contribution is quite significant. 
The admixture of the unnatural parity components is so large as $4.5\%$. 
We repeated a calculation by omitting the (2,\,2) channel. 
The resulting energy is approximately $-$1\,MeV, which is  
higher by about 5\,MeV than the full channel calculation. 
Though the importance of the tensor force 
in the $0^-_1$ state was pointed out 
many years ago~\cite{atms,barrett}, the present result 
indicates that not only the tensor force but also the kinetic energy and 
the central force are important factors to determine the energy of the 
$0^-_1$ state. 
Our calculation ignores another unnatural parity component 
(0,\,0). However, its contribution is probably negligible because 
it has no tensor coupling with the main channel (1,\,1).  

\begin{table}[t]
\caption{The Hamiltonian matrix elements, given in MeV, for 
the $0_1^-$ state of $^4$He. The column-row of the matrix 
is labeled by the channel $(L,\,S)$, which is arranged in the order of 
$(1,1)$ and (2,2). See text for the matrix elements in 
parentheses. The $P(L,\,S)$ values are 
$P(1,1)$=0.955 and $P(2,2)$=0.045. The G3RS potential is used.} 
\label{decom2}
\begin{tabular}{ccccc}
\hline
\multicolumn{2}{l}{$H$}&&\multicolumn{2}{l}{$T$}\\
\cline{1-2}\cline{4-5}
0.64&$-$13.50&&41.19 (43.13)&--  \\ 
&6.46    &&     &7.19 (159.7)\\
\hline
\multicolumn{2}{l}{$V_{\rm c}$}&&
\multicolumn{2}{l}{$V_\text{Coul}$}\\
\cline{1-2}\cline{4-5}
$-$27.82 ($-$29.13)&--&&0.46 (0.48)&--\\ 
&$-$1.10 ($-$24.42)&&&0.02 (0.45)\\
\hline
\multicolumn{2}{l}{$V_{\rm t}$}&&
\multicolumn{2}{l}{$V_{\rm b}$}\\
\cline{1-2}\cline{4-5}
$-$13.48 ($-$14.11)&$-$13.51 ($-$65.15)&&0.28 (0.30)&0.01 (0.04)\\
     &0.35 (7.81)&&     &$-$0.00 ($-$0.00)\\
\hline
\end{tabular}
\end{table}

\begin{figure}[t]
\epsfig{file=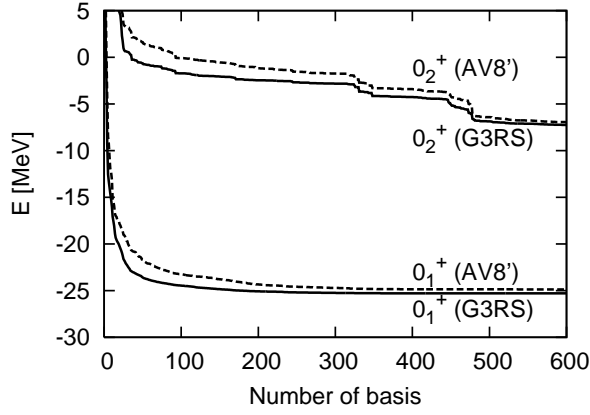,width=8.1cm,height=5.67cm}
\caption{The energy convergence of the ground and first excited states 
of $^4$He with the G3RS and AV8$^\prime$ potentials. The basis set 
is selected for the G3RS potential by the SVM, and the same set is 
used to calculate the energy with the AV8$^\prime$ potential.}
\label{diffpot.fig}
\end{figure}

It is interesting to ask the following question: How different are 
the G3RS and AV8$^\prime$ potentials? 
As seen in Table~\ref{spectra}, both potentials give similar 
energies for the few-nucleon systems. 
Figure~\ref{diffpot.fig} displays the energy change 
of the ground and first excited states of 
$^4$He as a function of the basis dimension. Here the basis set 
is the one selected using the G3RS Hamiltonian. 
The basis is increased to optimize the $0^+_2$ state after the dimension 
of 400. In the curves labeled G3RS the basis functions are ordered 
according to the importance criterion for the ground state. Now 
the curves labeled AV8$^\prime$ stand for the energy change 
obtained by just diagonalizing the AV8$^\prime$ Hamiltonian in the 
same basis set. 
The energy obtained in this way loses only 210\,keV 
compared to the energy listed in Table~\ref{spectra}. 
This means that the basis set determined with the G3RS potential 
already provides a fairly good basis set for the AV8$^\prime$ potential. 
Thus one may skip the basis search for AV8$^\prime$ but only needs 
to fine-tune the basis to reach the converged 
solution. This is particularly helpful in saving computer time 
because the G3RS potential is mild and simple enough to render 
numerical calculations fast. 

\begin{figure}[b]
\epsfig{file=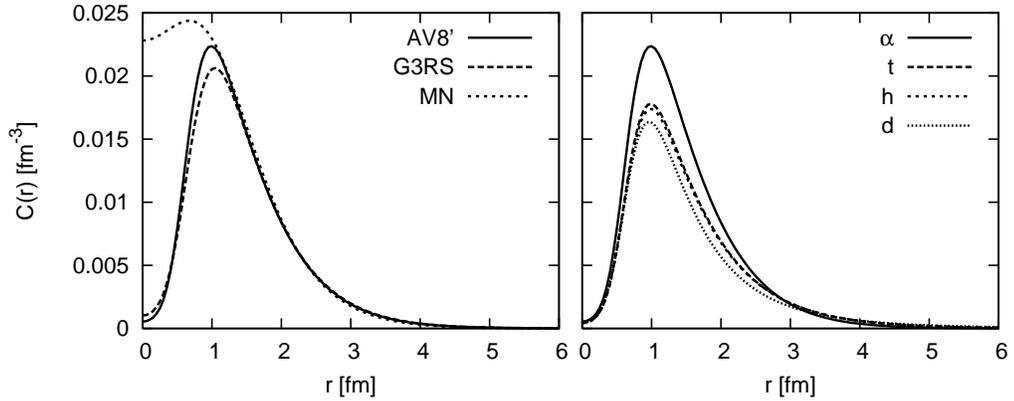,width=13.5cm,height=5.4cm}
\caption{Left: the correlation functions of $\alpha$-particle 
with the different potentials. Right: the correlation functions 
for the ground states of few-nucleon systems calculated with 
the AV8$^\prime$ potential.}
\label{corrfn.fig}
\end{figure}

Using Eq.~(\ref{me.multipole}) and noting the remark below it 
enables one to 
calculate the correlation function 
\begin{align}
C(r)=\frac{1}{4\pi r^2}\left<\Psi\right|
\delta(\left|\bm{r}_1-\bm{r}_2\right|-r)\left|\Psi\right>,
\label{correlation.function}
\end{align}
which has the normalization of $4\pi\int_0^\infty dr\, r^2C(r)$=1.
The left panel of Fig.~\ref{corrfn.fig} plots the correlation functions of 
the $\alpha$-particle calculated with 
the different potentials. The curve with 
AV8$^\prime$ agrees very well 
with that of the benchmark calculation~\cite{benchmark}. 
The curves calculated with the realistic forces show a strong depression 
at short distances due to the short-ranged repulsion, which is in 
sharp contrast with the MN potential case. 
One can see that the AV8$^\prime$ correlation function 
is more strongly suppressed than that of the G3RS potential. 
All the correlation functions are similar at large distances.  
The right panel of Fig.~\ref{corrfn.fig} compares the correlation
functions for few-nucleon systems calculated with the AV8$^\prime$ 
potential. We note that the peak position of $C(r)$ 
is almost independent of the systems. 
One can see that the correlation function of the 
$\alpha$-particle has larger amplitude around $r$=1\,fm and a 
shorter tail than those of deuteron and $^3$H. 
 
\begin{figure}[b]
\epsfig{file=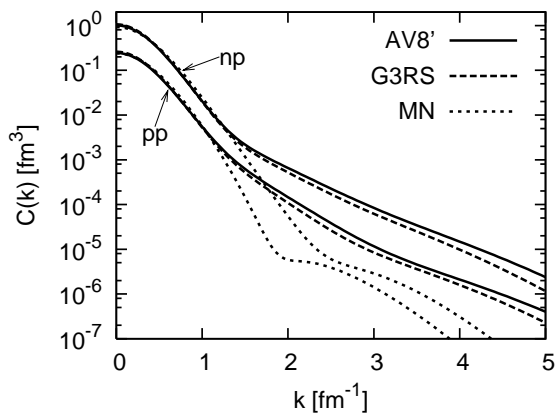,width=8.1cm,height=5.67cm}
\caption{Two-nucleon relative momentum distributions of 
$\alpha$-particle with the different potentials.}
\label{distk.fig}
\end{figure}

A two-nucleon momentum distribution also gives useful 
information on the correlation. 
The momentum distribution is easily calculated by transforming 
the coordinate-space 
wave function to momentum space (see Appendices A and B.3):
\begin{align}
C(k)=\frac{1}{4\pi k^2}
\left<\Psi\right|\delta(\tfrac{1}{2}\left|\bm{k}_1-\bm{k}_2\right|-k)
\{P_{1N}P_{2N}\}_+\left|\Psi\right>,
\label{momentum.correlation}
\end{align}
where the projection operator of a pair of nucleons, 
$\{P_{1N}P_{2N}\}_+$, stands for 
\begin{align}
\left(
\begin{array}{c}
P_{1\pi}P_{2\pi} \\
P_{1\nu}P_{2\nu} \\
\frac{1}{2}(P_{1\pi}P_{2\nu}+P_{1\nu}P_{2\pi}) \\
\end{array}
\right)
\quad {\rm for}
\left(
\begin{array}{c}
pp \\
nn \\
pn \\
\end{array}
\right).
\end{align}
The momentum distribution defined above has the following 
normalization 
\begin{align}
4\pi\int_0^\infty dk \, k^2C(k)=\frac{2}{A(A-1)}
\left<\Psi\right|\sum_{i<j}\{P_{iN}P_{jN}\}_+\left|\Psi\right>.
\label{mom.ratio}
\end{align}
We plot in Fig.~\ref{distk.fig} the $np$ and $pp$ 
momentum distributions of the $\alpha$-particle.  
The realistic forces produce similar distributions. 
The distribution calculated with the MN potential follows the 
realistic distribution up to $k<$\,1.3\,fm$^{-1}$, but then 
decreases rapidly as the momentum increases. This is because 
the MN potential has neither strong short-ranged repulsion nor 
tensor force~\cite{horiuchi}. As shown in ref.~\cite{schiavilla}, 
both $np$ and $pp$ distributions do not show very different 
behavior except that the former is roughly 
four times larger than the latter 
as expected from the relation (\ref{mom.ratio}). 
This is in sharp contrast to a specific momentum distribution 
such that the sum of the two momenta, $\bm{k}_1\!+\!\bm{k}_2$, 
is close to zero~\cite{horiuchi,schiavilla}, where the $nn$ ($pp$) 
momentum distribution, differently from the 
$np$ distribution, is characterized by a dip around 
2\,fm$^{-1}$. 

Calculations for other $J^{\pi}$ states of $^4$He are in progress. 
A detailed analysis will be reported elsewhere~\cite{he4}.

\subsection{$3\alpha$ system with a nonlocal RGM potential}

A second test of the CG basis calculation is 
a 3$\alpha$ system for $^{12}$C. That the $\alpha$$\alpha$ potential 
cannot be local for describing $^{12}$C is not surprising because the 
interaction between composite particles is intrinsically nonlocal.
Such interactions between two composite particles can be derived 
microscopically within the RGM. A first attempt to use an 
energy-dependent 
$\alpha\alpha$ RGM kernel has been developed in 
ref.~\cite{fuji} in a study of $^{12}$C. Using some average 
energy for this kernel 
provides fair results for $^{12}$C~\cite{fuji1} but not for other
three-cluster systems~\cite{theeten}. Hence the energy dependence 
is eliminated from the RGM equation, which enables us to 
obtain energy-independent nonlocal kernels. 
Recently, we have shown that this procedure 
leads to a consistent description of the two $0^+$ states of 
$^{12}$C~\cite{suzu} as well as the $3/2^-$ and $5/2^-$ states of 
$^9$Be and the ground state of $^6$He~\cite{theeten1}. 

The RGM equation for the two-cluster relative motion function $\chi$ reads
\begin{equation}
(T+V+\varepsilon K)\chi=\varepsilon \chi, 
\label{rgmeq}
\end{equation}
with a bare RGM potential $V$
\begin{align}
V=V_{\rm D}+V^{{\rm EX}}=V_{\rm D}+K_T+K_V.
\label{rgm.bare.pot}
\end{align}
Here $T$ is the intercluster kinetic energy, $K$ is the overlap 
kernel, $V_{\rm D}$ is local and called the direct potential, 
and the nonlocal potential $V^{{\rm EX}}$ comprises  
$K_T$ and $K_V$ which are the exchange nonlocal kernels for 
the kinetic and potential (including 
the Coulomb term) energies, respectively. 
Equation~(\ref{rgmeq}) can be converted to an energy-independent equation 
for $g=\sqrt{\cal N}\chi=\sqrt{1-K}\chi$ as follows:
\begin{align}
(T+V^{{\rm RGM}})g=\varepsilon g
\label{eq.motion.g}
\end{align}
with 
\begin{equation}
V^{{\rm RGM}}={\cal N}^{-1/2}(T+V){\cal N}^{-1/2}-T=V+W,
\label{def.vrgm}
\end{equation}
where $W$ is a difference between the
renormalized RGM potential $V^{{\rm RGM}}$ and the bare RGM potential $V$:
\begin{equation}
W={\cal N}^{-1/2}(T+V){\cal N}^{-1/2}-(T+V).
\label{def.w}
\end{equation}

The energy-independent potential $V^{{\rm RGM}}$ serves as 
an interaction between 
the composite particles. As shown in ref.~\cite{theeten1} 
the nonlocal interaction $W$ 
can very well be approximated with $\lambda W^{(1)}$, where  
\begin{equation}
W^{(1)}=\frac{1}{2}[K(T+V)+(T+V)K ],
\end{equation}
and $\lambda$ is an appropriate constant. The $\lambda$ value is set to 
be 1.30 for the $\alpha\alpha$ case. The $\alpha\alpha$ potential 
we use in the present study 
takes the form $V+\lambda W^{(1)}$. All of the needed 
$\alpha\alpha$ RGM kernels are explicitly given in ref.~\cite{theeten}. 
We use the harmonic-oscillator parameter 
$b\!=\!1.36\,$fm for the single-nucleon orbit in the 
$\alpha$ cluster, and the MN potential~\cite{minnesota} 
with $u\!=\!0.94687$ as the two-nucleon potential. 
The nonlocal $\alpha\alpha$ potential  
constructed in this way 
reproduces the $\alpha\alpha$ phase shifts very well. 

To obtain the ground and first excited $0^{+}$ states of $^{12}$C, 
we must have such a solution that does 
not contain Pauli-forbidden states $\varphi_{n\ell m}$. 
These states are 
defined as redundant states of the 2$\alpha$ RGM equation (\ref{rgmeq}). 
They are $0s$, $1s$ and $0d$ harmonic-oscillator states. We define 
an operator ${\it \Gamma}$ for the $3\alpha$ system by 
\begin{equation}
{\it \Gamma}=\sum_{l>k=1}^{3}{\it \Gamma}_{kl},
\end{equation}
where ${\it \Gamma}_{kl}$ is a separable nonlocal operator acting on 
the relative motion of the particles $k$ and $l$: 
\begin{equation}
{\it \Gamma}_{kl}={\it \Gamma}_{kl}(0s)+{\it \Gamma}_{kl}(1s)+{\it \Gamma}_{kl}(0d)
\end{equation}
with 
\begin{equation}
{\it \Gamma}_{kl}(n\ell)=\sum_{m=-\ell}^{\ell} 
\mid \varphi_{n\ell m}({\vi r}_k-{\vi r}_l)\rangle  
\langle  \varphi_{n\ell m}({\vi r}_k-{\vi r}_l)\mid.
\end{equation}
Eliminating the forbidden states 
is performed by adding a 
pseudopotential $\gamma {\it \Gamma}$ to the 3$\alpha$ Hamiltonian and  
finding such a solution that is stable for sufficiently 
large $\gamma$~\cite{enigma}. In practice, the $\gamma$ value is 
set around $10^4\,$MeV in the present calculation.  Since it is hard to
eliminate the forbidden states to high accuracy in a 
nonorthogonal basis set like the CG basis, the use
of larger $\gamma$ values leads to numerically unstable energies.  
The matrix elements of the nonlocal potentials, 
$V$, $W^{(1)}$ and ${\it \Gamma}(n\ell)$,  
can all be calculated using the formulas in Appendices C and D. 

Table~\ref{12C} compares the present result with that obtained by the 
HH approach~\cite{theeten1}. In the HH expansion method the 
size of the three-body basis depends on the maximum 
hypermomentum $K_{{\rm max}}$. The HH calculation in the table 
uses $K_{{\rm max}}\!=\!36$, 
and for this value there are 100 HH in the variational
expansion of the wave function~\cite{theeten2}. 
Also the hyperradius variable is 
discretized over 35 points and thus the Hamiltonian matrix has 
a size of 3500$\times$3500. In contrast to the HH expansion, 
the solution in the CG approach employs 200 basis functions. 
The energies of both the $0^+_1$ and $0^+_2$ states are in good
agreement between the two calculations. 
These values agree reasonably well with the experimental energies. 
In particular the energy of the $0^+_2$ state is 
in excellent agreement with the observed state called the Hoyle 
state. The matter rms radius of 
$^{12}$C is defined by adding the empirical value of the 
$\alpha$-particle ($\sqrt{\langle r^2_{\alpha}\rangle}\!=\!1.479\,$fm). 
The radius for the ground state agrees between the two
calculations. The comparison of the HH and CG results 
indicates that the CG basis is quite flexible 
to describe the different shapes of the wave functions in a 
small dimension. 

\begin{table}[t]
\caption{The energy $E$ (MeV) from the $3\alpha$ threshold and 
the rms radius $\sqrt{\langle r^2\rangle}$ (fm) for the ground state 
and the first excited $0^+$ state of $^{12}$C.  The energy-independent 
$\alpha\alpha$ RGM potential is used. The expectation value of 
each piece of the Hamiltonian is also shown.  
Experimental energies are $-$7.27 and 0.38\,MeV 
for the $0^+_1$ and $0^+_2$ states, respectively. 
The inputs are taken as ${\hbar^2}/{m_N}$=41.472\,MeV\,fm$^2$,
 $m_\alpha$=4$m_N$, $e^2$=1.44\,MeV\,fm$^{-2}$, where $m_\alpha$ is 
the mass of $\alpha$-particle.}
\label{12C}
\begin{tabular}{cccccccc}
\hline
Method  & $L^\pi$ & $E$  & $\langle T \rangle$ & $\langle V_{\rm D} \rangle$ & $\langle V^{{\rm EX}}+\lambda W^{(1)} \rangle$ & 
 $\lambda \langle W^{(1)} \rangle$ & $\sqrt{\langle r^2 \rangle}$ \\ 
\hline
CG&$0_1^+$&$-$9.83&79.06&$-$80.34&$-$8.55&7.75&2.19\\
  &$0_2^+$&   0.42&20.34&$-$18.32&$-$1.60&0.68&    \\
\hline
HH&$0_1^+$&$-$9.83&79.09&$-$80.36&$-$8.56&   &2.19 \\ 
  &$0_2^+$&   0.43&21.5 &$-$19.38&$-$1.69&   &     \\
\hline
\end{tabular}
\end{table}

The expectation value of 
each term of the Hamiltonian is also listed in Table~\ref{12C}. It 
agrees within 30\,keV for the $0^+_1$ 
state between the two methods, but the difference is 
still larger for the $0^+_2$ state. 
This may be due to that both methods obtain the latter state 
in a bound-state 
approximation without imposing a proper asymptotic 
condition, though it is located above the 3$\alpha$ threshold. 
To judge how well the elimination of the 
Pauli-forbidden components is achieved with the 200 
basis functions, we show the expectation value of ${\it \Gamma}$: It 
is 1.34$\times 10^{-6}$ for 
$0^+_1$ and 2.19$\times 10^{-7}$ for $0^+_2$, respectively. The 
contribution from the pseudopotential $\gamma {\it \Gamma}$ is 
subtracted in the energy listed in the table. 

\begin{figure}[b]
\epsfig{file=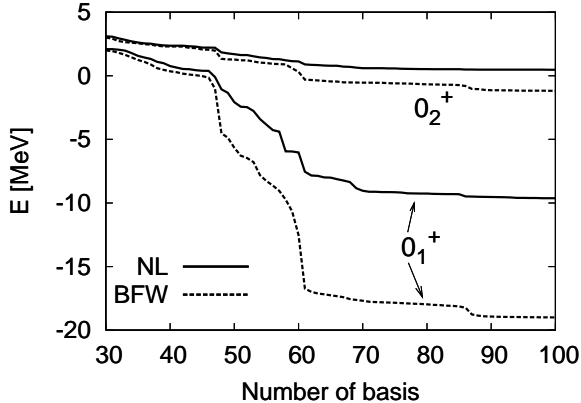,height=8cm,width=5.6cm,angle=-90}
\caption{The energy convergence of the $0^+$ states of $^{12}$C with the 
BFW and nonlocal RGM (NL) potentials. The basis set 
is determined by the SVM for the BFW potential.}
\label{NL-BFW}
\end{figure}

As discussed in refs.~\cite{enigma,suzu}, a local
$\alpha\alpha$ potential leads to unphysical results for the 3$\alpha$ 
energies even though the potential fits the $\alpha\alpha$ phase
shifts. For example, the well-known $\alpha\alpha$ potential~\cite{BFW} 
called the BFW potential gives an extremely deep energy 
of $-19.08\,$MeV for the $0^+_1$ state. 
This is the main reason 
why we advocate the importance of using the energy-independent 
RGM kernel as the interaction between the composite particles. 
Though the BFW potential cannot produce a physically acceptable 
result for the $0^+_1$ state, it may be worthwhile testing 
if this potential could be used for setting up 
a suitable basis set for the $3\alpha$ Hamiltonian 
with the nonlocal potential. This is so because the BFW potential 
whose nuclear part is given by a single Gaussian  
is by far economical than the nonlocal RGM potential from the 
point of view of computer time. Figure~\ref{NL-BFW} displays the 
energy convergence as the basis dimension increases. 
In this figure the basis set is selected to optimize the $0^+_1$ 
and $0^+_2$ states for the $3\alpha$ Hamiltonian employing 
the BFW $\alpha\alpha$ potential. 
This basis set 
is now just used to diagonalize the Hamiltonian with 
the nonlocal $\alpha\alpha$ potential. The curves labeled NL display 
the energy convergence of the latter calculation 
for the two $0^+$ states. 
The resulting energies are $-9.75\,$MeV for the $0^+_1$ state 
and $0.43\,$MeV for the $0^+_2$ state, which are 
close to those of Table~\ref{12C}. The rms radius of the 
$0^+_1$ state is reproduced as well. In fact the overlap 
between the wave functions obtained in the set of 
Table~\ref{12C} and the common set noted above is 
0.9995 for $0^+_1$ and 0.9987 for 
$0^+_2$. This finding is quite appealing for studying a 
multi-$\alpha$ system interacting via the nonlocal $\alpha\alpha$ 
potential. We just use the BFW potential to set up a suitable 
basis set and then use the refinement process to adapt the basis 
set to the nonlocal potential. This will be much more economical than 
selecting from the beginning the basis for the Hamiltonian 
with the nonlocal 
potential. A calculation along this line 
is in progress for a 4$\alpha$ description of $^{16}$O.

\section{Summary}
\label{sect.6}
We have extended our previous study on the global vector representation 
to cope with the orbital motion with an unnatural parity. The basic 
idea is to introduce two global vectors which are defined by a 
linear combination of the particle coordinates. Combining these 
angular parts with a spherical part of the orbital function provides 
a flexible basis function. We have shown that the matrix elements 
for most of physical operators of interest including nonlocal 
operators can be derived analytically 
if the spherical part is taken as a correlated Gaussian. The fact 
that its Fourier transform again takes the same functional 
form opens a wide 
applicability for studying correlations of a system 
in momentum space.  

Our numerical test examples included $A$=3, 4 nuclei interacting 
via realistic interactions of AV8 type. They have demonstrated the 
applicability and accuracy of the present method, in comparison 
with the partial-wave expansion results. We have shown 
that the $0^-_1$ state of $^4$He at the excitation energy of 21.01\,MeV 
offers a good example to indicate the importance of the tensor 
coupling between the natural parity and unnatural parity components. 
The correlated Gaussian basis was also tested for a nonlocal 
potential in the 3$\alpha$ system, and it succeeded to reproduce 
the results of the extensive hyperspherical harmonics calculation. 

A correlated Gaussian with the global vectors can easily be 
adapted to bound-state problems of a larger system 
as the tedious angular momentum couplings of partial-wave 
expansion are avoided. It will be interesting to devise a method 
of applying this flexibility to a continuum calculation 
of scattering and reactions of few-body systems. 

\bigskip

\appendix
\section{Momentum representation of correlated Gaussian basis}
\label{app.A}

The Fourier transform of the CG defines the corresponding 
basis function in momentum space. As is shown in Eq.~(\ref{cg.mom}) 
below, it is  
again a linear combination of the momentum space CG. The case with 
a single GV has recently been carried out in ref.~\cite{brac}.  
The momentum 
space CG is useful to evaluate matrix elements such as 
semi-relativistic kinetic energy or momentum 
distribution between the particles. See Appendix B.3. 

The transformation from the coordinate to momentum space is achieved 
by a function 
\begin{equation}
\Phi({\vi k},{\vi x})=\frac{1}{(2\pi)^{\frac{3}{2}(N-1)}}\, {\rm e}^{i\tilde{\vis
 k}{\vis x}},
\end{equation}
where ${\vi k}$ is an $(N\!-\!1)$-dimensional column 
vector whose $i$th element ${\vi k}_i$, multiplied by $\hbar$,  
is a momentum conjugate to ${\vi x}_i$.
The Fourier transform of the generating function $g$ reads 
\begin{equation}
\langle \Phi({\vi k},{\vi x})|g({\lambda}_1{\vi e}_1 u_1\!+\!{\lambda}_2{\vi
e}_2 u_2; A,{\vi x})\rangle
=\frac{1}{({\rm det} A)^{\frac{3}{2}}}\, {\rm e}^{\frac{1}{2}\tilde{\vis
v}A^{-1}{\vis v}},
\label{ftr.g}
\end{equation}
where ${\vi v}\!=\!{\lambda}_1{\vi e}_1 u_1\!+\!{\lambda}_2{\vi
e}_2 u_2\!-\!i{\vi k}$. 
Simplifying $\tilde{\vi v}A^{-1}{\vi v}$ by omitting 
both $\lambda_1^2$ and $\lambda_2^2$ terms, we have 
\begin{equation}
\frac{1}{2}\tilde{\vi v}A^{-1}{\vi v} \longrightarrow 
\lambda_1 \lambda_2 \rho \, {\vi e}_1\cdot{\vi e}_2-i\lambda_1 
{\vi e}_1\cdot\widetilde{A^{-1}u_1}{\vi k}-i\lambda_2 
{\vi e}_2\cdot\widetilde{A^{-1}u_2}{\vi k}-\frac{1}{2}\tilde{\vi k}A^{-1}{\vi k}
\label{vaiv}
\end{equation}
with
\begin{equation}
\rho=\widetilde{u_1}A^{-1}u_2.
\end{equation}
Here the symbol $\longrightarrow$ indicates that the $\lambda_i^2$ 
terms can safely be dropped.  
Substituting Eqs.~(\ref{ftr.g}) and (\ref{vaiv}) into Eq.~(\ref{gfn})
enables us to derive 
\begin{eqnarray}
& &\langle \Phi({\vi k},{\vi x})|F_{(L_1L_2)LM}(u_1,u_2,A,{\vi x})\rangle
=B_{L_1}B_{L_2}\frac{1}{({\rm det} A)^{\frac{3}{2}}}
{\rm exp}\left(-\frac{1}{2}\tilde{\vi k}A^{-1}{\vi k}\right)
\nonumber \\
& &\qquad \times \sum_{\ell=0}^{{\rm min}(L_1,L_2)}
\frac{(-i)^{L_1+L_2-2\ell}}{\ell!(L_1-\ell)!(L_2-\ell)!}\rho^{\ell}
\int\!\int d{\vi e}_1\, d{\vi e}_2 \, [Y_{L_1}({\vi e}_1) 
Y_{L_2}({\vi e}_2)]_{LM}
\nonumber \\
& & \qquad \times ({\vi e}_1\cdot{\vi e}_2)^{\ell}
({\vi e}_1\cdot\widetilde{A^{-1}u_1}{\vi k})^{L_1-\ell}
({\vi e}_2\cdot\widetilde{A^{-1}u_2}{\vi k})^{L_2-\ell}.
\end{eqnarray}

As noted below Eq.~(\ref{def.bl}), the above integration has a 
non-vanishing 
contribution provided the vectors ${\vi e}_i$ couple to their 
maximum value $L_i$. This can be done using the relation
\begin{equation}
({\vi e}_1\cdot{\vi e}_2)^{\ell}\Rightarrow
 \frac{\ell!}{B_{\ell}}(-1)^{\ell}\sqrt{2\ell+1}
[Y_{\ell}({\vi e}_1) Y_{\ell}({\vi e}_2)]_{00},
\label{exp.monomial}
\end{equation}
where the symbol $\Rightarrow$ indicates that the angular momentum 
coupling must be made to its maximum value for each ${\vi e}_i$. 
The terms $({\vi e}_1\cdot\widetilde{A^{-1}u_1}{\vi
k})^{L_1-\ell}$ and $({\vi e}_2\cdot\widetilde{A^{-1}u_2}{\vi
k})^{L_2-\ell}$ are expanded in a similar manner.  Combining 
these results leads to 
\begin{eqnarray}
& &\langle \Phi({\vi k},{\vi x})|F_{(L_1L_2)LM}(u_1,u_2,A,{\vi x})\rangle
\nonumber \\
& &\ =  
\frac{(-i)^{L_1+L_2}}{({\rm det} A)^{\frac{3}{2}}}
\sum_{\ell=0}^{\ell_M}
{\cal K}(L_1L_2L; \ell)(-\rho)^{\ell}
F_{(L_1-\ell\, L_2-\ell)LM}(A^{-1}u_1, A^{-1}u_2, A^{-1}, {\vi k}),
\label{cg.mom}
\end{eqnarray}
where $\ell_M\!=\!{\rm min}(L_1,L_2,[(L_1\!+\!L_2\!-\!L)/2])$. 
The coefficient ${\cal K}(L_1 L_2 L; \ell)$ is given by
\begin{eqnarray}
{\cal K}(L_1 L_2 L; \ell)\!&\!=\!&\! 
\frac{4\pi(-1)^{\ell+L-L_1-L_2}(2L_1\!+\!1)!!(2L_2\!+\!1)!!}{(2\ell\!+\!1)!!(2L_1\!-\!2\ell\!+\!1)!!(2L_2\!-\!2\ell\!+\!1)!!}
\sqrt{\frac{(2L_1\!+\!1)(2L_2\!+\!1)}{(2\ell\!+\!1)(2L\!+\!1)}}
\nonumber \\
&\!\times\!&\! C(\ell\, L_1\!-\!\ell; L_1)\, C(\ell\, L_2\!-\!\ell; L_2)\,
U(L_1\!-\!\ell\, L_1\, L_2\!-\!\ell\, L_2; \ell L),
\label{def.cal.K}
\end{eqnarray}
where $C$ is a coefficient which couples two spherical harmonics 
with the same argument
\begin{equation}
C(l_1l_2;l_3)=\sqrt{\frac{(2l_1\!+\!1)(2l_2\!+\!1)}{4\pi (2l_3\!+\!1)}}
\langle l_10l_20\vert l_30 \rangle,
\end{equation}
and $U$ is a unitary Racah coefficient~\cite{book}. 

The Fourier transforms of the CG of 
Eqs.~(\ref{cg1}) and (\ref{cg2}) are particularly simple:  
\begin{eqnarray}
&&\langle \Phi({\vi k},{\vi x})|F_{(L0)LM}(u_1,u_2,A,{\vi x})\rangle
=\frac{(-i)^{L}}{({\rm det} A)^{\frac{3}{2}}}
F_{(L0)LM}(A^{-1}u_1,A^{-1}u_2,A^{-1},{\vi k}), \nonumber \\
& &\langle \Phi({\vi k},{\vi x})|F_{(L1)LM}(u_1,u_2,A,{\vi x})\rangle
=\frac{(-i)^{L+1}}{({\rm det} A)^{\frac{3}{2}}}
F_{(L1)LM}(A^{-1}u_1,A^{-1}u_2,A^{-1},{\vi k}).
\end{eqnarray}

\bigskip

\section{Matrix elements of local operators}
\label{app.B}

\subsection{Overlap}
The overlap matrix element between the generating functions becomes 
\begin{eqnarray}
& &
\langle g({\lambda}_3{\vi e}_3 u_3\!+\!{\lambda}_4{\vi e}_4 u_4; A',{\vi x}) 
\vert g({\lambda}_1{\vi e}_1 u_1\!+\!{\lambda}_2{\vi e}_2 u_2; A,{\vi x})
\rangle = 
\left({\frac{(2\pi)^{N-1}} { {\rm det}B}}\right)^{\frac{3}{ 2}}
{\rm exp}\left({\frac{1}{2}}\tilde{\vi v}B^{-1}{\vi v}\right)
\nonumber \\
& &\quad \longrightarrow 
\left({\frac{(2\pi)^{N-1}} { {\rm det}B}}\right)^{\frac{3}{ 2}}
{\rm exp}\left(\sum_{j>i=1}^4
\rho_{ij}\lambda_i\lambda_j{\vi e}_i\cdot {\vi e}_j\right),
\label{gme}
\end{eqnarray}
where 
\begin{equation}
{\vi v}=\sum_{i=1}^4 \lambda_i {\vi e}_i u_i, 
\label{def.vector.v}
\end{equation}
and 
\begin{equation}
B=A+A',\ \ \ \ \ {\rho}_{ij}={\widetilde{{u_i}}}B^{-1}u_j.
\end{equation}

Following the procedure explained in Section~(\ref{sect.3}) and 
using Eq.~(\ref{exp.monomial}), we have 
\begin{eqnarray}
& &\prod_{j>i=1}^4{\frac{(\rho_{ij}{\vi e}_i\cdot{\vi e}_j)^{p_{ij}}}{p_{ij}!}}
 \, \Rightarrow \left(\prod_{j>i=1}^4\frac{(-1)^{p_{ij}}\sqrt{2p_{ij}+1}}{B_{p_{ij}}}(\rho_{ij})^{p_{ij}}
\right) \sum_{L}X(p_{13}p_{14}p_{23}p_{24};L)
\nonumber \\
& & \qquad \qquad \times \,
Y(p_{12}p_{12}p_{34}p_{34}\, L_1\!-\!p_{12}\, L_2\!-\!p_{12}\,
L_3\!-\!p_{34}\, L_4\!-\!p_{34}\, 0LL0;LL)
\nonumber \\
& & \qquad \qquad \times \, \Big[\big[Y_{L_1}({\vi e}_1)Y_{L_2}({\vi e}_2)\big]_{L}
\big[Y_{L_3}({\vi e}_3)Y_{L_4}({\vi e}_4)\big]_{L}\Big]_{00}.
\label{prod.YY}
\end{eqnarray}
Here the non-negative integer powers $p_{ij}$ are restricted by $L_i$ 
as follows: 
\begin{eqnarray}
& &p_{12}+p_{13}+p_{14}=L_1,\ \ \  p_{12}+p_{23}+p_{24}=L_2, \nonumber \\
& &p_{13}+p_{23}+p_{34}=L_3,\ \ \  p_{14}+p_{24}+p_{34}=L_4.
\label{eq.pij}
\end{eqnarray}
The coefficient $X$ needed for a stretched coupling for each ${\vi e}_i$, 
\begin{eqnarray}
&&\big[Y_{a}({\vi e}_1)Y_{a}({\vi e}_3)\big]_{00}
\big[Y_{b}({\vi e}_1)Y_{b}({\vi e}_4)\big]_{00}
\big[Y_{c}({\vi e}_2)Y_{c}({\vi e}_3)\big]_{00}
\big[Y_{d}({\vi e}_2)Y_{d}({\vi e}_4)\big]_{00}
\nonumber \\
&&\ \Rightarrow \sum_{L} X(abcd;L)
\Big[\big[Y_{a+b}({\vi e}_1)Y_{c+d}({\vi e}_2)\big]_{L}
\big[Y_{a+c}({\vi e}_3)Y_{b+d}({\vi e}_4)\big]_{L}\Big]_{00},
\end{eqnarray}
is given by 
\begin{eqnarray}
X(abcd;L) 
 &\!\!\!=\!\!\!& \sqrt{\frac{2L\!+\!1}{(2a\!+\!1)(2b\!+\!1)
(2c\!+\!1)(2d\!+\!1)}}\, 
C(a\, b;a\!+\!b)\, C(c \,d;c\!+\!d)
\nonumber \\
&\!\!\! \times \!\!\!& \, 
C(a\, c;a\!+\!c)\, C(b\, d;b\!+\!d)
\left[
\begin{array}{ccc}
a        &  b       &  a\!+\!b  \\
c        &  d       &  c\!+\!d  \\
a\!+\!c &  b\!+\!d&  L  
\end{array}\right],
\label{xcoef}
\end{eqnarray}
where the square bracket $\left[ \cdots \right]$ stands for 
a unitary 9-$j$ coefficient~\cite{book}, while the coefficient $Y$ appearing for the coupling 
\begin{eqnarray}
&&\Big[\Big[\big[Y_{a}({\vi e}_1)Y_{b}({\vi e}_2)\big]_{\ell}
\big[Y_{c}({\vi e}_3)Y_{d}({\vi e}_4)\big]_{\ell}\Big]_{0}
\Big[\big[Y_{\alpha}({\vi e}_1)Y_{\beta}({\vi e}_2)\big]_{\lambda}
\big[Y_{\gamma}({\vi e}_3)Y_{\delta}({\vi
e}_4)\big]_{\nu}\Big]_{\kappa}\Big]_{\kappa \mu}
\nonumber \\
&& \ \Rightarrow \sum_{LL'}
Y(abcd\alpha \beta \gamma \delta \ell \lambda \nu \kappa;LL')
\Big[\big[Y_{a+\alpha}({\vi e}_1)Y_{b+\beta}({\vi e}_2)\big]_{L}
\big[Y_{c+\gamma}({\vi e}_3)Y_{d+\delta}({\vi
e}_4)\big]_{L'}\Big]_{\kappa \mu},
\label{yyyy.coupl}
\end{eqnarray}
is given by    
\begin{eqnarray}
Y(abcd\alpha \beta \gamma \delta \ell \lambda \nu \kappa;LL')
&\!\!\!=\!\!\!&  
\left[
\begin{array}{ccc}
\ell     &  \ell &  0  \\
\lambda  &  \nu  &  \kappa  \\
L        &  L'   &  \kappa  
\end{array}\right]
\left[
\begin{array}{ccc}
a       &  b     &  \ell  \\
\alpha  & \beta  &  \lambda  \\
a\!+\!\alpha &  b\!+\!\beta &  L  
\end{array}\right]
\left[
\begin{array}{ccc}
c      &  d      &  \ell  \\
\gamma  &  \delta  &  \nu  \\
c\!+\!\gamma         &  d\!+\!\delta   &  L'    
\end{array}\right]
\nonumber \\
&\!\!\! \times \!\!\!& C(a \alpha;a\!+\!\alpha)\, C(b
 \beta;b\!+\!\beta)\, C(c \gamma;c\!+\!\gamma)\, 
C(d \delta;d\!+\!\delta).
\end{eqnarray}

The overlap matrix element reads  
\begin{eqnarray}
& &\langle F_{(L_3L_4)LM}({u_3},{u_4},A',{\vi x})
\vert F_{(L_1L_2)LM}(u_1,u_2,A,{\vi x})\rangle \nonumber \\ 
&=\!\!\!\!& \!\frac{(-1)^{L_1+L_2}}{\sqrt{2L\!+\!1}}
\left(\prod_{i=1}^{4}{B_{L_i}}\right)
\left({\frac{(2\pi)^{N-1}} {{\rm det}B}}\right)^{3 \over 2}
\sum_{p_{ij}}\left(\prod_{j>i=1}^4
{\frac{(-1)^{p_{ij}}\sqrt{2p_{ij}\!+\!1}}{B_{p_{ij}}}}
(\rho_{ij})^{p_{ij}}\right)
\label{overlapme} \\
&\times\!\!\!\!&\!\!  X(p_{13}p_{14}p_{23}p_{24};L)
Y(p_{12}p_{12}p_{34}p_{34}\, L_1\!-\!p_{12}\, L_2\!-\!p_{12}\,
L_3\!-\!p_{34}\,  L_4\!-\!p_{34}\,  0LL0;LL).
\nonumber 
\end{eqnarray}
As seen from Eq.~(\ref{eq.pij}), 
only two of $p_{ij}$, e.g., $p_{12}$ and $p_{13}$ are 
independent for given $L_i$ values. The sum in Eq.~(\ref{overlapme}) 
consists of only few terms:  just one term with $p_{13}\!=\!L$ for the 
natural parity, and  three 
terms with (i) $p_{13}\!=\!L\!-\!1, \, p_{14}\!=\!p_{23}\!=\! 1$, (ii)
$p_{13}\!=\!L\!-\!1, \, p_{12}\!=\!p_{34}\!=\! 1$, (iii) 
$p_{13}\!=\!L, \, p_{24}\!=\! 1$ for the unnatural parity. 
Other $p_{ij}$ values  all vanish.

\subsection{Kinetic energy and mean square distance}

The kinetic energy with the center of mass kinetic energy 
being subtracted is expressed as 
$T\!-\!T_{\rm cm}\!=\!({1}/{2})\widetilde{\vi \pi} \Lambda {\vi \pi}$, 
where ${\vi \pi}_j$ is $-{i}\hbar {\partial}/ {\partial {\vi x}_j}$ 
and $\Lambda$ is an appropriate 
$(N\!-\!1)\!\times \!(N\!-\!1)$ symmetric matrix. 
The matrix element between the 
generating functions is given by 
\begin{eqnarray}
& &\langle g({\lambda}_3{\vi e}_3 u_3\!+\!{\lambda}_4{\vi e}_4 u_4; A',{\vi x})
\vert {\frac{1}{2}}\widetilde{\vi \pi} \Lambda {\vi \pi}
\vert g({\lambda}_1{\vi e}_1 u_1\!+\!{\lambda}_2{\vi e}_2 u_2; A,{\vi x})
\rangle \nonumber \\
& &\ = {\frac{\hbar^2}{2}}(R-\widetilde{\vi z}\Lambda {\vi z})
\left({\frac{(2\pi)^{N-1}} { {\rm det}B}}\right)^{\frac{3}{ 2}}
{\rm exp}\left({\frac{1}{2}}\tilde{\vi v}B^{-1}{\vi v}\right), 
\label{kin.gme}
\end{eqnarray}
where 
\begin{equation}
{\vi z}=\sum_{i=1}^4\lambda_i{\vi e}_i A_iB^{-1}u_i,\ \ \ \ \ 
R=3{\rm Tr}(B^{-1}A'\Lambda A),
\label{def.z.R}
\end{equation}
with
\begin{equation}
A_1=A_2=A',\ \ \ \ \ A_3=A_4=-A.
\end{equation}
The desired matrix element is expressed using the 
overlap matrix element as follows: 
\begin{eqnarray}
&\langle&\!\!\!\!\!\!\! F_{(L_3L_4)LM}({u_3},{u_4},A',{\vi x})\vert
T-T_{\rm cm}\vert F_{(L_1L_2)LM}(u_1,u_2,A,{\vi x})\rangle
\nonumber \\
&=\!\!&\!\!\! {\frac{\hbar^2}{2}} \!
\left(\!R- 2\sum_{j>i=1}^4S_{ij}{\partial \over \partial \rho_{ij}}\right)\!
\langle F_{(L_3L_4)LM}({u_3},{u_4},A',{\vi x})
\vert F_{(L_1L_2)LM}(u_1,u_2,A,{\vi x})\rangle,
\label{kineticme}
\end{eqnarray}
where 
\begin{equation}
S_{ij}={\widetilde{{u_i}}}B^{-1}A_i\Lambda A_jB^{-1}u_j.
\end{equation}
Since ${\rho}_{ij}$ appears only as $(\rho_{ij})^{p_{ij}}$ in 
$\langle F' \vert F \rangle$, its differentiation is elementary. 

A mean square distance is a quantity to characterize 
the size of a system. What is meant by it is 
the expectation value of an operator such as 
$({\vi r}_k\!-\!{\vi r}_l)^2$ or $\sum_{k=1}^N({\vi r}_k\!-\!{\vi x}_N)^2$. 
Clearly such operators can be expressed as 
$\widetilde{\vi x}Q{\vi x}$ with an appropriate 
symmetric matrix $Q$. For example, Eq.~(\ref{def.rel.dist.vec}) 
enables us to obtain $Q\!=\!w\widetilde{w}$ for the case of 
$({\vi r}_k\!-\!{\vi r}_l)^2$. The 
matrix element of $\widetilde{\vi x}Q{\vi x}$ 
between the generating functions takes the form
\begin{eqnarray}
& &\langle g({\lambda}_3{\vi e}_3 u_3\!+\!{\lambda}_4{\vi e}_4 u_4; A',{\vi x})
\vert \widetilde{\vi x} Q {\vi x}
\vert g({\lambda}_1{\vi e}_1 u_1\!+\!{\lambda}_2{\vi e}_2 u_2; A,{\vi x})
\rangle \nonumber \\
& &\ = ({\cal R}+\widetilde{\vi v}B^{-1}QB^{-1}{\vi v})
\left({\frac{(2\pi)^{N-1}} { {\rm det}B}}\right)^{\frac{3}{ 2}}
{\rm exp}\left({\frac{1}{2}}\tilde{\vi v}B^{-1}{\vi v}\right)
\label{msd.gme}
\end{eqnarray}
with ${\cal R}\!=\!3{\rm Tr}(B^{-1}Q)$. 
Comparing to the kinetic energy matrix element~(\ref{kin.gme}) leads to 
\begin{eqnarray}
&\langle&\!\!\!\!\!\!\! F_{(L_3L_4)LM}({u_3},{u_4},A',{\vi x})\vert
\widetilde{\vi x} Q {\vi x} \vert F_{(L_1L_2)LM}(u_1,u_2,A,{\vi x})\rangle
\nonumber \\
&=\!\!&\!\!\! 
\left(\!{\cal R}+ 2\sum_{j>i=1}^4{\cal S}_{ij}{\partial \over \partial \rho_{ij}}\right)\!
\langle F_{(L_3L_4)LM}({u_3},{u_4},A',{\vi x})
\vert F_{(L_1L_2)LM}(u_1,u_2,A,{\vi x})\rangle
\end{eqnarray}
with ${\cal S}_{ij}\!=\!{\widetilde{{u_i}}}B^{-1}QB^{-1}u_j$.

\subsection{$\delta$-function or multipole operator}

A local potential $V({\vi r}_k\!-\!{\vi r}_l)$ can be expressed 
with use of ${\vi r}_k\!-\!{\vi r}_l\!=\!\widetilde{w}{\vi x}$ 
(see Eq.~(\ref{def.rel.dist.vec})) as 
\begin{equation}
V({\vi r}_k-{\vi r}_l)=\int d{\vi r}V({\vi r})
\delta(\widetilde{w}{\vi x}-{\vi r}).
\label{V.delta}
\end{equation} 
The matrix element of $V$ can thus be obtained through that of 
the $\delta$-function. 
The single-particle coordinate from the center of mass, 
${\vi r}_k\!-\!{\vi x}_N$, is also expressed in terms of ${\vi x}_i$ 
with appropriate coefficients $w_i$, and 
a single-particle operator $D({\vi r}_k\!-\!{\vi x}_N)$ is 
represented by the $\delta$-function as 
\begin{equation}
D({\vi r}_k-{\vi x}_N)=\int d{\vi r}D({\vi r})   
\delta(\widetilde{w}{\vi x}-{\vi r}),
\end{equation}
so that its matrix element again reduces to that of 
the $\delta$-function. 

The matrix element of the $\delta$-function between the generating 
functions is given by 
\begin{eqnarray}
& &\langle g({\lambda}_3{\vi e}_3 u_3\!+\!{\lambda}_4{\vi e}_4 u_4; A',{\vi x})
\vert \delta(\widetilde{w}{\vi x}-{\vi r}) 
\vert g({\lambda}_1{\vi e}_1 u_1\!+\!{\lambda}_2{\vi e}_2 u_2; A,{\vi x})
\rangle \nonumber \\
& & = \left({\frac{(2\pi)^{N-2}c}{{\rm det}B}}\right)^{\frac{3}{2}}
{\rm exp}\left({\frac{1}{2}}\widetilde{\vi v}B^{-1}{\vi v}
-{\frac{1}{2}}c({\vi r}-\widetilde{w}B^{-1}{\vi v})^2\right)
\nonumber \\
& & \longrightarrow \left({\frac{(2\pi)^{N-2}c}{{\rm det}B}}\right)^{\frac{3}{2}}
{\rm exp}\left(-{\frac{1}{2}}cr^2 +\sum_{j>i=1}^4\bar{\rho}_{ij}
\lambda_i\lambda_j {\vi e}_i\cdot{\vi e}_j 
+c\sum_{i=1}^4 \gamma_i\lambda_i {\vi e}_i\cdot{\vi r}\right), 
\label{me.delta.fn}
\end{eqnarray}
where 
\begin{equation}
c=(\widetilde{w}B^{-1}w)^{-1},\ \ \ \ \ 
\gamma_i=\widetilde{u_i}B^{-1}w=\widetilde{w}B^{-1}u_i,\ \ \ \ \  
\bar{\rho}_{ij}=\rho_{ij}-c\gamma_i\gamma_j=\widetilde{u_i}J^{-1}u_j.
\label{mebasic}
\end{equation}
The exponential part including ${\vi r}$ can be expanded in 
power series:
\begin{eqnarray}
& & \prod_{i=1}^4 \frac{(c\gamma_i {\vi e}_i\cdot{\vi r})^{q_i}}{q_i!} 
\, \Rightarrow \prod_{i=1}^4 \frac{(-1)^{q_i}\sqrt{2q_i+1}}{B_{q_i}}
(c\gamma_i)^{q_i}r^{q_i}\Big[Y_{q_i}({\vi e}_i)Y_{q_i}(\hat{\vi r})\Big]_{00}
\nonumber \\
& & \quad = \left(\prod_{i=1}^4 \frac{(-1)^{q_i}(c\gamma_i)^{q_i}}{B_{q_i}}
r^{q_i}\right)
\sum_{\mu \mu' \ell}\sqrt{2\ell+1}\, C(q_1q_2;\mu)\, C(q_3q_4;\mu')\, 
C(\mu \mu';\ell)
\nonumber \\
& & \quad \times \, \Big[\big[[Y_{q_1}({\vi e}_1)Y_{q_2}({\vi e}_2)]_{\mu}
[Y_{q_3}({\vi e}_3)Y_{q_4}({\vi e}_4)]_{\mu'}\big]_{\ell}
Y_{\ell}(\hat{\vi r})\Big]_{00}.
\label{four.e}
\end{eqnarray}

Using Eqs.~(\ref{prod.YY}), (\ref{yyyy.coupl}) and (\ref{four.e}) leads to 
the formula for the desired matrix element 
\begin{eqnarray}
& &\langle F_{(L_3L_4)L'M'}({u_3},{u_4},A',{\vi x})\vert
\delta(\widetilde{w}{\vi x}-{\vi r})
\vert F_{(L_1L_2)LM}(u_1,u_2,A,{\vi x})\rangle
\nonumber \\
& & \quad = \frac{(-1)^{L_1+L_2+L+M+L'+M'}}{\sqrt{2L'\!+\!1}}
\left(\prod_{i=1}^{4}{B_{L_i}}\right)
\left(\frac{(2\pi)^{N-2}c}{{\rm det}B}\right)^{\frac{3}{ 2}}
\nonumber \\
& &\quad  \times \sum_{p_{ij}}\left(\prod_{j>i=1}^4
{\frac{(-1)^{p_{ij}}\sqrt{2p_{ij}\!+\!1}}{B_{p_{ij}}}}
(\bar{\rho}_{ij})^{p_{ij}}\right)
\sum_{q_i}\left(\prod_{i=1}^4
{\frac{(-1)^{q_i}}{B_{q_i}}}
(c\gamma_i)^{q_i}\right)
\nonumber \\
& & \quad \times \,
\sum_{\kappa \mu \mu' \ell m} \langle LM \ell m \vert L' M' \rangle 
\sqrt{2\ell\!+\!1}X(p_{13}p_{14}p_{23}p_{24};\kappa)
\nonumber \\
& &\quad \times \  
Y(p_{12}p_{12}p_{34}p_{34}\,\, p_{13}\!+\!p_{14}\,\, p_{23}\!+\!p_{24}
\,\, p_{13}\!+\!p_{23}\,\, p_{14}\!+\!p_{24}\,0\kappa \kappa 0;\kappa \kappa)
\nonumber \\
& & \quad \times \ 
Y(L_1\!-\!q_1\, L_2\!-\!q_2\, L_3\!-\!q_3\, L_4\!-\!q_4\, q_1q_2q_3
q_4\kappa \mu \mu' \ell; LL')
\nonumber \\
& & \quad \times \ 
C(q_1q_2;\mu)\, C(q_3q_4;\mu')\, C(\mu \mu';\ell)\,
r^{q_1+q_2+q_3+q_4}{\rm e}^{-{\frac{1}{2}}cr^2}Y_{\ell\,-m}(\hat{\vi
r}).
\label{me.delta}
\end{eqnarray}
Here $p_{ij}$ and $q_i$ are non-negative integers, which  
must satisfy the following equations
\begin{eqnarray}
& &p_{12}+p_{13}+p_{14}+q_1=L_1,\ \ \  p_{12}+p_{23}+p_{24}+q_2=L_2, 
\nonumber \\
& &p_{13}+p_{23}+p_{34}+q_3=L_3,\ \ \  p_{14}+p_{24}+p_{34}+q_4=L_4.
\label{pqsum}
\end{eqnarray}
Using the expansion 
\begin{equation}
\delta(\widetilde{w}{\vi x}-{\vi r}) = \frac{\delta(\mid \widetilde{w}{\vi x} \mid -r)}
{r^2}\sum_{\ell m}Y_{\ell m}(\widehat{\widetilde{w}{\vi x}})Y_{\ell m}(\hat{\vi r})^*,
\end{equation}
we obtain the matrix element for the multipole operator 
\begin{eqnarray}
& &\langle F_{(L_3L_4)L'M'}({u_3},{u_4},A',{\vi x})\vert
V(\mid \!\widetilde{w}{\vi x}\! \mid) Y_{\ell m}(\widehat{\widetilde{w}{\vi x}})
\vert F_{(L_1L_2)LM}(u_1,u_2,A,{\vi x})\rangle
\nonumber \\
& & \quad = \frac{(-1)^{L_1+L_2+L+L'}}{\sqrt{2L'\!+\!1}}
\langle LM \ell m \vert L' M' \rangle \sqrt{2\ell\!+\!1}
\left(\prod_{i=1}^{4}{B_{L_i}}\right)
\left({\frac{(2\pi)^{N-2}c}{{\rm det}B}}\right)^{\frac{3}{2}} 
\nonumber \\
& &\quad  \times \sum_{p_{ij}}\left(\prod_{j>i=1}^4
{\frac{(-1)^{p_{ij}}\sqrt{2p_{ij}\!+\!1}}{B_{p_{ij}}}}
(\bar{\rho}_{ij})^{p_{ij}}\right)
\sum_{q_i}\left(\prod_{i=1}^4
{\frac{(-1)^{q_i}}{B_{q_i}}}
(c\gamma_i)^{q_i}\right)
\nonumber \\
& & \quad \times \,
\int_0^{\infty} dr\, r^{q_1+q_2+q_3+q_4+2}
{\rm e}^{-{\frac{1}{2}}cr^2}V(r)\nonumber \\
& & \quad \times \,
\sum_{\kappa \mu \mu'} X(p_{13}p_{14}p_{23}p_{24};\kappa)\,
C(q_1q_2;\mu)\,C(q_3q_4;\mu')\,C(\mu \mu';\ell)
\nonumber \\
& &\quad \times \ 
Y(p_{12}p_{12}p_{34}p_{34}\,\, p_{13}\!+\!p_{14}\,\, p_{23}\!+\!p_{24}
\,\, p_{13}\!+\!p_{23}\,\, p_{14}\!+\!p_{24}\, 0\kappa \kappa 0;\kappa \kappa)
\nonumber \\
& & \quad \times \ 
Y(L_1\!-\!q_1\, L_2\!-\!q_2\, L_3\!-\!q_3\, L_4\!-\!q_4\, q_1q_2q_3
q_4\kappa \mu \mu' \ell; LL').
\label{me.multipole}
\end{eqnarray}

Equation~(\ref{me.multipole}) 
with $\ell m\!=\!00$ has many important applications. 
The matrix element of the relative distance 
$|{\vi r}_k\!-\!{\vi r}_l|$ or $|{\vi r}_k\!-\!{\vi x}_N|$ is 
calculated by choosing $V(r)=r$ with appropriate $w$. 
The correlation function (\ref{correlation.function}) 
is also easily calculated by choosing $V(r)$=$\delta(r-a)$ (with $a$ being
replaced by $r$ later). Of particular importance in hadron spectroscopy 
is the matrix element for the semi-relativistic 
kinetic energy, $\sqrt{(\widetilde{w}{\vi \pi})^2+\mu^2}$. 
This matrix element reduces to that of 
the central matrix element with $V(r)\!=\!\sqrt{r^2+\mu^2}$ because 
the Fourier transform of the coordinate space CG is expressed 
in terms of the momentum space CG (see Eq.~(\ref{cg.mom})). 
The relative momentum distribution between the particles 
(\ref{momentum.correlation}) can also easily
be calculated in the momentum space by choosing $V(r)$ appropriately. 

The spin-orbit matrix element is obtained from the following result 
\begin{eqnarray}
& &\langle g({\lambda}_3{\vi e}_3 u_3\!+\!{\lambda}_4{\vi e}_4 u_4; A',{\vi x})
\vert \delta(\widetilde{w}{\vi x}-{\vi r}) (\widetilde{w}{\vi x}\times \widetilde{\xi}{\vi \pi})
\vert g({\lambda}_1{\vi e}_1 u_1\!+\!{\lambda}_2{\vi e}_2 u_2; A,{\vi x})
\rangle \nonumber \\
& & \quad = \frac{\hbar}{i}\left({\frac{(2\pi)^{N-2}c}{{\rm det}B}}\right)^{\frac{3}{2}}
{\vi r} \times (\widetilde{\xi}{\vi z}\!+\!c
\widetilde{\xi}AB^{-1}w\widetilde{w}B^{-1}{\vi v})
\nonumber \\
& & \quad \, \times \, 
{\rm exp}\left({\frac{1}{2}}\widetilde{\vi v}B^{-1}{\vi v}
\!-\!{\frac{1}{2}}c({\vi r}\!-\!\widetilde{w}B^{-1}{\vi v})^2\right),
\label{spin-orbit.generic}
\end{eqnarray}
where ${\vi z}$ is defined in Eq.~(\ref{def.z.R}). 
When the radial form of the spin-orbit potential is scalar, i.e. 
$V({\vi r})$ is a function of $r$, we may omit 
$c\widetilde{\xi}AB^{-1}w\widetilde{w}B^{-1}{\vi v}$ thanks to the relation
\begin{equation}
\int d{\vi r}\, V(r)({\vi r}\times {\vi a})\,  
{\rm exp}\left(-{\frac{1}{2}}c({\vi r}-{\vi a})^2\right)=0,
\end{equation}
which leads to 
\begin{eqnarray}
& &\langle g({\lambda}_3{\vi e}_3 u_3\!+\!{\lambda}_4{\vi e}_4 u_4; A',{\vi x})
\vert V(\vert\widetilde{w}{\vi x}\vert) 
(\widetilde{w}{\vi x}\times \widetilde{\xi}{\vi \pi})_m
\vert g({\lambda}_1{\vi e}_1 u_1\!+\!{\lambda}_2{\vi e}_2 u_2; A,{\vi x})
\rangle \nonumber \\
& &\longrightarrow \frac{\hbar}{i}\left({\frac{(2\pi)^{N-2}c}{{\rm
		    det}B}}\right)^{\frac{3}{2}} \int d{\vi r}\, V(r)
({\vi r} \times \widetilde{\xi}{\vi z})_m
\nonumber \\
& &\qquad \times 
{\rm exp}\left(-{\frac{1}{2}}cr^2 +\sum_{j>i=1}^4\bar{\rho}_{ij}
\lambda_i\lambda_j {\vi e}_i\cdot{\vi e}_j 
+c\sum_{i=1}^4 \gamma_i\lambda_i {\vi e}_i\cdot{\vi r}\right),
\label{meso.gfn}
\end{eqnarray}
where $({\vi a}\times {\vi b})_m$ stands for $-\sqrt{2}i[{\vi a}\times{\vi
b}]_{1m}\!=\!-(4\sqrt{2}\pi/3)iab[Y_1(\hat{\vi a})Y_1(\hat{\vi
b})]_{1m}$. 

Comparing Eq.~(\ref{meso.gfn}) with the matrix 
element~(\ref{me.delta.fn}) for the
$\delta$-function, it is possible to 
obtain an analytic expression for the spin-orbit matrix element though 
it is very much involved. 

\subsection{Potentials of Gaussian radial form}

When the radial form of $V(r)$ is Gaussian, ${\rm exp}\,{(-c'r^2/2)}$,
the formula 
turns out to be much more concise. Even when its form is not 
Gaussian, it may be approximated very well by a superposition of 
Gaussians with different $c'$ values. As we show below, the 
matrix element of operators with Gaussian radial form 
can be expressed using the overlap matrix element. 
\noindent
\begin{flushleft}
\underline{(i)\ \ Central potential}
\end{flushleft}
Using Eq.~(\ref{me.delta.fn}) yields 
\begin{eqnarray}
& &\langle g({\lambda}_3{\vi e}_3 u_3\!+\!{\lambda}_4{\vi e}_4 u_4; A',{\vi x})
\vert \exp\Big\{-\frac{1}{2}c'(\widetilde{w}{\vi x})^2\Big\} 
\vert g({\lambda}_1{\vi e}_1 u_1\!+\!{\lambda}_2{\vi e}_2 u_2; A,{\vi x})
\rangle \nonumber \\
& & \longrightarrow \left(\frac{c}{c+c'}\right)^{\frac{3}{2}}
\left({\frac{(2\pi)^{N-1}}{{\rm det}B}}\right)^{\frac{3}{2}}
{\rm exp}\left(\sum_{j>i=1}^4\Big[{\rho}_{ij}-
\frac{cc'}{c+c'}\gamma_i\gamma_j\Big]
\lambda_i\lambda_j {\vi e}_i\cdot{\vi e}_j \right). 
\end{eqnarray}
Comparing this result with Eq.~(\ref{gme}) confirms that 
$\rho_{ij}$ in the overlap is here replaced with 
${\rho}_{ij}-({cc'}/({c+c'}))\gamma_i\gamma_j$.  
The desired matrix element takes 
exactly the same form as the overlap 
\begin{eqnarray}
&\langle&\!\!\!\!\!\!\! F_{(L_3L_4)LM}({u_3},{u_4},A',{\vi x})\vert
\exp\Big\{-\frac{1}{2}c'(\widetilde{w}{\vi x})^2\Big\}
\vert F_{(L_1L_2)LM}(u_1,u_2,A,{\vi x})\rangle 
\nonumber \\
&=\!\! &\!\!\!\! \left(\!\frac{c}{c+c'}\!\right)^{\frac{3}{2}} 
\!\langle F_{(L_3L_4)LM}({u_3},{u_4},A',{\vi x})
\vert F_{(L_1L_2)LM}(u_1,u_2,A,{\vi x})\rangle\Big|_{\rho_{ij}
\to \rho_{ij}-\frac{cc'}{c+c'}\gamma_i\gamma_j}.
\end{eqnarray}

The Coulomb potential $V(r)\!=\!1/r$ is expressed as a
superposition of Gaussian potentials:  
\begin{equation}
\frac{1}{r}=\frac{2}{\sqrt{\pi}}\int_0^{\infty}dt\, {\rm e}^{-r^2t^2}.
\end{equation}
Thus we obtain
\begin{eqnarray}
& &\langle g({\lambda}_3{\vi e}_3 u_3\!+\!{\lambda}_4{\vi e}_4 u_4; A',{\vi x})
\vert \frac{1}{|\widetilde{w}{\vi x}|}
\vert g({\lambda}_1{\vi e}_1 u_1\!+\!{\lambda}_2{\vi e}_2 u_2; A,{\vi x})
\rangle \nonumber \\
& & =\frac{2}{\sqrt{\pi}}\int_0^{\infty}dt
\langle g({\lambda}_3{\vi e}_3 u_3\!+\!{\lambda}_4{\vi e}_4 u_4; A',{\vi x})
\vert \exp\left\{-t^2(\widetilde{w}{\vi x})^2\right\} 
\vert g({\lambda}_1{\vi e}_1 u_1\!+\!{\lambda}_2{\vi e}_2 u_2; A,{\vi x})
\rangle \nonumber \\
& & \longrightarrow \sqrt{\frac{2c}{\pi}}\int_0^1 du 
\left({\frac{(2\pi)^{N-1}}{{\rm det}B}}\right)^{\frac{3}{2}}
{\rm exp}\left(\sum_{j>i=1}^4\left[{\rho}_{ij}-cu^2 \gamma_i\gamma_j\right]
\lambda_i\lambda_j{\vi e}_i\cdot{\vi e}_j\right),
\end{eqnarray}
where a change of the integration variable is performed through  
$t$=$\sqrt{{c}/{2}}\,u/{\sqrt{1\!-\!u^2}}$.
The Coulomb matrix element reduces to the following  
integral of the overlap matrix element 
\begin{eqnarray}
&\langle&\!\!\!\!\!\!\! F_{(L_3L_4)LM}({u_3},{u_4},A',{\vi x})\vert
\frac{1}{|\widetilde{w}{\vi x}|}
\vert F_{(L_1L_2)LM}(u_1,u_2,A,{\vi x})\rangle
\nonumber \\
&=\!\!&\!\!\! \!\sqrt{\frac{2c}{\pi}}\!\int_0^1 \!du 
\langle F_{(L_3L_4)LM}({u_3},{u_4},A',{\vi x})
\vert F_{(L_1L_2)LM}(u_1,u_2,A,{\vi x})\rangle\Big|_{\rho_{ij}
\to \rho_{ij}-cu^2\gamma_i\gamma_j}.
\end{eqnarray}
The above integrand is a polynomial function of $u$ whose 
degree is at 
most $L_1\!+\!L_2\!+\!L_3\!+\!L_4$, so that the integration can be 
accurately performed with use of the Gauss quadrature. 

We note that the modified Coulomb potential
\begin{equation}
\frac{1}{r}\, {\rm erf}(\kappa r)=\frac{2\kappa}{\sqrt{\pi}}\int_0^1 dt \,
{\rm e}^{-\kappa^2 r^2t^2}
\label{mod.coulomb}
\end{equation}
can easily be calculated in the same way as above.  

Evaluating the matrix element of Yukawa potential is possible 
with use of Eq.~(\ref{me.multipole}). Another simple formula is,
however, obtained as below. Expressing the Yukawa potential as
\begin{equation}
\frac{1}{r}\, {\rm e}^{-\kappa r}=\frac{2}{\sqrt{\pi}}\int_0^{\infty} dt \,
{\rm exp}\left(-r^2t^2-\frac{\kappa^2}{4t^2}\right),
\end{equation}
which is a superposition of Gaussian potentials, 
we get an expression for the matrix element
\begin{eqnarray}
&\langle&\!\!\!\!\!\!\! F_{(L_3L_4)LM}({u_3},{u_4},A',{\vi x})\vert
\frac{1}{|\widetilde{w}{\vi x}|}\, {\rm e}^{-\kappa |\widetilde{w}{\vi x}|}
\vert F_{(L_1L_2)LM}(u_1,u_2,A,{\vi x})\rangle
\nonumber \\
&=\!\!&\!\!\!
 \!\sqrt{\frac{2c}{\pi}}\!\int_0^1 \!du \,
\exp\left(-\frac{\kappa^2}{2c}\frac{1-u^2}{u^2}\right)
\nonumber \\
&\!\!&\!\!\! \, \times 
\langle F_{(L_3L_4)LM}({u_3},{u_4},A',{\vi x})
\vert F_{(L_1L_2)LM}(u_1,u_2,A,{\vi x})\rangle\Big|_{\rho_{ij}
\to \rho_{ij}-cu^2\gamma_i\gamma_j}.
\end{eqnarray}

As the last example of the Gaussian central potential, 
we consider $V(r)=r\, {\exp}(-c'r^2/2)$, which is a 
derivative of the Gaussian potential, ${\exp}(-c'r^2/2)$. Though 
we can evaluate the matrix element for this potential 
from Eq.~(\ref{me.multipole}), it is possible to relate its matrix 
element to the overlap matrix element. Starting from
Eq.~(\ref{me.delta.fn}), we obtain
\begin{eqnarray}
& &\langle g({\lambda}_3{\vi e}_3 u_3\!+\!{\lambda}_4{\vi e}_4 u_4; A',{\vi x})
\vert |\widetilde{w}{\vi x}|\exp\Big\{-\frac{1}{2}c'(\widetilde{w}{\vi x})^2\Big\} 
\vert g({\lambda}_1{\vi e}_1 u_1\!+\!{\lambda}_2{\vi e}_2 u_2; A,{\vi x})
\rangle \nonumber \\
& & \longrightarrow \sqrt{\frac{2}{\pi c}}\left(\frac{c}{c+c'}\right)^{2}
\left({\frac{(2\pi)^{N-1}}{{\rm det}B}}\right)^{\frac{3}{2}}
{\rm exp}\left(\sum_{j>i=1}^4 \bar{{\rho}}_{ij}
\lambda_i\lambda_j {\vi e}_i\cdot{\vi e}_j \right)
\nonumber \\
& &\qquad \times \Bigg( 1+ \Big[1+\frac{2c^2}{c+c'}\sum_{j>i=1}^4 \gamma_i \gamma_j
\lambda_i\lambda_j {\vi e}_i\cdot{\vi e}_j
\Big]
\nonumber \\
& &\qquad \times \int_0^1 du
\exp\Big\{\frac{c^2}{c+c'}(1-u^2)\sum_{j>i=1}^4 \gamma_i \gamma_j
\lambda_i\lambda_j {\vi e}_i\cdot{\vi e}_j\Big\}
\Bigg). 
\end{eqnarray}
The required matrix thus reads 
\begin{eqnarray}
&\langle&\!\!\!\!\!\!\! F_{(L_3L_4)LM}({u_3},{u_4},A',{\vi x})\vert \, 
|\widetilde{w}{\vi x}|\, \exp\Big\{-\frac{1}{2}c'(\widetilde{w}{\vi x})^2\Big\} 
\vert F_{(L_1L_2)LM}(u_1,u_2,A,{\vi x})\rangle
\nonumber \\
&=\!\!&\!\!\!\sqrt{\frac{2}{\pi c}}\left(\frac{c}{c+c'}\right)^{2}
 \Bigg( 
\langle F_{(L_3L_4)LM}({u_3},{u_4},A',{\vi x})
\vert F_{(L_1L_2)LM}(u_1,u_2,A,{\vi x})\rangle\Big|_{\rho_{ij}
\to \bar{\rho_{ij}}}
\nonumber \\
& +\!\!\!&\!\!\!\int_0^1 du \, \Big[1+\frac{2c^2}{c+c'}\sum_{j>i=1}^4 \gamma_i \gamma_j
\frac{\partial}{\partial \rho_{ij}}
\Big]
\nonumber \\
&\times\!\!\! &\!\!\! \langle F_{(L_3L_4)LM}({u_3},{u_4},A',{\vi x})
\vert F_{(L_1L_2)LM}(u_1,u_2,A,{\vi x})\rangle\Big|_{\rho_{ij}
\to \rho_{ij}-\frac{c(c'+cu^2)}{c+c'}\gamma_i \gamma_j }
\Bigg).
\end{eqnarray}
 
\noindent
\begin{flushleft}
\underline{(ii)\ \ Tensor potential}
\end{flushleft}
For $V({\vi r})\!=\!{\rm exp}\,{(-c'r^2/2)}\,Y_{2m}(\hat{\vi r})$, we 
express it as Gaussian radial form by using a formula
\begin{equation}
{\rm e}^{-\frac{1}{2}c'r^2}Y_{2m}(\hat{\vi r})
={\rm e}^{-\frac{1}{2}c'r^2}{\cal Y}_{2m}({\vi r})
\int_0^{\infty} dt\, {\rm e}^{-tr^2}.
\label{tensor.no.r2} 
\end{equation}
Then use of Eq.~(\ref{me.delta.fn}) leads to 
\begin{eqnarray}
& &\langle g({\lambda}_3{\vi e}_3 u_3\!+\!{\lambda}_4{\vi e}_4 u_4; A',{\vi x})
\vert \exp\Big\{-\frac{1}{2}c'(\widetilde{w}{\vi x})^2\Big\}
Y_{2m}(\widehat{\widetilde{w}{\vi x}})
\vert g({\lambda}_1{\vi e}_1 u_1\!+\!{\lambda}_2{\vi e}_2 u_2; A,{\vi x})
\rangle \nonumber \\
& & \longrightarrow \int_0^{\infty} dt \int d{\vi r} \,
{\rm e}^{-\frac{1}{2}(c'+2t)r^2}{\cal Y}_{2m}({\vi r})
\nonumber \\
& & \quad \times 
\left({\frac{(2\pi)^{N-2}c}{{\rm det}B}}\right)^{\frac{3}{2}}
{\rm exp}\left(-{\frac{1}{2}}cr^2 +\sum_{j>i=1}^4\bar{\rho}_{ij}
\lambda_i\lambda_j {\vi e}_i\cdot{\vi e}_j 
+c\sum_{i=1}^4 \gamma_i\lambda_i {\vi e}_i\cdot{\vi r}\right)
\nonumber \\
& & \longrightarrow  
\left({\frac{(2\pi)^{N-1}}{{\rm det}B}}\right)^{\frac{3}{2}}
{\cal Y}_{2m}\Big(\sum_{i=1}^4\gamma_i\lambda_i{\vi e}_i\Big)
\nonumber \\
& & \quad \times \int_0^{\infty} dt 
\left(\frac{c}{c+c'+2t}\right)^{\frac{7}{2}}
{\rm exp}\left(\sum_{j>i=1}^4\Big[\bar{\rho}_{ij}+\frac{c^2}{c+c'+2t}
\gamma_i\gamma_j\Big]\lambda_i\lambda_j {\vi e}_i\cdot{\vi e}_j 
\right)
\nonumber \\
& &  = \frac{c}{2}
\left({\frac{(2\pi)^{N-1}}{{\rm det}B}}\right)^{\frac{3}{2}}
{\cal Y}_{2m}\Big(\sum_{i=1}^4\gamma_i\lambda_i{\vi e}_i\Big)
\nonumber \\
& & \quad \times 
\int_0^{\frac{c}{c+c'}} du \, u^{\frac{3}{2}}\, 
{\rm exp}\left(\sum_{j>i=1}^4\Big[\bar{\rho}_{ij}+cu
\gamma_i\gamma_j\Big]\lambda_i\lambda_j {\vi e}_i\cdot{\vi e}_j 
\right).
\label{tensor.aux}
\end{eqnarray}

The term ${\cal Y}_{2m}(\sum_{i=1}^4\gamma_i\lambda_i{\vi e}_i)$ 
is expanded as follows: 
\begin{eqnarray}
{\cal Y}_{2m}\Big(\sum_{i=1}^4\gamma_i\lambda_i{\vi e}_i\Big)\!
&\!\!\!=\!\!\!&\!\sum_{i=1}^4\gamma_i^2\lambda_i^2Y_{2m}({\vi e}_i)+
\sqrt{\frac{40\pi}{3}}\sum_{j>i=1}^4\gamma_i\gamma_j\lambda_i\lambda_j
[Y_{1}({\vi e}_i)Y_{1}({\vi e}_j)]_{2m}.
\label{quadrupole.op}
\end{eqnarray}
The exponential part of the integrand of Eq.~(\ref{tensor.aux}) 
is expanded in power series 
$(\lambda_i\lambda_j {\vi e}_i\cdot{\vi e}_j)^{\bar{p}_{ij}}$ as in
Appendix B.1. 
Suppose that the ${\bar p}_{ij}$ values give the following angular momenta 
\begin{eqnarray}
& &{\bar p}_{12}+{\bar p}_{13}+{\bar p}_{14}={\bar L}_1,\ \ \  {\bar p}_{12}+{\bar p}_{23}+{\bar p}_{24}={\bar L}_2, \nonumber \\
& &{\bar p}_{13}+{\bar p}_{23}+{\bar p}_{34}={\bar L}_3,\ \ \  {\bar
 p}_{14}+{\bar p}_{24}+{\bar p}_{34}={\bar L}_4.
\label{exponent.aux}
\end{eqnarray}
These terms contribute to the matrix element provided 
${\bar L}_k\!=\!L_k\!-\!2\delta_{ki}$ is met in the case of  
the coupling with $Y_{2m}({\vi e}_i)$, the first term 
of the right side of Eq.~(\ref{quadrupole.op}), while 
${\bar L}_k\!=\!L_k\!-\!\delta_{ki}\!-\!\delta_{kj}$ 
is met for the coupling with the second term, 
$[Y_{1}({\vi e}_i)Y_{1}({\vi e}_j)]_{2m}$.   
Let us define the corresponding coupling coefficients $Z_1$ and $Z_2$ by
\begin{eqnarray}
& &Y_{\kappa \mu}({\vi e}_i)
\Big[\big[Y_{{\bar L}_1}({\vi e}_1)Y_{{\bar L}_2}({\vi e}_2)\big]_{{\bar L}}
\big[Y_{{\bar L}_3}({\vi e}_3)Y_{{\bar L}_4}({\vi e}_4)\big]_{{\bar L}}\Big]_{00}
\nonumber \\
& & \Rightarrow 
\sum_{LL'} Z_1(\kappa {\bar L}_1{\bar L}_2{\bar L}_3{\bar L}_4 {\bar
L},LL'; i)
\Big[\big[Y_{L_1}({\vi e}_1)Y_{L_2}({\vi e}_2)\big]_{L}
\big[Y_{L_3}({\vi e}_3)Y_{L_4}({\vi e}_4)\big]_{L'}\Big]_{\kappa \mu},
\end{eqnarray}
and 
\begin{eqnarray}
& &[Y_{1}({\vi e}_k)Y_1({\vi e}_l)]_{\kappa \mu} 
\Big[\big[Y_{{\bar L}_1}({\vi e}_1)Y_{{\bar L}_2}({\vi e}_2)\big]_{{\bar L}}
\big[Y_{{\bar L}_3}({\vi e}_3)Y_{{\bar L}_4}({\vi e}_4)\big]_{{\bar L}}\Big]_{00}
\nonumber \\
& & \Rightarrow \sum_{LL'} Z_2(\kappa {\bar L}_1{\bar L}_2{\bar L}_3{\bar L}_4 
{\bar L},LL'; kl)
\Big[\big[Y_{L_1}({\vi e}_1)Y_{L_2}({\vi e}_2)\big]_{L}
\big[Y_{L_3}({\vi e}_3)Y_{L_4}({\vi e}_4)\big]_{L'}\Big]_{\kappa \mu}.
\label{z2.coef}
\end{eqnarray}
The $Z_1$ coefficients are given by
\begin{eqnarray}
& &Z_1(\kappa {\bar L}_1{\bar L}_2{\bar L}_3{\bar L}_4{\bar L}, LL'; 1)
=\delta_{{\bar L}L'}(-1)^{\kappa+{\bar L}-L}
\sqrt{\frac{2L\!+\!1}{(2\kappa\!+\!1)(2{\bar L}\!+\!1)}}
W(\kappa {\bar L}_1 {\bar L}_2 {\bar L} L),
\nonumber \\
& &Z_1(\kappa {\bar L}_1{\bar L}_2{\bar L}_3{\bar L}_4{\bar L}, LL'; 2)
=\delta_{{\bar L}L'}
\sqrt{\frac{2L+1}{(2\kappa\!+\!1)(2{\bar L}\!+\!1)}}
W(\kappa {\bar L}_2 {\bar L}_1 {\bar L} L),
\nonumber \\
& &Z_1(\kappa {\bar L}_1{\bar L}_2{\bar L}_3{\bar L}_4{\bar L}, LL'; 3)
=\delta_{{\bar L}L}
\sqrt{\frac{2L'\!+\!1}{(2\kappa\!+\!1)(2{\bar L}\!+\!1)}}
W(\kappa {\bar L}_3 {\bar L}_4 {\bar L} L'),
\nonumber \\
& &Z_1(\kappa {\bar L}_1{\bar L}_2{\bar L}_3{\bar L}_4{\bar L}, LL'; 4)
=\delta_{{\bar L}L}(-1)^{\kappa+{\bar L}-L'}
\sqrt{\frac{2L'\!+\!1}{(2\kappa\!+\!1)(2{\bar L}\!+\!1)}}
W(\kappa {\bar L}_4 {\bar L}_3 {\bar L} L')
\end{eqnarray}
with
\begin{equation}
W(\kappa abcd)=C(\kappa a; \kappa+a)U(\kappa a d b;\kappa\!+\!a \, c ).
\end{equation}
The $Z_2$ coefficients are given by 
\begin{eqnarray}
& &Z_2(\kappa {\bar L}_1{\bar L}_2{\bar L}_3{\bar L}_4{\bar L}, LL'; 12)
\nonumber \\
& & \quad =\delta_{{\bar L}L'}(-1)^{\kappa+{\bar L}-L}\sqrt{\frac{2L+1}
{(2\kappa+1)(2{\bar L}+1)}}
C(1{\bar L}_1;L_1)C(1{\bar L}_2;L_2)
\left[
\begin{array}{ccc}
1           &  1         & \kappa   \\
{\bar L}_1  & {\bar L}_2 & {\bar L} \\
L_1         &  L_2       & L  
\end{array}\right],
\nonumber \\
& &Z_2(\kappa {\bar L}_1{\bar L}_2{\bar L}_3{\bar L}_4{\bar L}, LL'; 13)
=W(1{\bar L}_1{\bar L}_2 {\bar L}L)W(1{\bar L}_3 {\bar L}_4 {\bar L} L')
\left[
\begin{array}{ccc}
1           &  1         & \kappa   \\
{\bar L}    & {\bar L}   & 0  \\
L           &  L'        & \kappa   
\end{array}\right],
\nonumber \\
& &Z_2(\kappa {\bar L}_1{\bar L}_2{\bar L}_3{\bar L}_4{\bar L}, LL'; 14)
=(-1)^{{\bar L}-L'+1}
W(1{\bar L}_1{\bar L}_2 {\bar L}L)W(1{\bar L}_4{\bar L}_3 {\bar L}L')
\left[
\begin{array}{ccc}
1           &  1         & \kappa   \\
{\bar L}    & {\bar L}   & 0  \\
L           &  L'        & \kappa   
\end{array}\right],
\nonumber \\
& &Z_2(\kappa {\bar L}_1{\bar L}_2{\bar L}_3{\bar L}_4{\bar L}, LL'; 23)
=(-1)^{{\bar L}-L+1}
W(1{\bar L}_2{\bar L}_1{\bar L}L)
W(1{\bar L}_3{\bar L}_4{\bar L}L')
\left[
\begin{array}{ccc}
1           &  1         & \kappa   \\
{\bar L}    & {\bar L}   & 0  \\
L           &  L'        & \kappa   
\end{array}\right],
\nonumber \\
& &Z_2(\kappa {\bar L}_1{\bar L}_2{\bar L}_3{\bar L}_4{\bar L}, LL'; 24)
=(-1)^{L+L'}
W(1{\bar L}_2{\bar L}_1{\bar L}L)W(1{\bar L}_4{\bar L}_3{\bar L}L')
\left[
\begin{array}{ccc}
1           &  1         & \kappa   \\
{\bar L}    & {\bar L}   & 0  \\
L           &  L'        & \kappa   
\end{array}\right],
\nonumber \\
& &Z_2(\kappa {\bar L}_1{\bar L}_2{\bar L}_3{\bar L}_4{\bar L}, LL'; 34)
\nonumber \\
& & \quad 
=\delta_{{\bar L}L}\sqrt{\frac{2L'+1}{(2\kappa+1)(2{\bar L}+1)}}
C(1{\bar L}_3 ;L_3)C(1{\bar L}_4 ;L_4)
\left[
\begin{array}{ccc}
1           &  1         & \kappa   \\
{\bar L}_3  & {\bar L}_4 & {\bar L}  \\
L_3         &  L_4       & L'  
\end{array}\right].
\end{eqnarray}

The tensor matrix element of the Gaussian radial form reads 
\begin{eqnarray}
& &\langle F_{(L_3L_4)L'M'}({u_3},{u_4},A',{\vi x})\vert
\exp\Big\{-\frac{1}{2}c'(\widetilde{w}{\vi x})^2\Big\}
Y_{2m}(\widehat{\widetilde{w}{\vi x}})
\vert F_{(L_1L_2)LM}(u_1,u_2,A,{\vi x})\rangle
\nonumber \\
& & \quad = \frac{(-1)^{L_1+L_2+L+L'} \sqrt{5}}{\sqrt{2L'\!+\!1}}
\langle LM 2 m \vert L' M' \rangle\, \frac{c}{2}
\int_0^{\frac{c}{c+c'}} du \, u^{\frac{3}{2}}\,
\nonumber \\
& & \quad \times \Bigg\{
\sum_{k=1}^4 \gamma_k^2(-1)^{{\bar L}_1+{\bar L}_2}
\left(\prod_{i=1}^4 \frac{B_{L_i}}{B_{{\bar L}_i}}\right)
\sum_{\bar L}\sqrt{2{\bar L}+1}
Z_1(2 {\bar L}_1{\bar L}_2{\bar L}_3{\bar L}_4{\bar L}, LL'; k)
\nonumber \\
& & \quad \qquad  \times 
\langle F_{({\bar L}_3 {\bar L}_4){\bar L}{\bar M}}({u_3},{u_4},A',{\vi x})
\vert F_{({\bar L}_1 {\bar L}_2){\bar L}{\bar M}}(u_1,u_2,A,{\vi x})
\rangle\Big|_{\rho_{ij}\to {\bar {\rho}}_{ij}+cu\gamma_i\gamma_j}
\nonumber \\
& & \quad \quad + 
\sqrt{\frac{40\pi}{3}}\sum_{l>k=1}^4 \gamma_k
\gamma_l (-1)^{{\bar L}_1+{\bar L}_2}
\left(\prod_{i=1}^4 \frac{B_{L_i}}{B_{{\bar L}_i}}\right)
\sum_{\bar L}\sqrt{2{\bar L}+1}
Z_2(2 {\bar L}_1{\bar L}_2{\bar L}_3{\bar L}_4{\bar L}, LL'; kl)
\nonumber \\
& & \quad \qquad  \times 
\langle F_{({\bar L}_3 {\bar L}_4){\bar L}{\bar M}}({u_3},{u_4},A',{\vi x})
\vert F_{({\bar L}_1 {\bar L}_2){\bar L}{\bar M}}(u_1,u_2,A,{\vi x})
\rangle\Big|_{\rho_{ij}\to {\bar {\rho}}_{ij}+cu\gamma_i\gamma_j}
\Bigg\}.
\label{tensor.me}
\end{eqnarray}
It should be noted here that ${\bar L}_i\!=\!L_i\!-\!2\delta_{ik}$ 
in the first sum of the curly bracket, whereas 
${\bar L}_i\!=\!L_i\!-\!\delta_{ik}\!-\!\delta_{il}$ in the second sum. 
The integral appearing in Eq.~(\ref{tensor.me}) can be reduced to 
\begin{equation}
\int_0^{\frac{c}{c+c'}} du \, u^{\frac{3}{2}}\,f(u)
=2\left(\frac{c}{c+c'}\right)^{\frac{5}{2}}\int_0^1 dx\, x^4\, 
f\Big(\frac{c}{c+c'}x^2\Big),
\end{equation}
where $f(u)$ is a polynomial function of $u$ whose degree is at most  
$({\bar L}_1\!+\!{\bar L}_2\!+\!{\bar L}_3\!+\!{\bar L}_4)/2$. This
integral can therefore be accurately evaluated using the Gauss quadrature. 

We note that the calculation of the tensor matrix element of the 
following radial form 
\begin{equation}
\langle F_{(L_3L_4)LM}({u_3},{u_4},A',{\vi x})\vert
\exp\Big\{-\frac{1}{2}c'(\widetilde{w}{\vi x})^2\Big\}
{\cal Y}_{2m}({\widetilde{w}{\vi x}})
\vert F_{(L_1L_2)LM}(u_1,u_2,A,{\vi x})\rangle
\label{tensor.with.r2}
\end{equation}
is easier. In this case the integration of Eq.~(\ref{tensor.no.r2}) 
is not needed, and the above matrix 
element (\ref{tensor.with.r2}) is 
obtained simply by replacing  $u$ with ${c}/(c+c')$ in Eq.~(\ref{tensor.me}). 
\noindent
\begin{flushleft}
\underline{(iii)\ \ Spin-orbit potential}
\end{flushleft}
Using Eq.~(\ref{meso.gfn}), we obtain 
\begin{eqnarray}
& &\langle g({\lambda}_3{\vi e}_3 u_3\!+\!{\lambda}_4{\vi e}_4 u_4; A',{\vi x})
\vert \exp\left\{-\frac{1}{2}c'(\widetilde{w}{\vi x})^2\right\}
 (\widetilde{w}{\vi x}\times \widetilde{\xi}{\vi \pi})_m
\vert g({\lambda}_1{\vi e}_1 u_1\!+\!{\lambda}_2{\vi e}_2 u_2; A,{\vi x})
\rangle \nonumber \\
& &\  \longrightarrow \left(\frac{c}{c+c'}\right)^{\frac{5}{2}}
\left({\frac{(2\pi)^{N-1}}{{\rm det}B}}\right)^{\frac{3}{2}}
\sum_{l>k=1}^4
(\gamma_k\tau_l-\gamma_l\tau_k)\lambda_k\lambda_l \frac{\hbar}{i}({\vi e}_k
\times {\vi e}_l)_m
\nonumber \\
& & \quad \times \, 
{\rm exp}\left(\sum_{j>i=1}^4\left[{\rho}_{ij}-
\frac{cc'}{c+c'}\gamma_i\gamma_j\right]
\lambda_i\lambda_j {\vi e}_i\cdot{\vi e}_j \right) 
\end{eqnarray}
with
\begin{equation}
\tau_i=\widetilde{\xi}A_iB^{-1}u_i.
\end{equation}
The difference in the matrix elements between the overlap~(\ref{gme}) 
and the above spin-orbit potential is that the latter contains an extra factor 
$-i ({\vi e}_k\times {\vi e}_l)_m$.

Following the procedure similar to the one described in 
Eqs.~(\ref{quadrupole.op}), (\ref{exponent.aux}) and 
(\ref{z2.coef}), we obtain the spin-orbit matrix element 
for the Gaussian radial form as follows:
\begin{eqnarray}
& &\langle F_{(L_3L_4)L'M'}({u_3},{u_4},A',{\vi x})\vert
\exp\left\{-\frac{1}{2}c'(\widetilde{w}{\vi x})^2\right\} 
({\widetilde{w}{\vi x}}\times {\widetilde{\xi}{\vi \pi}})_m
\vert F_{(L_1L_2)LM}(u_1,u_2,A,{\vi x})\rangle
\nonumber \\
& & \quad = \frac{(-1)^{L_1+L_2+L+L'}}{\sqrt{2L'\!+\!1}}
\langle LM 1m \vert L' M' \rangle \left(4\pi\sqrt{\frac{2}{3}}\hbar\right)
\left(\frac{c}{c+c'}\right)^{\frac{5}{2}}
\nonumber \\
& & \quad \, \times 
\sum_{l>k=1}^4(\gamma_k\tau_l-\gamma_l\tau_k) (-1)^{\bar{L}_1+\bar{L}_2}
\left(\prod_{i=1}^{4}\frac{B_{L_i}}{B_{{\bar L}_i}}\right)
\sum_{\bar L}\sqrt{2{\bar L}+1}
Z_2(1{\bar L}_1{\bar L}_2{\bar L}_3{\bar L}_4{\bar L}, LL'; kl)
\nonumber \\
& & \quad \, \times 
\, \langle F_{({\bar L}_3 {\bar L}_4){\bar L}{\bar M}}({u_3},{u_4},A',{\vi x})
\vert F_{({\bar L}_1 {\bar L}_2){\bar L}{\bar M}}(u_1,u_2,A,{\vi x})
\rangle\Big|_{\rho_{ij}\to {\rho}_{ij}-\frac{cc'}{c+c'}\gamma_i\gamma_j},
\end{eqnarray}
where ${\bar L}_i$ is defined by $L_i\!-\!\delta_{ik}\!-\!\delta_{il}$ 
depending on the summation labels $k$ and $l$.
\noindent
\begin{flushleft}
\underline{(iv)\ \ Multipole moment}
\end{flushleft}
We calculate the matrix element for a multipole moment operator. 
Using Eqs.~(\ref{V.delta}) and (\ref{me.delta.fn}) with 
$V({\vi r})={\cal Y}_{\ell m}({\vi r})$, we obtain 
\begin{eqnarray}
& &\langle g({\lambda}_3{\vi e}_3 u_3\!+\!{\lambda}_4{\vi e}_4 u_4; A',{\vi x})
\vert {\cal Y}_{\ell m}({\widetilde{w}{\vi x}})
\vert g({\lambda}_1{\vi e}_1 u_1\!+\!{\lambda}_2{\vi e}_2 u_2; A,{\vi x})
\rangle \nonumber \\
& & \longrightarrow \int d{\vi r} \,{\cal Y}_{\ell m}({\vi r})
\left({\frac{(2\pi)^{N-2}c}{{\rm det}B}}\right)^{\frac{3}{2}}
{\rm exp}\left(-{\frac{1}{2}}cr^2 +\sum_{j>i=1}^4\bar{\rho}_{ij}
\lambda_i\lambda_j {\vi e}_i\cdot{\vi e}_j 
+c\sum_{i=1}^4 \gamma_i\lambda_i {\vi e}_i\cdot{\vi r} \right)
\nonumber \\
& &  = \left({\frac{(2\pi)^{N-1}}{{\rm det}B}}\right)^{\frac{3}{2}}
{\cal Y}_{\ell m}\Big(\sum_{i=1}^4\gamma_i\lambda_i{\vi e}_i\Big)\, 
{\rm exp}\left(\sum_{j>i=1}^4\rho_{ij}\lambda_i\lambda_j {\vi e}_i\cdot{\vi
	  e}_j \right).
\end{eqnarray}
In the case of $\ell\!=\!2$, the above result is compared to
Eq.~(\ref{tensor.aux}). That is, it is simply obtained by 
dropping ${c}/{2}$ and by setting $u\!=\!1$. Therefore the matrix element 
\[
\langle F_{(L_3L_4)LM}({u_3},{u_4},A',{\vi x})\vert
{\cal Y}_{2m}({\widetilde{w}{\vi x}})
\vert F_{(L_1L_2)LM}(u_1,u_2,A,{\vi x})\rangle 
\]
is obtained from Eq.~(\ref{tensor.me}) by the same procedure as 
noted above. 

It is easy to 
obtain the matrix element for the dipole operator ($\ell\!=\!1$) in a
similar way: 
\begin{eqnarray}
& &\langle F_{(L_3L_4)L'M'}({u_3},{u_4},A',{\vi x})\vert
{\cal Y}_{1m}({\widetilde{w}{\vi x}})
\vert F_{(L_1L_2)LM}(u_1,u_2,A,{\vi x})\rangle
\nonumber \\
& & \quad = \frac{(-1)^{L_1+L_2+L+L'+1}\sqrt{3}}{\sqrt{2L'\!+\!1}}
\langle LM 1 m \vert L' M' \rangle \, 
\nonumber \\
& & \quad \times 
\sum_{k=1}^4 \gamma_k(-1)^{{\bar L}_1+{\bar L}_2}
\left(\prod_{i=1}^4 \frac{B_{L_i}}{B_{{\bar L}_i}}\right)
\sum_{\bar L}\sqrt{2{\bar L}+1}\,
Z_1(1 {\bar L}_1{\bar L}_2{\bar L}_3{\bar L}_4{\bar L}, LL'; k)
\nonumber \\
& & \quad  \times 
\langle F_{({\bar L}_3 {\bar L}_4){\bar L}{\bar M}}({u_3},{u_4},A',{\vi x})
\vert F_{({\bar L}_1 {\bar L}_2){\bar L}{\bar M}}(u_1,u_2,A,{\vi x})
\rangle,
\end{eqnarray}
where ${\bar L}_i$ is defined by ${\bar L}_i\!=\!L_i\!-\!\delta_{ik}$ 
depending on the summation label $k$. 

\subsection{Many-particle correlation function}

A many-particle correlation function is useful to visualize the 
structure of a 
system~\cite{few.electron,takahashi,nemura1,cluster.correl,exotic}. 
The function is defined by the 
matrix element of the product of $n$ $\delta$-functions $(n\le
N\!-\!1)$, e.g.,  
$\delta({\vi r}_i\!-\!{\vi r}_j\!-\!{\vi d}_1)
\delta({\vi r}_k\!-\!{\vi r}_l\!-\!{\vi d}_2)\cdots=
\delta(\widetilde{w^{(1)}}{\vi x}-{\vi d}_{1})
\delta(\widetilde{w^{(2)}}{\vi x}-{\vi d}_{2})\cdots$.
The basic matrix element is the one between the generating functions:
\begin{eqnarray}
& &\langle g(\lambda_3{\vi e}_3u_3\!+\!\lambda_4{\vi e}_4u_4; A',{\vi x})\vert \prod_{\alpha=1}^n 
\delta(\widetilde{w^{(\alpha)}}{\vi x}-{\vi d}_{\alpha}) \vert
g(\lambda_1{\vi e}_1u_1\!+\!\lambda_2{\vi e}_2u_2; A,{\vi x})\rangle
\nonumber \\
& & \quad =\left({\frac{(2\pi)^{N-1}}{{\rm det}B}}\right)^{\frac{3}{2}}
\left({\frac{{\rm det}{\cal C}}{(2\pi)^n}}\right)^{\frac{3}{2}} 
{\rm exp}\left(\frac{1}{2}\widetilde{\vi v}B^{-1}{\vi v}-\frac{1}{2}
{\vi Q}{\cal C}{\vi Q}\right),
\end{eqnarray}
which is obtained by expressing the $\delta$-functions as
Fourier integrals, and where 
\begin{equation}
({\cal C}^{-1})_{\alpha \beta}=\widetilde{w^{(\alpha)}}
B^{-1}w^{(\beta)},\ \ \ \ \ 
{\vi Q}_{\alpha}=\widetilde{w^{(\alpha)}}B^{-1}{\vi v}-{\vi d}_{\alpha},
\ \ \ \ (\alpha,\beta=1,...,n).
\end{equation}
Omitting the $\lambda_i^2$ terms leads to 
\begin{equation}
\frac{1}{2}\widetilde{\vi v}B^{-1}{\vi v}-\frac{1}{2}
{\vi Q}{\cal C}{\vi Q}\longrightarrow
\sum_{j>i=1}^4{\bar{\varrho}}_{ij}\lambda_i\lambda_j{\vi e}_i\cdot{\vi e}_j
+\sum_{i=1}^4\lambda_i{\vi e}_i\cdot{\vi D}_i-\frac{1}{2}
{\widetilde{\vi d}}{\cal C}{\vi d},
\end{equation}
where
${\vi d}$ is a 
one-column matrix consisting of ${\vi d}_1,...,{\vi d}_n$, and 
\begin{equation}
{\bar{\varrho}}_{ij}=\rho_{ij}-(\widetilde{\Gamma}{\cal C}\Gamma)_{ij},\ \ \ \ \ 
{\vi D}_i=(\widetilde{\Gamma}{\cal C}{\vi d})_{i}
\label{DEFRHOD}
\end{equation}
with  
\begin{equation} 
\Gamma_{\alpha i}=\widetilde{w^{(\alpha)}}B^{-1}u_i, 
\ \ \ \ \ \ (\alpha=1,...,n;\ i=1,...,4).
\end{equation}

Following the same procedure as in Appendix B.3 (cf. Eq.~(\ref{me.delta})) 
leads to 
\begin{eqnarray}
& &\langle  F_{(L_3L_4)L'M'}({u_3},{u_4},A',{\vi x}) \vert \prod_{\alpha=1}^n 
\delta(\widetilde{w^{(\alpha)}}{\vi x}-{\vi d}_{\alpha}) 
\vert F_{(L_1L_2)LM}(u_1,u_2,A,{\vi x}) \rangle \nonumber \\
& & \quad = \frac{(-1)^{L_1+L_2+L+M+L'+M'}}{\sqrt{2L'\!+\!1}}
\left(\prod_{i=1}^{4}{B_{L_i}}\right)
\left({\frac{(2\pi)^{N-1}}{{\rm det}B}}\right)^{\frac{3}{2}}
\left({\frac{{\rm det}{\cal C}}{(2\pi)^n}}\right)^{\frac{3}{2}} 
{\rm exp}\left(-{\frac{1}{2}}{\widetilde{\vi d}}{\cal C}{\vi d}\right)
\nonumber \\
& &\quad  \times \sum_{p_{ij}}\left(\prod_{j>i=1}^4
{\frac{(-1)^{p_{ij}}\sqrt{2p_{ij}\!+\!1}}{B_{p_{ij}}}}
({\bar{\varrho}}_{ij})^{p_{ij}}\right)
\sum_{q_i}\left(\prod_{i=1}^4
{\frac{(-1)^{q_i}}{B_{q_i}}}\right)
\nonumber \\
& & \quad \times \,
\sum_{\kappa \mu \mu' \ell m} \langle LM \ell m \vert L' M' \rangle 
\sqrt{2\ell\!+\!1}X(p_{13}p_{14}p_{23}p_{24};\kappa)
\nonumber \\
& &\quad \times \  
Y(p_{12}p_{12}p_{34}p_{34}\,\, p_{13}\!+\!p_{14}\,\, p_{23}\!+\!p_{24}
\,\, p_{13}\!+\!p_{23}\,\, p_{14}\!+\!p_{24}\, 0\kappa \kappa 0;\kappa \kappa)
\nonumber \\
& & \quad \times \ 
Y(L_{1}\!-\!q_{1}\,L_{2}\!-\!q_{2}\, 
L_{3}\!-\!q_{3}\,L_{4}\!-\!q_{4}\,q_1q_2q_3
q_4\kappa \mu \mu' \ell; LL')
\nonumber \\
& & \quad \times \ 
[[{\cal Y}_{q_1}({\vi D}_1) {\cal Y}_{q_2}({\vi D}_2)]_{\mu}
 [{\cal Y}_{q_3}({\vi D}_3) 
{\cal Y}_{q_4}({\vi D}_4)]_{{\mu}^{\prime}}]_{\ell\, -m}.
\label{MTDELTAFN}
\end{eqnarray}
The values of $p_{ij}$ and $q_i$ are restricted by the condition 
of Eq.~(\ref{pqsum}). 

Note that the matrix element of a many-body force is readily evaluated 
using Eq~(\ref{MTDELTAFN}).

\bigskip

\section{Integral transform of the correlated Gaussian basis}
\label{app.C}

\subsection{Integral transform of the generating function}
\label{app.C1}

An integral transform of the CG 
is needed to adapt nonlocal RGM kernels 
into a variational calculation. A typical form of the kernel 
we consider is 
\begin{equation}
K({\vi r}',{\vi r})=\exp\left(-{\frac{1}{2}}p'{\vi r}'^2-
{\frac{1}{2}}p{\vi r}^2-q{\vi r}'\cdot{\vi r}\right),
\label{def.K}
\end{equation}
where ${\vi r}$ is one of the relative distance vectors,  
say, between particles $k$ and $l$. Thus ${\vi r}$ denotes  
$\widetilde{w}{\vi x}$ (see Eq.~(\ref{def.rel.dist.vec})). 
The vector ${\vi r}'$ is defined in exactly the same way as ${\vi r}$. 

In order to calculate the integral transform acted by $K$ on the 
basis function (\ref{cg}), 
we start from its action on the generating function
\begin{equation}
[Kg]({\vi x})
=\int d{\vi r}\, K({\vi r}',{\vi r})g({\vi s}; A, {\vi x}). 
\label{def.kg}
\end{equation}
The  operator $K$ acting on the function $g$ 
changes it to a new function $Kg$, which is again a 
function of ${\vi x}$ because ${\vi r}'$ denotes ${\widetilde{w}{\vi x}}$.  

The first step of obtaining $[Kg]({\vi x})$ 
is to transform the coordinate set ${\vi x}$ to 
a set of coordinates 
${\vi y}\!=\!\{{\vi y}_1,...,{\vi y}_{N-1}\}$:
\begin{equation}
{\vi y}_i=\sum_{k=1}^{N-1}(w^{(i)})_k{\vi x}_k=
\widetilde{w^{(i)}}{\vi x},\ \ \ \ \ w^{(1)}=w.
\label{transf}
\end{equation}
The vector $w^{(i)}$ is chosen in such a way that the 
first coordinate ${\vi y}_1$ reduces to 
$\widetilde{w}{\vi x}\!=\!{\vi r}_k\!-\!{\vi r}_l$. 
The choice of other relative coordinates 
$\{{\vi y}_2,...,{\vi y}_{N-1}\}$, namely 
$w^{(i)}\, (i\!=\!2, \ldots, N\!-\!1)$, is not unique. 
Their choice is, however, subject to the 
condition that any of $\{{\vi y}_2,...,{\vi y}_{N-1}\}$ must be 
independent of ${\vi y}_1$, that is, it 
may contain ${\vi r}_k$ and ${\vi r}_l$ only as a 
combination of $(m_k{\vi r}_k\!+\!m_l{\vi r}_l)/(m_k\!+\!m_l)$ but not 
as ${\vi r}_k$ or ${\vi r}_l$ alone; otherwise 
the integration over ${\vi r}\!=\!{\vi r}_k\!-\!{\vi r}_l$ in Eq.~(\ref{def.kg}) cannot be performed as it is meant. 
The coordinate transformation (\ref{transf}) 
from ${\vi x}$ to ${\vi y}$ 
is expressed as ${\vi y}\!=\!T^{-1}{\vi x}$ by the matrix $T^{-1}$: 
\begin{equation}
T^{-1}=\left(
\begin{array}{c}
\widetilde{w^{(1)}} \\
\widetilde{w^{(2)}} \\
\vdots \\
\widetilde{w^{(N-1)}} \\
\end{array}
\right).
\label{def.Tinv}
\end{equation}
The transformation from ${\vi y}$ to ${\vi x}$, ${\vi x}\!=\!T{\vi y}$, 
reads as 
\begin{equation}
T=\left(\zeta^{(1)}\ \zeta^{(2)}\cdots \zeta^{(N-1)}
\right),\ \ \ \ \ \zeta^{(1)}=\zeta.
\label{def.T}
\end{equation}
Since $T^{-1}T\!=\!TT^{-1}\!=\!1$, we have  
\begin{equation}
\widetilde{w^{(i)}}\zeta^{(j)}=\widetilde{\zeta^{(i)}}w^{(j)}=\delta_{ij},\ \ \ \ \ 
\sum_{k=1}^{N-1}\zeta^{(k)}_i w^{(k)}_j=\delta_{ij}.
\label{TTinv}
\end{equation}
The second equation is rewritten in a matrix form as  
\begin{equation}
\sum_{k=2}^{N-1}\zeta^{(k)}\widetilde{w^{(k)}}=1-\zeta
\widetilde{w}, \ \ \ \ {{\rm or}}\ \ \ \ 
\sum_{k=2}^{N-1}w^{(k)}\widetilde{\zeta^{(k)}}=1-w \widetilde{\zeta}.
\label{completeness}
\end{equation}

The next step is to substitute ${\vi x}\!=\!T{\vi y}$ to 
$g$ and to separate the part depending on ${\vi y}_1$:
\begin{eqnarray}
g({\vi s}; A, {\vi x})
&=&g(\widetilde{T}{\vi s};\widetilde{T}AT,{\vi y})
\nonumber \\
&=&{\rm exp}\left(-{\frac{1}{2}}a{\vi y}_1^2
+{\vi t}\!\cdot\!{\vi y}_1-\widetilde{a^{(1)}}{\vi y}^{(1)}\cdot
{\vi y}_1 \right)
g({\vi t}^{(1)}; A^{(1)}, {\vi y}^{(1)}),
\label{grel.x-y}
\end{eqnarray}
where we introduce short-hand notations for 
the matrix $\widetilde{T}AT$ and  
the vectors, $\widetilde{T}{\vi s}$ and ${\vi y}$, by 
\begin{equation}
\widetilde{T}AT
=\left(
\begin{array}{cc}
a & \widetilde{a^{(1)}} \\
a^{(1)} & A^{(1)} \\
\end{array}
\right),\ \ \ \ \ 
\widetilde{T}{\vi s}
=\left(
\begin{array}{c}
{\vi t} \\
{\vi t}^{(1)} \\
\end{array}
\right),\ \ \ \ \ 
{\vi y}
=\left(
\begin{array}{c}
{\vi y}_1 \\
{\vi y}^{(1)} \\
\end{array}
\right).
\label{def.param}
\end{equation}
Here $a^{(1)}$ is an $(N\!-\!2)$-dimensional column vector, 
$A^{(1)}$ an $(N\!-\!2)\!\times \!(N\!-\!2)$ symmetric matrix, and 
${\vi t}^{(1)}$ is an $(N\!-\!2)$-dimensional column vector whose 
element is a 3-dimensional vector. More explicitly they are given by 
\begin{eqnarray}
& &a= \widetilde{\zeta}A\zeta,\ \ \ \ \ 
a^{(1)}_i=\widetilde{\zeta^{(i+1)}}A\zeta,\ \ \ \ \  
A^{(1)}_{ij}=\widetilde{\zeta^{(i+1)}}A\zeta^{(j+1)},
\nonumber \\
& &
{\vi t}=\widetilde{\zeta}{\vi s},\ \ \ \ \ 
{\vi t}^{(1)}_i=\widetilde{\zeta^{(i+1)}}{\vi s}, 
\label{def.at}
\end{eqnarray}
where $i,j\!=\!1,2,\ldots, N\!-\!2$. 
Substituting Eqs.~(\ref{grel.x-y}) and (\ref{def.K}) 
into Eq.~(\ref{def.kg}), integrating over 
${\vi r}\!=\!{\vi y}_1$ and then renaming ${\vi r}'\!=\!{\vi y}_1$ lead to 
\begin{equation}
[Kg]({\vi x})
=\left(\frac{2\pi}{a+p}\right)^{\frac{3}{2}}
\exp\left(\frac{1}{2(a+p)}{\vi t}^2\right)
g(\bar{\vi s}; \bar{A}, {\vi y}),
\label{int.kg}
\end{equation}
where
\begin{equation}
\bar{A}
=\left(
\begin{array}{cc}
p'-\frac{q^2}{a+p} & -\frac{q}{a+p}\widetilde{a^{(1)}} \\
-\frac{q}{a+p}a^{(1)} & A^{(1)}-\frac{1}{a+p}a^{(1)}\widetilde{a^{(1)}} \\
\end{array}
\right),\ \ \ \ \ 
\bar{\vi s}
=\left(
\begin{array}{c}
-\frac{q}{a+p}{\vi t} \\
{\vi t}^{(1)} -\frac{1}{a+p}{\vi t}a^{(1)} \\
\end{array}
\right).
\end{equation}

The last step is to express the above function 
$g(\bar{\vi s}; \bar{A}, {\vi y})$ in terms of ${\vi x}$. 
This is achieved by replacing ${\vi y}$ with $T^{-1}{\vi x}$ (see 
Eqs.~(\ref{trans.cg}) and (\ref{aup.relation})): 
\begin{equation}
g(\bar{\vi s}; \bar{A}, {\vi y})=g(\widetilde{T^{-1}}\bar{\vi s}; 
\widetilde{T^{-1}}\bar{A}T^{-1}, {\vi x})
=g(Q_K{\vi s}; A_K, {\vi x}),
\label{gfn.tmp}
\end{equation}
where the matrices $Q_K$ and $A_K$ turn out to be given by
\begin{eqnarray}
&&Q_K=1-\frac{p+q}{a+p}w\widetilde{\zeta}
-\frac{1}{a+p}A\zeta\widetilde{\zeta},
\nonumber \\
&&A_K=A+\Big(a+p'-\frac{(a-q)^2}{a+p}\Big)w\widetilde{w}
-\frac{p+q}{a+p}\Big(w\widetilde{\zeta}A+
                      A\zeta\widetilde{w}\Big) 
                      -\frac{1}{a+p}A\zeta\widetilde{\zeta}A.
\label{def.q.a}
\end{eqnarray}
Substitution of Eqs.~(\ref{def.at}) and (\ref{gfn.tmp}) 
in Eq.~(\ref{int.kg}) 
leads to the desired result  
\begin{equation}
[Kg]({\vi x})=\left(\frac{2\pi}{a+p}\right)^{\frac{3}{2}}
\exp\left(\frac{1}{2(a+p)}
\widetilde{\vi s}\zeta\widetilde{\zeta}{\vi s}\right)
g(Q_K{\vi s}; A_K, {\vi x}).
\label{g.actedby.K}
\end{equation}
Using this result (\ref{g.actedby.K}) with 
${\vi s}\!=\!\lambda_1{\vi e}_1u_1\!+\!\lambda_2{\vi e}_2u_2$ 
in Eq.~(\ref{gfn}), we can derive 
\begin{eqnarray}
& &[KF_{(L_1L_2)LM}(u_1,u_2,A)]({\vi x})\! =\! \left(\frac{2\pi}{a+p}\right)^{\frac{3}{2}}
\sum_{\ell=0}^{\ell_M}{\cal K}(L_1 L_2 L; \ell)\!
\left(\frac{\widetilde{u_1}\zeta \widetilde{\zeta}u_2}
{a+p}\right)^{\ell}
\nonumber \\
& &\qquad \qquad \qquad \qquad \qquad \qquad 
\times \, F_{(L_1-\ell \, L_2-\ell)LM}(Q_Ku_1,Q_Ku_2,A_K,{\vi x}),
\label{glvactedK}
\end{eqnarray}
where $\ell_M\!=\!{\rm min}(L_1,L_2,[({L_1+L_2-L})/{2}])$ and 
${\cal K}(L_1 L_2 L; \ell)$ is defined in Eq.~(\ref{def.cal.K}). 
The action of the integral kernel on the CG 
with two GV leads to that of the 
same kinds. In particular we obtain a very simple result for 
the CG of types of  Eqs.~(\ref{cg1}) and (\ref{cg2}):   
\begin{eqnarray}
& &[KF_{(L0)LM}(u_1,u_2,A)]({\vi x})\!=\!
\left(\frac{2\pi}{a\!+\!p}\right)^{\frac{3}{2}}
\!F_{(L0)LM}(Q_Ku_1,Q_Ku_2,A_K,{\vi x}), \nonumber \\
& &[KF_{(L1)LM}(u_1,u_2,A)]({\vi x})\!=\!
\left(\frac{2\pi}{a\!+\!p}\right)^{\frac{3}{2}}
\!F_{(L1)LM}(Q_Ku_1,Q_Ku_2,A_K,{\vi x}).
\end{eqnarray}

Apparently both $[Kg]({\vi x})$ and 
$[KF_{(L_1L_2)LM}(u_1,u_2,A)]({\vi x})$ should not depend on 
the choice of $w^{(i)}\, (i\!=\!2,\ldots,N\!-\!1)$. This is equivalent 
to the statement that 
both $Q_K$ and $A_K$ are independent of the choice of 
$w^{(i)}\, (i\!=\!2,\ldots,N\!-\!1)$. For this to hold true, it is 
sufficient to show that $\zeta$ is independent of that choice. 
In order to prove this, 
let $w'^{(i)}\, (i\!=\!2,\ldots,N\!-\!1)$ denote a set of other choice. 
They are 
related to the original set $w^{(i)}\, (i\!=\!2,\ldots,N\!-\!1)$ by 
a non-singular linear transformation $W$ as 
\begin{equation}
\left(w'^{(2)}\cdots w'^{(N-1)}\right)=\left(w^{(2)}\cdots w^{(N-1)}\right)W.
\label{coord.trans}
\end{equation}
The coordinate transformation from ${\vi x}$ to ${\vi y}'$ 
(${\vi y}'_i\!=\!\widetilde{w'^{(i)}}{\vi x}$) is 
performed by a matrix $T'^{-1}$:  
\begin{equation}
T'^{-1}=\left(
\begin{array}{c}
\widetilde{w} \\
\widetilde{w'^{(2)}} \\
\vdots \\
\widetilde{w'^{(N-1)}} \\
\end{array}
\right). 
\label{def.Tprime}
\end{equation}
Using Eqs.~(\ref{def.Tinv}), (\ref{coord.trans}) and (\ref{def.Tprime}) 
enables one to show 
$T'^{-1}=\left(
\begin{array}{cc}
1 & 0 \\
0 & \widetilde{W} \\
\end{array}
\right)T^{-1}$,
leading to 
\begin{equation}
T'=T\left(
\begin{array}{cc}
1 & 0 \\
0 & \widetilde{W}^{-1} \\
\end{array}
\right).
\end{equation}
This equation indicates that the first 
column of the matrix $T'$ is the same as that of $T$, namely 
$\zeta$, so that $\zeta$ remains the same against 
any choice of $W$. 

\subsection{Determination of $\zeta$}

The vector $\zeta$ is determined by the condition 
$\widetilde{w^{(i)}}\zeta\!=\!\delta_{i1}$. See Eq.~(\ref{TTinv}). 
As shown below, it is possible to determine 
$\zeta$ without specifying $w^{(i)}\, (i\!=\!2,\ldots,N\!-\!1)$. 

Any of $\{{\vi y}_2,...,{\vi y}_{N-1}\}$ is expressed as a linear 
combination of ${\vi r}_i$ as 
\begin{equation}
\sum_{i=1}^N a_i{{\vi r}_i},
\label{arb.vector}
\end{equation}
where $a_i$ may be taken arbitrarily provided that they satisfy 
the following conditions
\begin{equation}
a_k=\frac{m_k}{m_k+m_l}\lambda,\ \ \ \ \ 
a_l=\frac{m_l}{m_k+m_l}\lambda,\ \ \ \ \ 
\sum_{i=1}^{N}a_i=0.
\label{akal}
\end{equation}
The last equation assures that ${\vi y}_i$, as one of the 
relative coordinates, 
has no dependence on the center of mass coordinate. 
The $\lambda$ value is determined as 
\begin{equation}
\lambda=-\sum_{i\ne k,l}a_i.
\label{par.lambda}
\end{equation}
Substitution of ${\vi r}_i=\sum_{j=1}^N (U^{-1})_{ij}{\vi x}_j$ (see 
Eq.~(\ref{def.matU})) 
enables one to express the vector (\ref{arb.vector}) 
in terms of the coordinate $\vi x$ as 
\begin{equation}
\sum_{i=1}^N a_i{{\vi r}_i}=\sum_{j=1}^N (\widetilde{U^{-1}}a)_j 
{\vi x}_j,
\end{equation}
which indicates that the $j$-th element of $w^{(i)}$ is given by 
$(\widetilde{U^{-1}}a)_j$. 
The condition to determine $\zeta$ is thus expressed as 
\begin{equation}
\sum_{j=1}^{N-1}w_j\zeta_j
=\sum_{j=1}^{N-1}\Big[(U^{-1})_{kj}-(U^{-1})_{lj}\Big]\zeta_j=1,\ \ \ \ \ 
\sum_{j=1}^{N-1} (\widetilde{U^{-1}}a)_j \zeta_j=0.
\label{cond.zeta}
\end{equation}
Using Eqs.~(\ref{akal}) and
(\ref{par.lambda}) reduces the second equation in 
Eq.~(\ref{cond.zeta}) to 
\begin{eqnarray}
& &\sum_{i=1}^N \sum_{j=1}^{N-1} (U^{-1})_{ij}\zeta_j a_i
=\sum_{i\ne k,l}(U^{-1}\zeta)_ia_i+(U^{-1}\zeta)_ka_k
+(U^{-1}\zeta)_la_l
\nonumber \\
& & =\sum_{i\ne k,l}\left[(U^{-1}\zeta)_i
-\frac{m_k}{m_k+m_l}(U^{-1}\zeta)_k
-\frac{m_l}{m_k+m_l}(U^{-1}\zeta)_l\right]a_i=0.
\end{eqnarray}
Since $a_i\, (i\ne k,l)$ can be taken arbitrarily, 
the coefficients of $a_i$ must vanish, that is, 
\begin{equation}
\sum_{j=1}^{N-1}\left[(U^{-1})_{ij}
-\frac{m_k}{m_k+m_l}(U^{-1})_{kj}
-\frac{m_l}{m_k+m_l}(U^{-1})_{lj}\right]\zeta_j=0\ \ \ \ \ 
{\rm for} \ {i\ne k,l}.
\end{equation}
The solution of the first equation in Eq.~(\ref{cond.zeta}) 
together with 
the above equation determines $\zeta$.

\subsection{${\vi \ell}^2$-dependent potential of Gaussian radial form}

As a useful application of the decomposition of Eq.~(\ref{grel.x-y}), we 
calculate the matrix element of an ${\vi \ell}^2$-dependent potential of 
Gaussian radial form, where the angular momentum ${\vi \ell}$ is 
defined as ${\vi \ell}\!=\!{\vi y}_1\!\times \!(-i{\hbar})
{\partial}/{\partial {\vi y}_1}$ with 
${\vi y}_1\!=\!\widetilde{w}{\vi x}\!=\!{\vi r}_k\!-\!{\vi r}_l$.
Using Eq.~(\ref{grel.x-y}) we obtain 
\begin{eqnarray}
{\vi \ell}\, g({\vi s}; A, {\vi x})
=\frac{\hbar}{i}\widetilde{w}{\vi x}\times \left({\vi
t}-\widetilde{a^{(1)}}{\vi y}^{(1)}\right)\, g({\vi s}; A, {\vi x})
=\frac{\hbar}{i}\widetilde{w}{\vi x} \times (\widetilde{\zeta}
{\vi s}-\widetilde{\zeta}A{\vi x})\, g({\vi s}; A, {\vi x}),
\end{eqnarray}
where use is made of ${\vi t}\!=\!\widetilde{\zeta}{\vi s}$ and 
$\widetilde{a^{(1)}}{\vi y}^{(1)}\!=\!\widetilde{\zeta}A{\vi x}-a
\widetilde{w}{\vi x}$ in the last step. 

Using the property of ${\vi \ell }^2\!=\!{\vi \ell}
\cdot{\vi \ell}\!=\!{\vi \ell}^{\dagger}
\cdot{\vi \ell}$ and the fact that $\vi \ell$ commutes with a scalar  
function leads to an expression for the 
matrix element between the generating functions as follows:
\begin{eqnarray}
&&\langle g({\vi s}'; A',{\vi x})
\vert \exp\left\{-\frac{1}{2}c'(\widetilde{w}{\vi x})^2\right\}
{\vi \ell }^2 \vert g({\vi s}; A,{\vi x})\rangle \nonumber \\
& &=\hbar^2 \langle g({\vi s}'; A',{\vi x})
\vert \exp\left\{-\frac{1}{2}c'(\widetilde{w}{\vi x})^2\right\}
{\cal L}^2
\vert g({\vi s}; A,{\vi x})\rangle 
\label{lsquarepot}
\end{eqnarray}
with
\begin{eqnarray}
{\cal L}^2&=&\left(\widetilde{w}{\vi x} \times (\widetilde{\zeta}
{\vi s}'-\widetilde{\zeta}A'{\vi x})\right)\cdot 
\left(\widetilde{w}{\vi x} \times (\widetilde{\zeta}
{\vi s}-\widetilde{\zeta}A{\vi x})\right)
\nonumber \\
&=&\widetilde{\vi x}w\widetilde{w}{\vi x}
\left(\widetilde{{\vi s}'}\zeta\widetilde{\zeta}{\vi s}
-\widetilde{{\vi s}'}\zeta\widetilde{\zeta}A{\vi x}
-\widetilde{{\vi s}}\zeta\widetilde{\zeta}A'{\vi x}
+\widetilde{\vi x}A'\zeta \widetilde{\zeta}A{\vi x}\right)
\nonumber \\
& &\!\!\!\!\!-\, 
\big(\widetilde{{\vi s}'}\zeta\widetilde{w}{\vi x}-
\widetilde{{\vi x}}A'\zeta\widetilde{w}{\vi x}\big)
\big(\widetilde{{\vi s}}\zeta\widetilde{w}{\vi x}-
\widetilde{{\vi x}}A\zeta\widetilde{w}{\vi x}\big).
\end{eqnarray}
The integration in Eq.~(\ref{lsquarepot}) is elementary. It reduces 
to the following result 
\begin{eqnarray}
&\langle & \!\!\!\!\!\!\!\!\! 
g({\lambda}_3{\vi e}_3 u_3\!+\!{\lambda}_4{\vi e}_4 u_4; A',{\vi x})
\vert \exp\left\{-\frac{1}{2}c'(\widetilde{w}{\vi x})^2\right\}{\vi \ell}^2 
\vert g({\lambda}_1{\vi e}_1 u_1\!+\!{\lambda}_2{\vi e}_2 u_2; A,{\vi x})
\rangle \nonumber \\
&\longrightarrow &\!\!\!\! \hbar^2\left(\frac{c}{c+c'}\right)^{\frac{3}{2}}
\left({\frac{(2\pi)^{N-1}}{{\rm det}B}}\right)^{\frac{3}{2}}
{\rm exp}\left(\sum_{j>i=1}^4\left[{\rho}_{ij}-
\frac{cc'}{c+c'}\gamma_i\gamma_j\right]
\lambda_i\lambda_j {\vi e}_i\cdot{\vi e}_j \right)
\nonumber \\
&\times & \!\!\!\!\!\!\!
\left(L^{(0)}+2\sum_{j>i=1}^4L^{(1)}_{ij}\lambda_i\lambda_j 
{\vi e}_i\cdot{\vi e}_j 
+4\sum_{j>i=1}^4 \sum_{l>k=1}^4 L^{(2)}_{ij,kl}
\lambda_i\lambda_j {\vi e}_i\cdot{\vi e}_j 
\lambda_k\lambda_l {\vi e}_k\cdot{\vi e}_l
\right).
\label{lsquare.gg.me}
\end{eqnarray}
Here the constants $L^{(0)}$,  $L^{(1)}_{ij}$ and $L^{(2)}_{ij,kl}$ 
are 
determined through $w$, $\zeta$, $u_1$, $u_2$, $u_3$, $u_4$, 
$A$, $A'$, $c'$. Their expressions are lengthy and not written here. 
Substitution of Eq.~(\ref{lsquare.gg.me}) into
Eq.~(\ref{grandformula}) makes it possible to relate the 
desired matrix element to the overlap matrix element by 
\begin{eqnarray}
& &\langle F_{(L_3L_4)LM}({u_3},{u_4},A',{\vi x})\vert
\exp\Big\{-\frac{1}{2}c'(\widetilde{w}{\vi x})^2\Big\}{\vi \ell}^2
\vert F_{(L_1L_2)LM}(u_1,u_2,A,{\vi x})\rangle 
\nonumber \\
&& = \hbar^2 \left(\frac{c}{c+c'}\right)^{\frac{3}{2}} 
\! \left(L^{(0)}+2\sum_{j>i=1}^4L^{(1)}_{ij}\frac{\partial}{\partial \rho_{ij}}	
+4\sum_{j>i=1}^4 \sum_{l>k=1}^4 L^{(2)}_{ij,kl}\frac{\partial^2}{\partial \rho_{ij}
\partial \rho_{kl}}
\right)
\nonumber \\
&& \quad \times 
\langle F_{(L_3L_4)LM}({u_3},{u_4},A',{\vi x})
\vert F_{(L_1L_2)LM}(u_1,u_2,A,{\vi x})\rangle\Big|_{\rho_{ij}
\to \rho_{ij}-\frac{cc'}{c+c'}\gamma_i\gamma_j}.
\end{eqnarray}

\bigskip

\section{Matrix elements of nonlocal operators}
\label{app.D}

\subsection{Nonlocal operator matrix element for 
the generating function}
\label{app.D1}

To prove Eq.~(\ref{gndformula}), we start from the matrix element of 
the nonlocal operator~(\ref{def.Vnl}) for the generating function. 
With use of Eq.~(\ref{grel.x-y}) on the bra-ket generating functions, we 
obtain~\cite{takahashi} 
\begin{eqnarray}
& &\langle g({\vi s}'; A',{\vi x}) \mid V_{{\rm NL}}\mid 
g({\vi s}; A,{\vi x}) \rangle
\!\!=\!\!\int\!\int  d{\vi r}'d{\vi r}V({\vi r}',{\vi r})\,
{\rm exp}\left(-{\frac{1}{2}}a'{\vi r}'^2
+{\vi t}'\!\cdot\!{\vi r}'-{\frac{1}{2}}a{\vi r}^2
+{\vi t}\!\cdot\!{\vi r} \right)
\nonumber \\
& &\qquad \qquad \qquad \times \mid {\rm det}T\mid ^3
\left(\frac{(2\pi)^{N-2}}{{\rm det}(A^{(1)}+A'^{(1)})}
\right)^{\frac{3}{2}}
{\rm exp}\left(\frac{1}{2}\widetilde{\vi d}(A^{(1)}+A'^{(1)})^{-1}
{\vi d}\right),
\label{gfn.VNL.me}
\end{eqnarray}
where $|{\rm det}T|^3$ is the Jacobian, 
${\rm det}(\partial{\vi x}/\partial{\vi y})$, for the 
transformation from ${\vi x}$ to ${\vi y}$, and 
\begin{equation}
{\vi d}={\vi t}^{(1)}-a^{(1)}{\vi r}
+{\vi t}'^{(1)}-a'^{(1)}{\vi r}'=
\left(
\begin{array}{c}
\widetilde{\zeta^{(2)}} \\
\widetilde{\zeta^{(3)}} \\
\vdots \\
\widetilde{\zeta^{(N-1)}} \\
\end{array}
\right)
{\vi D},
\end{equation}
with 
\begin{equation}
{\vi D}={\vi s}+{\vi s}'-A\zeta{\vi r}-A'\zeta{\vi r}'.
\end{equation}
Here $a', \, a'^{(1)}, \, A'^{(1)}, \, {\vi t}', \, {\vi t}'^{(1)}$ 
are 
defined in exactly the same way as the corresponding quantities 
given in Eq.~(\ref{def.at}) by replacing $A$ and ${\vi s}$ with 
$A'$ and ${\vi s}'$. 

The matrix element (\ref{gfn.VNL.me}) 
must be independent of the choice of 
$(w^{(2)},\ldots,w^{(N-1)})$ from its construction. 
This is indeed assured by a theorem 
\begin{eqnarray}
& &{\rm det}(A^{(1)}+A'^{(1)})=\frac{1}{c}\,
({\rm det}T)^2 {\rm det}B,\nonumber \\
& &\widetilde{\vi d}(A^{(1)}+A'^{(1)})^{-1}{\vi d}
=\widetilde{\vi D}J^{-1}{\vi D},
\label{theorem}
\end{eqnarray}
where $c$ and $J^{-1}$ are given in Eq.~(\ref{def.c.and.J}). 
Substituting these results in Eq.~(\ref{gfn.VNL.me}) leads to 
the result (\ref{gndformula}), which is manifestly independent 
of the choice of $w^{(i)}$.

\subsection{Proof of a theorem~(\ref{theorem})}

With a slight change of notation the above theorem is stated as follows. \\

\noindent
({Theorem}) Let $A$ be an $(N\!-\!1)\!\times\!(N\!-\!1)$ 
real, symmetric, positive-definite matrix and 
$T$ be an $(N\!-\!1)\!\times\!(N\!-\!1)$ non-singular matrix given 
by Eq.~(\ref{def.T}).  Let a matrix $\widetilde{T}AT$ be decomposed into  
\begin{equation}
\widetilde{T}AT
=\left(
\begin{array}{cc}
a & \widetilde{a^{(1)}} \\
a^{(1)} & A^{(1)} \\
\end{array}
\right),
\label{def.TAT}
\end{equation}
where $a$, $a^{(1)}$ and $A^{(1)}$ are defined by Eq.~(\ref{def.at}). 
The theorem reads 
\begin{eqnarray}
& &{\rm det}A^{(1)}=\frac{1}{\alpha} ({\rm det}T)^2 {\rm det}A,
\label{theorem1} \\
& &\left(\zeta^{(2)} \zeta^{(3)}\cdots \zeta^{(N-1)}\right)(A^{(1)})^{-1}
\left(
\begin{array}{c}
\widetilde{\zeta^{(2)}} \\
\widetilde{\zeta^{(3)}} \\
\vdots \\
\widetilde{\zeta^{(N-1)}} \\
\end{array}
\right)
=A^{-1}-\alpha A^{-1}w\widetilde{w}A^{-1},
\label{theorem2}
\end{eqnarray}
with
\begin{equation}
\alpha=({\widetilde{w}A^{-1}w})^{-1},
\end{equation}
and a row vector $\widetilde{w}$ is the first row of $T^{-1}$ as defined in 
Eq.~(\ref{def.Tinv}). \\

\noindent
({Proof}) The heart of this theorem lies in that the quantities 
on the left sides of Eqs.~(\ref{theorem1}) and (\ref{theorem2}) 
are determined by $w$ alone. The theorem was proved 
by one of the authors (Y.S.)~\cite{takahashi}.  
We here show its proof as the application 
of nonlocal operators becomes increasingly important. 

We can make use of the following identity
\begin{equation}
\left(
\begin{array}{cc}
a & \widetilde{a^{(1)}} \\
a^{(1)} & A^{(1)} \\
\end{array}
\right)
\left(
\begin{array}{cc}
1 & -\frac{1}{a}\widetilde{a^{(1)}} \\
0 & 1 \\
\end{array}
\right)
=\left(
\begin{array}{cc}
a & 0 \\
a^{(1)} & A^{(1)}-\frac{1}{a}a^{(1)}\widetilde{a^{(1)}} \\
\end{array}
\right),
\label{proof.1}
\end{equation}
because $a\!=\!\widetilde{\zeta}A\zeta$ is not zero 
thanks to the positive-definiteness of $A$. 
Taking the determinant of the above equation, we have
\begin{equation}
{\rm det}
\left(
\begin{array}{cc}
a & \widetilde{a^{(1)}} \\
a^{(1)} & A^{(1)} \\
\end{array}
\right)=({\rm det} T)^2 {\rm det}A
=a\,{\rm det}\left(A^{(1)}-\frac{1}{a}a^{(1)}\widetilde{a^{(1)}}\right).
\label{proof.2}
\end{equation}
Using the Sherman-Morrison formula~\cite{book,recipes} reduces 
the right side of Eq.~(\ref{proof.2}) to 
\begin{equation}
a\left(1-\frac{1}{a}\widetilde{a^{(1)}}(A^{(1)})^{-1}a^{(1)}\right)
{\rm det}A^{(1)}.
\label{aux.proof.2}
\end{equation}

To calculate $x\!=\!\widetilde{a^{(1)}}(A^{(1)})^{-1}a^{(1)}$, 
we invert Eq.~(\ref{proof.1}), obtaining 
\begin{equation}
\left(
\begin{array}{cc}
1 & \frac{1}{a}\widetilde{a^{(1)}} \\
0 & 1 \\
\end{array}
\right)
T^{-1}A^{-1}{\widetilde{T}}^{-1}
=\left(
\begin{array}{cc}
\frac{1}{a} & 0 \\
-\frac{1}{a}\left(A^{(1)}-\frac{1}{a}a^{(1)}\widetilde{a^{(1)}}
\right)^{-1}a^{(1)}\  & \left(A^{(1)}-\frac{1}{a}a^{(1)}
\widetilde{a^{(1)}}\right)^{-1} \\
\end{array}
\right).
\label{proof.3}
\end{equation}
Let $G$ denote an $(N\!-\!2)\!\times\!(N\!-\!2)$ matrix 
which is obtained by removing the first row and first column from 
the above matrix. Comparing both sides of Eq.~(\ref{proof.3}) 
leads to  
\begin{eqnarray}
G&\equiv& \left(
\begin{array}{c}
\widetilde{w^{(2)}} \\
\widetilde{w^{(3)}} \\
\vdots \\
\widetilde{w^{(N-1)}} \\
\end{array}
\right)A^{-1}
\left(w^{(2)} w^{(3)}\cdots w^{(N-1)}\right)
\nonumber \\
& =& \left(A^{(1)}-\frac{1}{a}a^{(1)}
\widetilde{a^{(1)}}\right)^{-1}
=(A^{(1)})^{-1}+\frac{1}{a-x}(A^{(1)})^{-1}a^{(1)}
\widetilde{a^{(1)}}(A^{(1)})^{-1},
\label{proof.4}
\end{eqnarray}
where the Sherman-Morrison formula is used in the last step.  
Multiplying $\widetilde{a^{(1)}}$ from the left and $a^{(1)}$ from 
the right, we obtain 
\begin{equation}
\widetilde{a^{(1)}}Ga^{(1)}=x+\frac{x^2}{a-x}
=\frac{ax}{a-x}.
\label{proof.5}
\end{equation}
Using the definition of $G$ and Eq.~(\ref{def.at}) for $a^{(1)}$ 
makes it possible to obtain 
\begin{eqnarray}
\widetilde{a^{(1)}}Ga^{(1)}=\widetilde{\zeta}A(1-\zeta\widetilde{w})
A^{-1}(1-{w}\widetilde{\zeta})A\zeta=a^2\widetilde{w}A^{-1}w-a
=\frac{a^2}{\alpha}-a
\end{eqnarray}
owing to Eqs.~(\ref{TTinv}) and (\ref{completeness}). 
Thus we obtain $x$=$a\!-\!\alpha$. 
From Eqs.~(\ref{proof.2}) and (\ref{aux.proof.2}), we have 
${\rm det}A^{(1)}$=$({1}/(a\!-\!x)) ({\rm det}T)^2 {\rm det}A$, 
which is the first equality (\ref{theorem1}). 

To prove the second equality (\ref{theorem2}), we start from 
Eq.~(\ref{proof.4}), yielding the equation   
$G^{-1}$=$A^{(1)}\!-\!\frac{1}{a}a^{(1)}\widetilde{a^{(1)}}$. 
Inverting $A^{(1)}$ with use of the Sherman-Morrison formula leads to 
\begin{equation}
(A^{(1)})^{-1}=
(G^{-1}+\frac{1}{a}a^{(1)}\widetilde{a^{(1)}})^{-1}=
G-\frac{1}{a+\widetilde{a^{(1)}}Ga^{(1)}}
Ga^{(1)}\widetilde{a^{(1)}}G,
\end{equation}
where $a\!+\!\widetilde{a^{(1)}}Ga^{(1)}\!=\!{a^2}/
{\alpha}$. The left side of Eq.~(\ref{theorem2}) is thus expressed as 
\begin{eqnarray}
& &\left(\zeta^{(2)} \zeta^{(3)}\cdots\zeta^{(N-1)}\right)(A^{(1)})^{-1}
\left(
\begin{array}{c}
\widetilde{\zeta^{(2)}} \\
\widetilde{\zeta^{(3)}} \\
\vdots \\
\widetilde{\zeta^{(N-1)}} \\
\end{array}
\right)=\left(\zeta^{(2)} \zeta^{(3)}\cdots\zeta^{(N-1)}\right)G
\left(
\begin{array}{c}
\widetilde{\zeta^{(2)}} \\
\widetilde{\zeta^{(3)}} \\
\vdots \\
\widetilde{\zeta^{(N-1)}} \\
\end{array}
\right)
\nonumber \\
& &\qquad \qquad -\frac{\alpha}{a^2}
\left(\zeta^{(2)} \zeta^{(3)}\cdots\zeta^{(N-1)}\right)
Ga^{(1)}\widetilde{a^{(1)}}G
\left(
\begin{array}{c}
\widetilde{\zeta^{(2)}} \\
\widetilde{\zeta^{(3)}} \\
\vdots \\
\widetilde{\zeta^{(N-1)}} \\
\end{array}
\right).
\label{proof.6}
\end{eqnarray}
By using Eqs.~(\ref{TTinv}) and (\ref{completeness}) 
together with the definition of $a$ and $\alpha$,  
the first and second terms on the right side of Eq.~(\ref{proof.6}) 
turn out to be  
\begin{eqnarray}
& &(1-\zeta\widetilde{w})A^{-1}(1-w\widetilde{\zeta})=
A^{-1}-\zeta\widetilde{w}A^{-1}-A^{-1}w\widetilde{\zeta}+
\frac{1}{\alpha}\zeta\widetilde{\zeta},
\nonumber \\
& &-\frac{\alpha}{a^2}
(1-\zeta\widetilde{w})A^{-1}(1-w\widetilde{\zeta})A
\zeta\widetilde{\zeta}A(1-\zeta\widetilde{w})A^{-1}(1-w\widetilde{\zeta})
\nonumber \\
&& \quad \ = -\frac{\alpha}{a^2}
\left(A^{-1}-\zeta\widetilde{w}A^{-1}-A^{-1}w\widetilde{\zeta}+
\frac{1}{\alpha}\zeta\widetilde{\zeta}\right)
A\zeta\widetilde{\zeta}A
\left(A^{-1}-\zeta\widetilde{w}A^{-1}-A^{-1}w\widetilde{\zeta}+
\frac{1}{\alpha}\zeta\widetilde{\zeta}\right)
\nonumber \\
&&\quad \ = -\alpha\left(\frac{1}{\alpha^2}\zeta\widetilde{\zeta}
-\frac{1}{\alpha}A^{-1}w\widetilde{\zeta}
-\frac{1}{\alpha}\zeta\widetilde{w}A^{-1}
+A^{-1}w\widetilde{w}A^{-1}
\right).
\end{eqnarray}
Putting together these results completes the second equality 
(\ref{theorem2}).

\subsection{Examples of nonlocal kernels}

We here collect the matrix elements of typical nonlocal operators which 
appear in the RGM treatment for light nuclei. See Eq.~(\ref{def.Vnl})
for the definition of the nonlocal operator. 
\noindent
\begin{flushleft}
\underline{(i)\ \ $V({\vi r}',{\vi r})=\exp\left(-{\frac{1}{2}}p'{\vi r}'^2-
{\frac{1}{2}}p{\vi r}^2-q{\vi r}'\cdot{\vi r}\right)$}
\end{flushleft}
The nonlocal operator is nothing but $K({\vi r}',{\vi r})$ of
Eq.~(\ref{def.K}). Utilizing 
Eq.~(\ref{glvactedK}) we obtain 
\begin{eqnarray}
& &\langle F_{(L_3L_4)L'M'}({u_3},{u_4},A',{\vi x}) 
\mid V_{{\rm NL}} \mid F_{(L_1 L_2)LM}(u_1,u_2,A,{\vi x})
 \rangle 
\nonumber \\
& & \, =
\langle F_{(L_3L_4)L'M'}({u_3},{u_4},A',{\vi x}) \mid 
[KF_{(L_1 L_2)LM}(u_1,u_2,A)]({\vi x}) \rangle.
\label{Kme}
\end{eqnarray}
As another method, it is possible to start from Eq.~(\ref{gndformula}), 
yielding 
\begin{eqnarray}
& &\langle g({\lambda}_3{\vi e}_3 u_3\!+\!{\lambda}_4{\vi e}_4 u_4; A',{\vi x})
\vert V_{{\rm NL}} 
\vert g({\lambda}_1{\vi e}_1 u_1\!+\!{\lambda}_2{\vi e}_2 u_2; A,{\vi x})
\rangle \nonumber \\
& &\, = 
 \left({\frac{(2\pi)^{N-2}c}{{\rm det}B}}\right)^{\frac{3}{2}}
\left({\frac{(2\pi)^{2}}{{\rm det}M}}\right)^{\frac{3}{2}}
\, {\rm exp}\left({\frac{1}{2}}\tilde{\vi v}J^{-1}{\vi v}
\!+\!{\frac{1}{2}}\widetilde{\vi Z}M^{-1}{\vi Z}\right),
\end{eqnarray}
where $M$ is a $2\!\times \!2$ symmetric matrix defined by
\begin{equation}
M= 
\left(
\begin{array}{cc}
\alpha+p & \beta+q \\
\beta+q & \alpha'+p' \\
\end{array}
\right)
\label{def.matrixM}
\end{equation}
with $\alpha, \alpha', \beta$ being given in Eq.~(\ref{def.alpha.beta}) 
and ${\vi Z}$ is a 2-dimensional vector whose elements are 
usual vectors defined by 
\begin{equation}
{\vi Z}_1=(\widetilde{\zeta}-\widetilde{\eta}){\vi s}
-\widetilde{\eta}{\vi s}'=\sum_{i=1}^4 \lambda_i{\vi e}_if_{1}^{(i)}, 
\ \ \ \ \ 
{\vi Z}_2=(\widetilde{\zeta}-\widetilde{\eta'}){\vi s}'
-\widetilde{\eta'}{\vi s}=\sum_{i=1}^4 \lambda_i{\vi e}_if_{2}^{(i)}.
\label{def.ZZ}
\end{equation}
Here $f_{1}^{(i)}$ and $f_{2}^{(i)}$ are components of a 
2-dimensional vector $f^{(i)}$:
\begin{equation}
f^{(i)}=
\left(
\begin{array}{c}
\widetilde{\zeta}u_{i}-\widetilde{\eta}u_{i} \\
-\widetilde{\eta'}u_{i}\\
\end{array}
\right)\ \ {\rm for \ }i=1,2, \ \ \ \ \ \ 
f^{(i)}=
\left(
\begin{array}{c}
-\widetilde{\eta}u_{i} \\
\widetilde{\zeta}u_{i}-\widetilde{\eta'}u_{i}\\
\end{array}
\right)\ \ {\rm for \ }i=3,4.
\end{equation}
The desired matrix element is expressed in terms 
of the overlap matrix element (\ref{overlapme}) as 
\begin{eqnarray}
& &\langle F_{(L_3L_4)L'M'}({u_3},{u_4},A',{\vi x}) 
\mid V_{{\rm NL}} \mid F_{(L_1 L_2)LM}(u_1,u_2,A,{\vi x})
 \rangle 
\nonumber \\
& & = 
\left({\frac{2\pi c}{{\rm det}M}}\right)^{\frac{3}{2}} 
 \langle F_{(L_3L_4)L'M'}({u_3},{u_4},A',{\vi x}) 
\mid F_{(L_1 L_2)LM}(u_1,u_2,A,{\vi x})\rangle \Big\vert_{{\rho}_{ij} \to
\chi_{ij}},
\label{Kme.2}
\end{eqnarray}
where
\begin{equation}
\chi_{ij}=\widetilde{u_i}J^{-1}u_j+ \widetilde{f^{(i)}}M^{-1}f^{(j)}
=\bar{\rho}_{ij}+\widetilde{f^{(i)}}M^{-1}f^{(j)}.
\end{equation}

As a straightforward application of the above formula, we consider 
a nonlocal operator  
$V({\vi r}',{\vi r})$=$\{{\rm erf}({\tau \mid \sigma {\vi r}
\!+\!\sigma'{\vi r}'\mid} )/{\mid \sigma {\vi r}\!+\!\sigma'{\vi r}'\mid}\}$ $
\exp\left(-p'{\vi r}'^2/2\!-\!
p{\vi r}^2/2\!-\!q{\vi r}'\cdot{\vi r}\right)$ which often 
appears as the RGM kernel for the Coulomb potential. 
Using the relation~(\ref{mod.coulomb}), 
the operator is expressed as an integral of the nonlocal 
operator of type (i), so that its matrix element reduces
to Eq.~(\ref{Kme}) with the replacement of $p,\, p', \,q$ by 
\begin{equation}
p\to p+2\sigma^2\tau^2t^2,\ \ \ \ \ 
p'\to p'+2{\sigma}'^2\tau^2t^2,\ \ \ \ \ 
q\to q+2\sigma \sigma'\tau^2t^2.
\end{equation}
\noindent
\begin{flushleft}
\underline{(ii)\ \ $V({\vi r}',{\vi r})=\Big({\cal Q}_{11}{\vi r}^2+{\cal Q}_{22}{{\vi r}'}^2+2{\cal Q}_{12}{\vi
 r}\cdot{\vi r}'\Big)\exp\left(-{\frac{1}{2}}p'{\vi r}'^2-
{\frac{1}{2}}p{\vi r}^2-q{\vi r}'\cdot{\vi r}\right)$}
\end{flushleft}
Using Eq.~(\ref{gndformula}) leads to the following expression
\begin{eqnarray}
& &\langle g({\lambda}_3{\vi e}_3 u_3\!+\!{\lambda}_4{\vi e}_4 u_4; A',{\vi x})
\vert V_{{\rm NL}} 
\vert g({\lambda}_1{\vi e}_1 u_1\!+\!{\lambda}_2{\vi e}_2 u_2; A,{\vi x})
\rangle \nonumber \\
& &\, = \Big(3{\rm Tr}M^{-1}{\cal Q}+\widetilde{\vi Z}M^{-1}{\cal Q}M^{-1}{\vi Z}\Big) \nonumber \\ 
& & \, \times \left({\frac{(2\pi)^{N-2}c}{{\rm det}B}}\right)^{\frac{3}{2}}
\left({\frac{(2\pi)^{2}}{{\rm det}M}}\right)^{\frac{3}{2}}
\, {\rm exp}\left({\frac{1}{2}}\tilde{\vi v}J^{-1}{\vi v}
\!+\!{\frac{1}{2}}\widetilde{\vi Z}M^{-1}{\vi Z}\right),
\end{eqnarray}
with
\begin{equation}
{\cal Q}=
\left(
\begin{array}{cc}
{\cal Q}_{11} & {\cal Q}_{12} \\
{\cal Q}_{12} & {\cal Q}_{22} \\
\end{array}
\right).
\end{equation}
We can follow the steps of Eqs.~(\ref{gme}), (\ref{kin.gme}) and 
(\ref{kineticme}), and find that 
the desired matrix element is given by 
\begin{eqnarray}
& &\langle F_{(L_3L_4)L'M'}({u_3},{u_4},A',{\vi x}) 
\mid V_{{\rm NL}} \mid F_{(L_1 L_2)LM}(u_1,u_2,A,{\vi x})
 \rangle 
\nonumber \\
& & = 
\left({\frac{2\pi c}{{\rm det}M}}\right)^{\frac{3}{2}} 
\Big(3{\rm Tr}M^{-1}{\cal Q}+2\sum_{j>i=1}^4h_{ij}\frac{\partial}{\partial
\rho_{ij}}\Big)
\nonumber \\
& & \ \times \langle F_{(L_3L_4)L'M'}({u_3},{u_4},A',{\vi x}) 
\mid F_{(L_1 L_2)LM}(u_1,u_2,A,{\vi x})\rangle \Big\vert_{{\rho}_{ij} \to \chi_{ij}},
\end{eqnarray}
where
\begin{equation}
h_{ij}=\widetilde{f^{(i)}}M^{-1}{\cal Q}M^{-1}f^{(j)}.
\end{equation}

\par\noindent
\begin{flushleft}
\underline{(iii)\ \ $V({\vi r}',{\vi r})=(-1)^{\ell}\sqrt{2\ell +1}
[{\cal Y}_{\ell }({\vi r}'){\cal Y}_{\ell }({\vi r})]_{00}
\exp\left(-{\frac{1}{2}}p'{\vi r}'^2
-{\frac{1}{2}}p{\vi r}^2\right)$}
\end{flushleft}
As the last example, we consider a separable kernel 
which often appears in eliminating redundant states from the relative
motion between the particles. Expressing $[{\cal Y}_{\ell }({\vi r}')
{\cal Y}_{\ell }({\vi r})]_{00}$ as 
\begin{equation}
[{\cal Y}_{\ell }({\vi r}')
{\cal Y}_{\ell }({\vi r})]_{00}=
\left(\frac{B_{\ell}}{\ell !}\right)^2\int\!\int d{\vi e}'d{\vi e}
[Y_{\ell}({\vi e}')Y_{\ell}({\vi e})]_{00}
\frac{\partial ^{2\ell}}{\partial \lambda^{\ell} \lambda'^{\ell}}
\exp(\lambda {\vi e}\cdot{\vi r}+\lambda' {\vi e}'\cdot{\vi r}')
\Big\vert_{\lambda=\lambda'=0},
\end{equation}
we start from the 
matrix element between the generating functions 
\begin{eqnarray}
& &\langle g({\lambda}_3{\vi e}_3 u_3\!+\!{\lambda}_4{\vi e}_4 u_4; A',{\vi x})
\vert V_{{\rm NL}} 
\vert g({\lambda}_1{\vi e}_1 u_1\!+\!{\lambda}_2{\vi e}_2 u_2; A,{\vi x})
\rangle \nonumber \\
& &\, = (-1)^{\ell}\sqrt{2\ell +1}
\left(\frac{B_{\ell}}{\ell !}\right)^2\int\!\int d{\vi e}'d{\vi e}
[Y_{\ell}({\vi e}')Y_{\ell}({\vi e})]_{00}\frac{\partial ^{2\ell}}
{\partial \lambda^{\ell} \lambda'^{\ell}}
\nonumber \\
& &\ \  \times 
\left({\frac{(2\pi)^{N-2}c}{{\rm det}B}}\right)^{\frac{3}{2}}
\left({\frac{(2\pi)^{2}}{{\rm det}M_0}}\right)^{\frac{3}{2}}
\, {\rm exp}\left({\frac{1}{2}}\tilde{\vi v}J^{-1}{\vi v}
\!+\!{\frac{1}{2}}\widetilde{\vi W}M_0^{-1}{\vi W}\right)\Big\vert_{\lambda=\lambda'=0}
\nonumber \\
& &\, \longrightarrow 
(-1)^{\ell}\sqrt{2\ell +1}
\left(\frac{B_{\ell}}{\ell !}\right)^2
\left({\frac{(2\pi)^{N-1}}{{\rm det}B}}\right)^{\frac{3}{2}}
\left({\frac{2\pi c}{{\rm det}M_0}}\right)^{\frac{3}{2}}
\exp\left(\sum_{j>i=1}^4\lambda_i\lambda_j\chi_{ij}{\vi e}_i\cdot{\vi e}_j
\right) 
\nonumber \\
& & \ \  \times 
\int\!\int d{\vi e}'d{\vi e}
[Y_{\ell}({\vi e}')Y_{\ell}({\vi e})]_{00}\frac{\partial ^{2\ell}}
{\partial \lambda^{\ell} \lambda'^{\ell}}
\nonumber \\
& & \ \  \times  \exp\left(\sum_{i=1}^4 \lambda \lambda_i \chi_{\, 1}^{(i)}
{\vi e}\cdot{\vi e}_i+\sum_{i=1}^4 \lambda'\lambda_i \chi_{\, 2}^{(i)}
{\vi e}'\cdot{\vi e}_i+(M_0^{-1})_{12}\lambda \lambda'{\vi e}
\cdot{\vi e}'
\right)\Big\vert_{\lambda=\lambda'=0},
\end{eqnarray}
where $M_0$ is obtained by putting $q\!=\!0$ in $M$ of 
Eq.~(\ref{def.matrixM}), and 
2-component vectors $\chi^{(i)}$ and ${\vi W}$ are defined by 
\begin{equation}
\chi^{(i)}=M_0^{-1}f^{(i)}\ \ \ (i=1,\dots,4),\ \ \ \ \ 
{\vi W}=
\left(
\begin{array}{c}
{\vi Z}_1+\lambda{\vi e}  \\
{\vi Z}_2+\lambda'{\vi e}' \\
\end{array}
\right),
\end{equation}
with ${\vi Z}_1$ and ${\vi Z}_2$ being defined in Eq.~(\ref{def.ZZ}).

The desired matrix element is obtained from Eq.~(\ref{grandformula}). 
To simplify the needed calculation, we introduce the following notations
\begin{eqnarray}
& &L_5=L_6=\ell,\ \ \ \ \ 
{\vi e}_5={\vi e},\ \ \ \ \ {\vi e}_6={\vi e}',\ \ \ \ \ 
\lambda_5=\lambda,\ \ \ \ \ \lambda_6=\lambda',
\nonumber \\
& &\chi_{i5}=\chi_{\, 1}^{(i)},\ \ \ \ \ \chi_{i6}=\chi_{\, 2}^{(i)},\ \ \ \ \
 \chi_{56}=(M_0^{-1})_{12},
\end{eqnarray}
which make it possible to extend Eq.~(\ref{grandformula}) to 
\begin{eqnarray}
& &\langle F_{(L_3L_4)L'M'}({u_3},{u_4},A',{\vi x}) 
\mid V_{{\rm NL}} \mid F_{(L_1 L_2)LM}(u_1,u_2,A,{\vi x})
 \rangle 
\nonumber \\
& &= (-1)^{\ell}\sqrt{2\ell +1}
\left(\prod_{i=1}^{6}{\frac{B_{L_i}}{L_i!}}\right)
\left({\frac{(2\pi)^{N-1}}{{\rm det}B}}\right)^{\frac{3}{2}}
\left({\frac{2\pi c}{{\rm det}M_0}}\right)^{\frac{3}{2}}
\nonumber \\
& &\ \times \int\!\cdots\!\int \prod_{i=1}^6\, d{\vi e}_i
([Y_{L_3}({\vi e}_3) Y_{L_4}({\vi e}_4)]_{L'M'})^{*}
[Y_{L_1}({\vi e}_1)  Y_{L_2}({\vi e}_2)]_{LM} 
[Y_{L_5}({\vi e}_5)Y_{L_6}({\vi e}_6)]_{00}
\nonumber \\
& &\ \times \left(\prod_{i=1}^6 {\frac{\partial^{L_i}}{
 \partial{{\lambda}_i}^{L_i}}}\right)\, 
\exp\left(\sum_{j>i=1}^6\lambda_i\lambda_j\chi_{ij}{\vi e}_i\cdot{\vi e}_j
\right)\Big|_{\lambda_i=0}.
\label{nl.sepa.me}
\end{eqnarray}
The above operation can be performed similarly 
to the overlap case, yielding a formula 
\begin{eqnarray}
& &\langle F_{(L_3L_4)L'M'}({u_3},{u_4},A',{\vi x}) 
\mid V_{{\rm NL}} \mid F_{(L_1 L_2)LM}(u_1,u_2,A,{\vi x})
 \rangle 
\nonumber \\
\nonumber \\
& &= \delta_{L L'}\delta_{M M'}(-1)^{L_1+L_2+\ell}\sqrt{\frac{2\ell +1}{2L+1}}
\left(\prod_{i=1}^{6}B_{L_i}\right)
\left({\frac{(2\pi)^{N-1}}{{\rm det}B}}\right)^{\frac{3}{2}}
\left({\frac{2\pi c}{{\rm det}M_0}}\right)^{\frac{3}{2}}
\nonumber \\
& & \, \times \sum_{p_{ij}}\left(\prod_{j>i=1}^6
{\frac{(-1)^{p_{ij}}\sqrt{2p_{ij}\!+\!1}}{B_{p_{ij}}}}
(\chi_{ij})^{p_{ij}}\right)
\nonumber \\
& & \, \times 
\sqrt{\frac{(2L\!+\!1)(2L_5\!+\!1)}{(2p_{56}\!+\!1)(2L_5\!-\!2p_{56}\!+\!1)}}
\left[
\begin{array}{ccc}
p_{12}    &  p_{12}  &  0  \\
L_1\!-\!p_{12}   &  L_2\!-\!p_{12}   & L \\
L_1 & L_2 &  L  
\end{array}\right]
\left[
\begin{array}{ccc}
p_{34}    &  p_{34}  &  0  \\
L_3\!-\!p_{34}   &  L_4\!-\!p_{34}   & L \\
L_3 & L_4 &  L  
\end{array}\right]
\nonumber \\
& & \, \times \, C(p_{12}\ L_1\!-\!p_{12};L_1)\, 
C(p_{13}\!+\!p_{14}\ p_{15}\!+\!p_{16};L_1\!-\!p_{12})
\nonumber \\
& & \, \times \, C(p_{12}\ L_2\!-\!p_{12};L_2)\, 
C(p_{23}\!+\!p_{24}\  p_{25}\!+\!p_{26}; L_2\!-\!p_{12})
\nonumber \\
& & \, \times \, C(p_{34}\ L_3\!-\!p_{34};L_3)\, C(p_{13}\!+\!p_{23}\ 
 p_{35}\!+\!p_{36};L_3\!-\!p_{34})
\nonumber \\
& & \, \times \, C(p_{34}\  L_4\!-\!p_{34};L_4)\, C(p_{14}\!+\!p_{24}\ 
 p_{45}\!+\!p_{46};L_4\!-\!p_{34})
\nonumber \\
& & \, \times \, C(p_{56}\  L_5\!-\!p_{56};L_5)\, C(p_{15}\!+\!p_{25}\ 
 p_{35}\!+\!p_{45};L_5\!-\!p_{56})
\nonumber \\
& & \, \times \, C(p_{56}\  L_6\!-\!p_{56};L_6)\, C(p_{16}\!+\!p_{26}\ 
 p_{36}\!+\!p_{46};L_6\!-\!p_{56})
\nonumber \\
& & \, \times 
\sum_{\lambda \lambda'}\frac{1}{\sqrt{(2\lambda\!+\!1)(2\lambda'\!+\!1)^3}}
X(p_{13}p_{14}p_{23}p_{24};\lambda)X(p_{15}p_{16}p_{25}p_{26};\lambda')
X(p_{35}p_{36}p_{45}p_{46};\lambda')
\nonumber \\
& & \, \times 
\left[
\begin{array}{ccc}
p_{13}\!+\!p_{14}  &  p_{23}\!+\!p_{24}  &  \lambda  \\
p_{15}\!+\!p_{16}  &  p_{25}\!+\!p_{26}  & \lambda' \\
L_1\!-\!p_{12}     &  L_2\!-\!p_{12}     &  L  
\end{array}\right]
\left[
\begin{array}{ccc}
p_{13}\!+\!p_{23}  &  p_{14}\!+\!p_{24}  &  \lambda  \\
p_{35}\!+\!p_{36}  &  p_{45}\!+\!p_{46}  & \lambda' \\
L_3\!-\!p_{34}     &  L_4\!-\!p_{34}     &  L  
\end{array}\right]
\nonumber \\
& & \, \times 
\left[
\begin{array}{ccc}
p_{15}\!+\!p_{25}  &  p_{16}\!+\!p_{26}  & \lambda'  \\
p_{35}\!+\!p_{45}  &  p_{36}\!+\!p_{46}  & \lambda' \\
L_5\!-\!p_{56}     &  L_6\!-\!p_{56}     &  0  
\end{array}\right],
\end{eqnarray}
where non-negative integers $p_{ij}$ satisfy the following equations
\begin{eqnarray}
& &p_{12}+p_{13}+p_{14}+p_{15}+p_{16}=L_1,\ \ \ \ \ 
   p_{12}+p_{23}+p_{24}+p_{25}+p_{26}=L_2,
\nonumber \\
& &p_{13}+p_{23}+p_{34}+p_{35}+p_{36}=L_3,\ \ \ \ \ 
   p_{14}+p_{24}+p_{34}+p_{45}+p_{46}=L_4,
\nonumber \\
& &p_{15}+p_{25}+p_{35}+p_{45}+p_{56}=L_5,\ \ \ \ \ 
   p_{16}+p_{26}+p_{36}+p_{46}+p_{56}=L_6.
\end{eqnarray}
There are 15 $p_{ij}$, which satisfy 6 conditions. Thus 
we have in general 9 independent $p_{ij}$ in contrast to 2 
in the overlap case. In the case of 
natural parity states, however, because of $L_2\!=\!L_4\!=\!0$ 
only two of $p_{ij}$, e.g., $p_{13}$ and $p_{15}$ are independent.

\begin{acknowledge}
This work was in part supported by a Grant for Promotion of Niigata
 University Research Projects (2005-2007), and a Grant-in Aid for
 Scientific Research for Young Scientists (No. 19$\cdot$3978).
One of the authors (Y.S.) thanks the JSPS core-to-core program 
(Exotic Femto Systems) for partial support and the Institute 
for Nuclear Theory at the University of Washington for its 
hospitality which enabled him to have useful discussions during 
the INT Workshop on Correlations in Nuclei, November 26-29, 2007.

\end{acknowledge}

\end{document}